\documentclass[numbers=enddot]{scrreprt}
\usepackage{titlesec}
\usepackage{manfnt}

\usepackage{graphicx}
\usepackage{color} 
\usepackage[usenames,dvipsnames]{xcolor}
\usepackage{url}
\usepackage{geometry}
\usepackage[colorlinks=true,filecolor=red,linkcolor=blue,urlcolor=cyan, hyperfootnotes=false,unicode=true]{hyperref}
\hypersetup{linktocpage}
\usepackage[all]{hypcap}
\usepackage{amsmath}

\definecolor{blue}{rgb}{0,0,1}
\definecolor{red}{rgb}{1,0,0.2}
\definecolor{green}{rgb}{0,0.5,0.1}
\begin{document}

\def\la{\mathrel{\mathchoice {\vcenter{\offinterlineskip\halign{\hfil
$\displaystyle##$\hfil\cr<\cr\sim\cr}}}
{\vcenter{\offinterlineskip\halign{\hfil$\textstyle##$\hfil\cr
<\cr\sim\cr}}}
{\vcenter{\offinterlineskip\halign{\hfil$\scriptstyle##$\hfil\cr
<\cr\sim\cr}}}
{\vcenter{\offinterlineskip\halign{\hfil$\scriptscriptstyle##$\hfil\cr
<\cr\sim\cr}}}}}
\def\ga{\mathrel{\mathchoice {\vcenter{\offinterlineskip\halign{\hfil
$\displaystyle##$\hfil\cr>\cr\sim\cr}}}
{\vcenter{\offinterlineskip\halign{\hfil$\textstyle##$\hfil\cr
>\cr\sim\cr}}}
{\vcenter{\offinterlineskip\halign{\hfil$\scriptstyle##$\hfil\cr
>\cr\sim\cr}}}
{\vcenter{\offinterlineskip\halign{\hfil$\scriptscriptstyle##$\hfil\cr
>\cr\sim\cr}}}}}

\newcommand{\rd}{{\rm d}}
\newcommand{\beq}{\begin{equation}}
\newcommand{\eeq}{\end{equation}}
\newcommand{\bea}{\begin{eqnarray}}
\newcommand{\eea}{\end{eqnarray}}
\newcommand{\parr}{\parallel}
\newcommand{\pa}{\partial}
\newcommand{\vp}{\varpi}
\newcommand{\vh}{\varphi}
\newcommand{\nb}{\bs{\nabla}}
\newcommand{\cs}{c_{\rm s}}
\newcommand{\va}{v_{\rm A}}
\newcommand{\curb}{\bs{\nb}{\bf\times B} }  
\newcommand{\bs}{\boldsymbol}
\newcommand{\rmd}{{\rm d}}
\newcommand{\ncd}{\kern -1pt\cdot\kern -2pt }
\newcommand{\ecd}{\kern -2pt\cdot\kern -1pt }
\newcommand{\pb}{}
\newcommand{\can}{\color{Red}} 
\newcommand{\cpr}{\color{green}}
\newcommand{\bk}{\color{black}}

\pagenumbering{gobble}
\centerline{\Large\bf Essential magnetohydrodynamics for astrophysics}

\vspace{0.5cm}
\centerline{H.C. Spruit} 
\centerline{Max Planck Institute for Astrophysics}
\centerline{\tt henk$@\,$mpa-garching.mpg.de} 
\vspace{2cm}
\centerline{v3.0, May 2016}

\vspace{2cm}
\noindent
The most recent version of this text, including small animations of elementary MHD processes, is located at \url{http://www.mpa-garching.mpg.de/~henk/mhd12.zip} (25 MB). 
 
\bigskip
\leftline{\bf Links}
To navigate the text, use the bookmarks bar of your pdf viewer, and/or the links highlighted in color. Links to pages, sections and equations are in {\color{blue}blue}, those to the animations {\can{red}}, external links such as urls in {\color{cyan} cyan}. When using these links, you will need a way of returning to the location where you came from. This depends on your pdf viewer, which typically does not provide an html-style backbutton. On a Mac,  the key combination cmd-[  works with Apple Preview and Skim, cmd-left-arrow with Acrobat. In Windows and Unix Acrobat has a way to add a back button to the menu bar. The appearance of the animations depends on the default video viewer of your system. On the Mac,  the animation links work well with Acrobat Reader, Skim and with the default pdf viewer in the Latex distribution (tested  with VLC and Quicktime Player), but not with Apple preview. 

\pagenumbering{roman}

\tableofcontents
\vfill\eject

\pagenumbering{arabic}

{\large\bf Introduction}

\vspace{0.5cm}

This text is intended as an introduction to magnetohydrodynamics in astrophysics, emphasizing 
a fast path to the elements essential for physical understanding. It assumes experience with concepts 
from fluid mechanics\,: the fluid equation of motion and the Lagrangian and Eulerian descriptions of 
fluid flow\footnote{For an introduction to fluid mechanics  \hyperlink{land}{Landau \& Lifshitz} is recommended}. In addition, the basics of vector calculus and elementary special relativity are needed. Not much knowledge of electromagnetic theory is required. In fact, since MHD is much closer in spirit to fluid mechanics than to electromagnetism, an important part of the learning curve is to overcome intuitions based on the vacuum electrodynamics of one's high school days.

The first chapter (only 39 pp)  is meant as a practical introduction. This is the `essential' part.  The  exercises included are important as illustrations of the points made in the text (especially the less intuitive ones). Almost all are mathematically unchallenging. The supplement in chapter 2 contains further explanations,
more specialized topics and connections to the occasional topic somewhat outside MHD. 

The basic astrophysical applications of MHD were developed from the 1950s through the 1980's. The experience with MHD that developed in this way has tended to remain confined to somewhat specialized communities in stellar astrophysics. The advent of powerful tools for numerical simulation of the MHD equations has enabled application to much more realistic astrophysical problems than could be addressed before, making magnetic fields attractive to a wider community. In the course of this numerical development, familiarity with the basics of MHD appears to have suffered somewhat.

This text aims to show how MHD can be used more convincingly when armed with
a good grasp of its intrinsic power and peculiarities, as distinct from those of vacuum
electrodynamics or plasma physics. The emphasis is on physical understanding by the visualization of MHD processes, as opposed to more formal approaches. This allows one to formulate good questions more quickly, and to interpret computational results more meaningfully. For more comprehensive introductions to astrophysical MHD, see \hyperlink{park1}{Parker (1979)}, \hyperlink{kuls}{Kulsrud (2005)} and \hyperlink{mest}{Mestel (2012)}.

In keeping with common astrophysical practice Gaussian units are used.

\vfill\pagebreak
\chapter{Essentials}\label{Essen}
Magnetohydrodynamics describes electrically conducting fluids\footnote{~In astrophysics `fluid' is used as a generic  term for a gas, liquid or plasma} in which a magnetic field is present. A high electrical conductivity is ubiquitous in astrophysical objects. Many astrophysical  phenomena are influenced by the presence of magnetic fields, or even explainable only in terms of magnetohydrodynamic processes. The  atmospheres  of planets are an exception. Much of the intuition we have for ordinary earth-based fluids is relevant for MHD as well, but  more theoretical experience is needed to develop a feel for what is specific to MHD. The aim of  this text is  to provide the means to develop this intuition, illustrated with a number of simple examples and warnings for common pitfalls. 

\section{Equations}
\label{Eqs}
\subsection{The MHD approximation}\label{MHDapp}
The equations of magnetohydrodynamics are a reduction of the equations of fluid mechanics coupled with Maxwell's equations. Compared with plasma physics in general, MHD is a strongly reduced theory. Of the formal apparatus of vacuum electrodynamics with its two EM vector fields, currents and charge densities, MHD can be described with only a single additional vector\,: the magnetic field.  The `MHD approximation' that makes this possible involves some assumptions\,:\\
{\bf 1.} The fluid approximation\,: local thermodynamic quantities can be meaningfully defined in the plasma, and variations in these quantities are slow compared with the time scale of the microscopic processes in the plasma. This is the essential approximation.\\
{\bf 2.} In the plasma there is a local, instantaneous relation between electric field and current density (an `Ohm's law').\\
{\bf 3.} The plasma is electrically neutral.\\
This statement of the approximation is somewhat imprecise. I return to it in some of the supplementary sections of the text (chapter \ref{epicyc}).
The first of the assumptions involves the same approximation as used in deriving the equations of fluid mechanics and thermodynamics from statistical physics.  It is assumed that a sufficiently large number of particles is present so that local fluid properties, such as pressure, density and velocity can be defined.  It is  sufficient that particle distribution functions can be defined properly on the length and time scales of interest.  It is, for example, not necessary that the distribution functions are thermal, isotropic, or in equilibrium, as long as they change sufficiently slowly.

The second assumption can be relaxed. The third is closely related to the second 
(cf.\ {sect.~\ref{chargedd}}). For the moment we consider these as separate.
In {\bf 2.}\  it is assumed that whatever plasma physics processes take place on small scales,  they average out to an instantaneous, mean relation (not necessarily linear) between the local electric field and current density, on the length and time scales of interest.  The third  assumption of electrical neutrality is satisfied in most astrophysical environments, but it excludes near-vacuum conditions such as the magnetosphere of a pulsar  ({sect.~\ref{pulsar}}). 

Electrical conduction, in most cases, is due to the (partial) ionization of a plasma. The degree of ionization needed for  {\bf 2.}\ to hold is generally not large  in astrophysics. The approximation that the density of charge carriers is large enough that the fluid has very little electrical resistance\,: the assumption of {\em perfect conductivity},  is usually a good first step. Exceptions are, for example, pulsar magnetospheres, dense molecular clouds or the atmospheres of planets. 

\subsection{Ideal MHD}\label{ideal}
Consider  the MHD of a perfectly conducting fluid, i.e.\ in the limit of zero electrical resistance. This limit is also called {\em ideal MHD}. Modifications when the conductivity is finite are discussed in {sections \ref{diffusion}} and {\ref{hall-am}}.

The electric field in a perfect conductor is trivial\,: it vanishes, since the electric current would become arbitrarily large if it did not. However, the fluid we are considering is generally in motion. Because of the magnetic field present, the electric field vanishes only in a frame of reference moving with the flow; in any other frame there is an electric field to be accounted for.

Assume the fluid to move with velocity ${\bf v}({\bf r})$ relative to the observer. Let ${\bf E}^\prime$ and ${\bf B}^\prime$ be the electric and magnetic field strengths measured in an instantaneous inertial frame where the fluid is at rest (locally at the point $\bf r$, at time $t$). We call this the {\em comoving} frame or {\em fluid} frame. They are related to the fields ${\bf E}$, ${\bf B}$ measured in the observer's frame  by a Lorentz transformation (e.g.\ Jackson E\&M Chapter 11.10). Let $E_\parr={\bf v\cdot E}/v$ be the component of ${\bf E}$ parallel to the flow, ${\bf E}_\perp={\bf E}-{\bf v\,E}\cdot{\bf v}/v^2$ the perpendicular component of ${\bf E}$, and similar for ${\bf B}$. The transformation is then\footnote{~This transformation is reproduced incorrectly in some texts on MHD.}

\bea
E^\prime_\parr&=&E_\parr,\label{epar}\cr
{\bf E}^\prime_\perp&=&\gamma({\bf E_\perp}+{\bf v\times B}/c),\label{etra}\cr
B^\prime_\parr&=&B_\parr,\cr
{\bf B}^\prime_\perp&=&\gamma({\bf B_\perp}-{\bf v\times E}/c),\label{bperp}
\eea
where $\gamma$ is the Lorentz factor $\gamma=(1-v^2/c^2)^{-1/2}$, and $c$ the speed of light.
By the assumption of infinite conductivity, ${\bf E}^\prime=0$. The electric field measured by the observer then follows from (\ref{epar}) as:
\beq {\bf E}=-{\bf v\times B}/c.\label{ee}\eeq

Actually measuring this electric field would require some planning, since it can be observed only in an environment that is not itself conducting (or else the electric field would be shunted out there as well). The experimenter's electroscope would have to be kept in an insulating environment separated from the plasma. In astrophysical application, this means that electric fields in ideal MHD become physically significant only {\em at a boundary with a non-conducting medium}. The electric fields associated with differential flow speeds {\em within} the fluid are of no consequence, since the fluid elements themselves do not sense them. 

A  useful  assumption is that the magnetic permeability and dielectrical properties of the fluid can be ignored, a good approximation for many astrophysical applications.
This allows the distinction between magnetic field strength and  magnetic induction, and between electric field and displacement to be ignored. This is not an essential assumption. Maxwell's equations are then

\bea 
\label{amp}
4\pi{\bf j}+{\pa{\bf E}/\pa t}&=&~~c{\bs{\nb}\bf\times B}, \\
 {\pa{\bf B}/\pa t}&=&-c{\bs{\nb}\bf\times E},\label{inM} \\
 {\bs{\nb}\bf\cdot E}&=&4\pi\sigma,\label{chargeM} \\
 {\bs{\nb}\bf\cdot B}&=&0, 
\eea
where ${\bf j}$ is the electrical current density and $\sigma$ the charge density.   Taking the divergence of Maxwell's equation (\ref{amp}) yields the conservation of charge\,: 
\beq {\pa\sigma\over\pa t}+{\bs{\nb}\bf\cdot j}=0. \label{charge}\eeq

\subsection{The induction equation}\label{indu}

Using expression (\ref{ee}) for the electric field, the induction equation (\ref{inM}) becomes
\beq {\pa{\bf B}\over\pa t}={\bs{\nb}\bf\times(v\times B)}.\label{ind}\eeq

This is known as the {\em induction equation of ideal MHD}, or MHD induction equation for short. It describes how the magnetic field in a perfectly conducting fluid changes with time under the influence of a velocity field $\bf v$ (see {section \ref{diffusion}} for its extension to cases when conductivity is finite). 

By the MHD approximation made, the original electromagnetic induction equation has changed flavor drastically. From something describing the generation of voltages by changing magnetic fields in coils, it has become an evolution equation for a magnetic field embedded in a fluid flow. For historical reasons, it has retained the name induction equation even though it actually is better understood as something new altogether.
The divergence of (\ref{ind}) yields
\beq {\pa\over\pa t}{\bs{\nb}\kern-1pt\cdot\kern-1pt\bf B}=0. \label{Bn}\eeq
The MHD induction equation thus incorporates the condition ${\bs{\nb}\kern-1pt\cdot\kern-1pt\bf B}=0$. It need not be considered explicitly in solving the MHD equations, except as a requirement to be satisfied by the initial conditions. 

\begin{figure}
\hfil{\includegraphics[width=0.9\hsize]{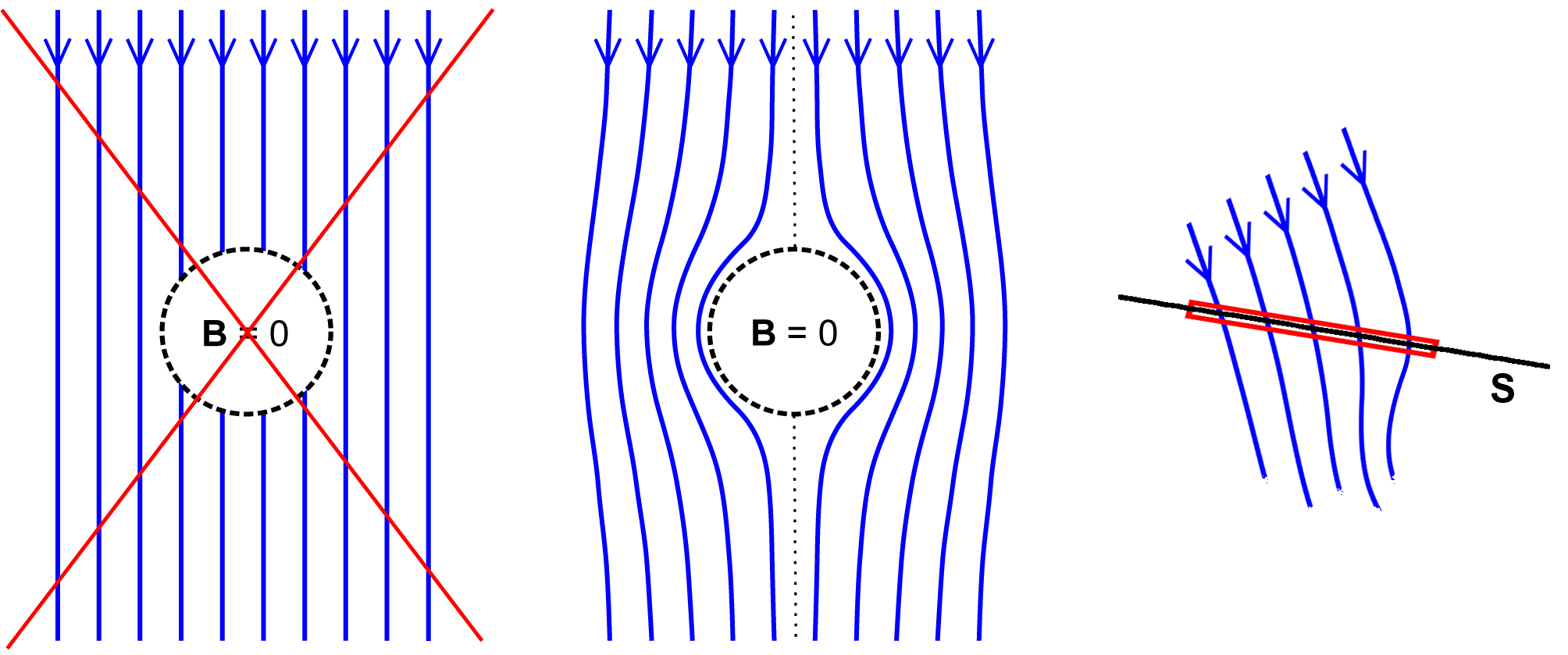}}\hfil
\caption{\label{gap} \small Field lines near a field-free inclusion (dashed contour). The configuration on the left contains field lines ending on the surface of the inclusion. $\bs{\nb}\kern-1pt\cdot{\bf B}=0$ excludes such configurations. Regions of reduced field strength distort the field in their surroundings (middle). Right\,: pill box (red) to derive continuity of $B_{\rm n}$ from $\bs{\nb}\kern-1pt\cdot{\bf B}=0$ .}
\end{figure}

\subsection{Geometrical meaning of{ \rm div}\,${\bf B}=0$} \label{geomm}
\label{divb0}
The field lines of a divergence-free vector such as $\bf B$  (also called a {\em solenoidal} vector) `have no ends'. While electric field lines can be seen as starting and ending on charges, a magnetic field line (a path in space with tangent vector everywhere parallel to the magnetic field vector) wanders around without meeting such monopolar singularities.  As a consequence, regions of reduced field strength cannot be local\,: magnetic field lines must `pass around'  them. This is illustrated in Fig.\ \ref{gap}. In contrast with scalars such as temperature or density, a change in field strength must be accommodated by changes in field line shape and strength in the surroundings (for an example and consequences see {\cpr problem \ref{diamag})}. 

A bit more formally, let $S$ be a surface in the field configuration, at an arbitrary angle to the field lines, and let $\bf n$ be a unit vector normal to $S$. Apply Gauss's theorem to ${\rm div}\, {\bf B}=0$ in a thin box of thickness $\epsilon\rightarrow 0$ oriented parallel to $S$. The integral over the volume being equal to the integral of the normal component over the box then yields that the component of $\bf B$ normal to the surface,
\beq B_n={\bf B\cdot n}\label{bnorm}\eeq
is continuous across  any surface $S$. Hence also across the dashed surface in the left panel of Fig.\,\ref{gap}.

\subsection{Electrical current}
\label{ecur}
Up to this point, the derivation is still valid for arbitrary fluid velocities. In particular the induction equation (\ref{ind}) is valid relativistically, i.e.\ at  all velocities $v<c$ (though only in ideal MHD, not with finite resistivity). We now specialize to the nonrelativistic limit $v\ll c$.  
Quantities of first order in $v/c$ have to be kept in taking the limit, since the electric field is of this order, but higher orders are omitted. Substituting (\ref{ee}) into (\ref{bperp}), one finds that
\beq {\bf B}^\prime={\bf B}[1+{\cal O}(v^2/c^2)], \eeq
i.e.\ the magnetic field strength does not depend on the frame of reference, in the nonrelativistic limit.
Substituting (\ref{ee}) into (\ref{amp}) yields:
\beq 
4\pi{\bf j}-{\pa\over\pa t}({\bf v\times B})/c=c{\bs{\nb}\bf\times B}. \label{a0}
\eeq
The second term on the left, the displacement current, can be ignored if $v\ll c$. To see this, choose a length scale $L$ that is of interest for the phenomenon to be studied. Then $\vert{\bs{\nb}\bf\times B}\vert$ is of the order $B/L$. Let $V$ be a typical value of the fluid velocities relevant for the problem; the typical time scales of interest are then of order $\tau=L/V$. An upper limit to the displacement current for such length and time scales is thus of order $\vert {\bf v\times B}/\tau\vert/c\sim B(V/L)(V/c)$, which vanishes to second order in $v/c$ compared to the right hand side. Thus (\ref{a0}) reduces to\footnote{~Perfect conductivity has been assumed here, but the result also applies at finite conductivity. See {\cpr problem \ref{curdif}\bk}.}
\beq {\bf j}={c\over 4\pi}{\bs{\nb}\bf\kern -1pt \times B}. \label{curr} \eeq
Taking the divergence:
\beq {\bs{\nb}\kern -2pt \bf\cdot j}=0.\label{divj}\eeq

Equation (\ref{divj}) shows that in MHD currents have no sources or sinks. As a consequence it is not necessary to worry `where the currents  close' in any particular solution of the MHD equations.
The equations are automatically consistent with charge conservation\footnote{~The fact $\bs{\nb}\kern-1pt\cdot{\bf j}=0$ is stated colloquially as `in MHD currents always close'. The phrase stems from the observation that lines of a solenoidal (divergence-free) vector in two dimensions have two choices\,: either they extend to infinity in both directions, or they form closed loops. In three dimensions it is more complicated\,: the lines of a solenoidal field enclosed in a finite volume are generally ergodic. A field line can loop around the surface of a torus, for example, never to return to the same point but instead filling the entire 2-dimensional surface. It is more accurate to say that since currents are automatically divergence-free in MHD, the closing of currents is not an issue.}. It is not even necessary that the field computed is an accurate solution of the equations of motion and induction. As long as the MHD approximation holds and the field is physically realizable, i.e.\ ${\bs{\nb}\kern -2pt \bf\cdot B}=0$, the current is just the curl of $\bf B$ (eq.\ \ref{curr}), and its divergence consequently vanishes. 

Of course, this simplification only holds as long as the MHD approximation itself is valid.  But whether that is the case  or not is a different question altogether\,: it  depends on things like the microscopic processes determining the conductivity of the plasma, not on global properties like the topology of the currents.

\subsection{Charge density}\label{chd}
With the equation for charge conservation (\ref{charge}), eq.\ (\ref{divj}) yields
\beq {\pa\sigma\over\pa t}=0. \label{dsigma}\eeq 
We conclude that it is sufficient to specify $\sigma=0$ in the initial state to guarantee that charges will remain absent, consistent  with our assumption of a charge-neutral plasma. Note, however, that we have derived this only in the non-relativistic limit. The charge density needs closer attention in relativistic MHD, see {section \ref{chargedd}}.

Eq.\ (\ref{dsigma}) only shows that a charge density cannot change in MHD, and one might ask what happens when a charge density is present in the initial conditions.  In practice, such a charge density cannot not last very long. Due to the electrical conductivity of the plasma assumed in MHD, charge densities are quickly neutralized. They  appear only at the boundaries of the volume in which MHD holds. See {\ref{clouds}} and {\ref{applic}}.

\subsection{Lorentz force, equation of motion}\label{eqmot}
With relation (\ref{curr}) between field strength and current density, valid in the non-relativistic limit, the Lorentz force acting per unit volume on the fluid carrying the current is
\beq 
{\bf F}_{\rm L}={1\over c}{\bf j\times B}=
{1\over 4\pi}({\bs{\nb}\bf\kern -1pt \times B}){\bf\times B}.\label{FL}
\eeq
This looks very different from the Lorentz force as explained in wikipedia. From a force acting on a charged particle orbiting in a magnetic field, it has become the force per unit volume exerted by a magnetic field on an {\em electrically neutral}, but conducting fluid. 

In  many  astrophysical applications viscosity can be ignored;  we restrict attention here to such {\em inviscid} flow, since extension to a viscous fluid can be done in the same way as in ordinary fluid mechanics. Gravity is often important as an external force, however. Per unit volume, it is 
\beq {\bf F}_{\rm g}=\rho{\bf g}=-\rho\bs{\nb}\phi,\eeq
where $\bf g$ is the acceleration of gravity, $\phi$ its potential and $\rho$ the mass per unit volume. If $p$ is the gas pressure, the equation of motion thus becomes
\beq 
\rho {\rd {\bf v}\over\rd t}=-\bs{\nb} p+{1\over 4\pi}({\bs{\nb}\bf\times B}){\bf\times B}+\rho{\bf g},\label{eqm}
\eeq
where $\rd/\rd t$ is the total or Lagrangian time-derivative, 
\beq \rd/\rd t={\pa/\pa t}+{\bf v\kern -2pt \cdot\kern -3pt \bs{\nb}~}.\eeq

Eq.\ (\ref{FL}) shows that he Lorentz force in MHD is quadratic in ${\bf B}$ and does not depend on its sign. The induction equation (\ref{ind}):
\beq {\pa{\bf B}\over\pa t}={\bs{\nb}\bf\times(v\times B)}\label{indn}\eeq
 is also invariant under a change of sign of $\bf B$. The ideal MHD equations are therefore  invariant under a change of sign of  $\bf B$\,: the fluid `does not  sense  the sign of the magnetic field'\footnote{~In non-ideal MHD this can be different, for example when Hall drift is important ({sect.~\ref{hall-am}}).}. Electrical forces do not appear in the equation of motion since charge densities are negligible in the non-relativistic limit ({sect.~\ref{chargedd}}).

The remaining equations of fluid mechanics are as usual. In particular the continuity equation, which expresses the conservation of mass:
\beq {\pa\rho\over\pa t}+{\bs{\nb}\bf\cdot}(\rho{\bf v})=0,\label{cty}\eeq
or
\beq {\rd\rho\over\rd t}+\rho{\bs{\nb}\bf\cdot}{\bf v}=0.\label{ctya}\eeq
In addition to this an equation of state is needed\,: a relation $p(\rho,T)$ between pressure, density, and temperature $T$. Finally an energy equation is needed if sources or sinks of thermal energy are present. It can be regarded as the equation determining the variation in time of temperature (or another convenient thermodynamic function of $p$ and $\rho$).  It will not be needed explicitly here (but see \hyperlink{kuls}{Kulsrud} or \hyperlink{mest}{Mestel} for details).

The equation of motion (\ref{eqm}) and the induction equation (\ref{indn}) together determine the two vectors ${\bf B}$ and ${\bf v}$. Compared with ordinary fluid mechanics, there is a new vector field, ${\bf B}$. There is an additional equation for the evolution of this field\,: the MHD induction equation, and an additional force appears in the equation of motion. 

These equations can be solved without reference to the other quantities appearing in Maxwell's equations. This reduction vastly simplifies the understanding of magnetic fields in astrophysics. The price to be paid is that one has to give up most of the intuitive notions acquired from classical examples of electromagnetism, because MHD does not behave like EM anymore. It is a fluid theory, close in spirit to ordinary fluid mechanics and to the theory of elasticity.

\subsection{The status of currents in MHD}\label{statcur}
Suppose $\bf v$ and ${\bf B}$ have been obtained as a solution of the equations of motion  and induction (\ref{eqm}, \ref{indn}), for a particular problem. Then Amp\`ere's law ({\ref{curr}}) can be used to calculate the current density at any point in the solution by taking the curl of the magnetic field. 

This shows how the {\em nature} of Amp\`ere's law has changed\,: from an equation for the magnetic field produced by a current distribution, as in vacuum electrodynamics, it has been demoted to the status of an operator for evaluating a secondary quantity, the current. 

As will become apparent from the examples further on in the text, the secondary nature of currents in MHD is not just a mathematical curiosity.  Currents are also rarely useful for physical understanding in MHD. They appear and disappear as the  magnetic  field geometry changes in the course of its interaction with the fluid flow.   An example illustrating the transient nature of currents in MHD is given in {\cpr problem \ref{pr.0}}.  Regarding the currents as the source of the magnetic field, as is standard practice in laboratory electrodynamics and plasma physics, is counterproductive in MHD. 

When familiarizing oneself with MHD one must set aside intuitions based on batteries, current wires, and induction coils.  Thinking in terms of currents as the sources of $\bf B$ leads astray; `there are no batteries in MHD'.  (For the origin of currents in the absence of batteries see {\ref{origj}}). 
Another source of confusion is that currents are not tied to the fluid in the  way household and laboratory currents are linked to copper wires. A popular mistake is to think of currents as entities that are carried around with the fluid. For the currents there is no equation like the continuity equation or the induction equation, however. They are not conserved in displacements of the fluid. 

\subsection{Consistency of the MHD approximation}
\label{consist}
In arriving at the MHD equations, we have so far accounted for 3 of the 4 of Maxwell's equations. The last one,  eq.\ ({\ref{chargeM}}), is not needed anymore. It has been bypassed by the fact that the electric field in MHD follows directly from a frame transformation,  eq.\ ({\ref{ee}}). Nevertheless, it is useful to check that the procedure followed has not introduced an inconsistency with Maxwell's equations, especially at relativistic velocities. This is done in {section \ref{chargedd}}. 

\section{The motion of field lines}\label{motline}
\label{motion}
In vacuum electrodynamics, field lines do not `move' since they do not have identity that can be traced from one moment to the next. In ideal MHD they become traceable as if they had an individual identity, because of their tight coupling to fluid elements, which do have individual identity. (See {\ref{freeze}} about this coupling at the microscopic level.)

This coupling is described by the induction equation ({\ref{indn}}). It does for the magnetic field (the flux density) what the continuity equation does for the mass density, but there are important differences because of the divergence-free vector nature of the field. To explore these differences, write the continuity equation (\ref{cty}) as
\beq {\pa\rho\over\pa t}=-\rho{\bs{\nb}\bf}\cdot{\bf v}-{\bf v}\cdot\bs{\nb}\rho.\label{conti}\eeq
 
The first term describes how the mass density $\rho$ varies in time at some point in space as fluid contracts or expands. The second term, called the {\em advection} of the density $\rho$, describes the change of $\rho$  at a point in space as fluid of varying density passes by it. The induction equation can be written by expanding its right hand side, using the standard vector identities ({sect.~\ref{idents}}):
\beq 
{\pa{\bf B}\over\pa t}=-{\bf B}\,\bs{\nb}\kern-1pt\cdot{\bf v}-({\bf v}\cdot\bs{\nb}){\bf B}+ ({\bf B\cdot\bs{\nb}}){\bf v}, \label{ind1}
\eeq
where ${\bs{\nb}\bf\cdot B}=0$ has been used\footnote{Note the form of expressions like ${\bf a}\kern-1pt \cdot\kern-2pt \nb {\bf b}$ when working in curvilinear coordinates.}. In this form it has a tempting similarity to the continuity equation (\ref{conti}). The first term looks like it describes the effect  of compression and expansion of the fluid, like the first term in (\ref{conti}) does for the mass density. The second term similarly suggests the effect of advection. There is, however, a third term, and as a consequence {\em neither} the effects of compression {\em nor} the advection  of a magnetic field   are properly described by the first and second terms alone.

A form that is sometimes useful is obtained by combining the equations of continuity and induction:
\beq 
{\rd\over\rd t}\left({{\bf B}\over\rho}\right)=\left({{\bf B}\over\rho}\bf\cdot \bs{\nb}  \right){\bf v},\label{brho}
\eeq
called Wal\'en's equation. It describes how the ratio of magnetic flux to mass density changes when the fluid velocity varies along a field line (see {\cpr problems \ref{pr.3}a, \ref{pr.stretch}}). 
\medskip

\begin{figure}[h]
\hfil{\includegraphics[width=0.4\hsize]{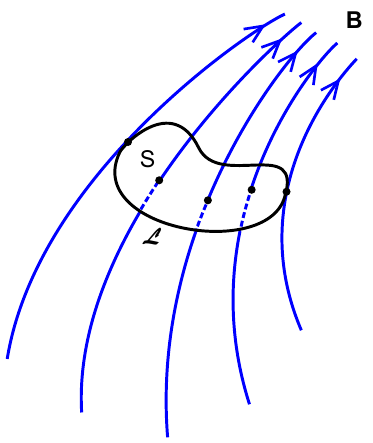}}\hfil
\caption{\label{loop} \small A closed loop $\cal L$ of fluid elements carried by the flow, with field lines passing through it. }
\end{figure}

\subsection{Magnetic flux}\label{mflux}
While the induction equation does not work for $\bf B$ in the same way as the continuity equation does for the gas density, there is conservation of something playing a similar role, namely magnetic flux. First we need to define magnetic flux in this context. 

Consider a closed loop $\cal L$ of infinitesimal fluid elements (Fig.\,\ref{loop}). It moves with the fluid, changing its  length  and shape. The magnetic flux of this loop is now defined as the `number of field lines passing through' it. A bit more formally, let $S$ be a surface bounded by the loop. There are many such surfaces, it does not matter which one we take ({\cpr problem \ref{pr.1}}). Then define the magnetic flux of the loop as
\beq \Phi(\cal L)=\int_S {\bf B}\cdot\rd {\bf S}, \label{Phi}\eeq
where $\rd {\bf S}={\bf n}\,\rd S$, with $\rd S$ an element of the surface $S$ and ${\bf n}$ the normal to $S$. 
In a perfectly conducting fluid the value of $\Phi$ is then a property of the loop, constant in time, for any loop moving with the flow (Alfv\'en's theorem)\,: 
\beq {\rd \Phi\over\rd t} =0.\label{fluxc}\eeq
 The equivalence of eq.\ (\ref{fluxc}) with the ideal MHD induction equation (\ref{indn}) is derived (slightly intuitively) in {\ref{der.1a}}.
 
The flux $\Phi(S)$ of the loop also defines a {\em flux bundle}\,: on account of ${\bs{\nb}\bf\cdot B}=0$ the field lines enclosed by the loop can be extended in both directions away from the loop, tracing out a volume in space filled with magnetic field lines. Like the loop $\cal L$ itself, this bundle of flux $\Phi$ moves with the flow, as if it had a physical identity. By dividing the loop into infinitesimally small sub-loops, we can think of the flux bundle as consisting of `single flux lines', and go on to say that the induction equation describes how such flux lines move with the flow. This is also described by saying that field lines are `frozen-in' the fluid. Each of these field lines can be labeled in a time-independent way, for example by a labeling of the fluid elements on the surface $S$ at some point in time $t_0$. 

If we define a `fluid element' intuitively as a microscopic {\em volume} carrying a fixed amount of mass (in the absence of diffusion of particles), a flux line is a macroscopic {\em string} of such elements. It carries a fixed (infinitesimal) amount of magnetic flux (in the absence of magnetic diffusion) through a cross section that varies along its length and in time.

When the conductivity is finite, (\ref{fluxc}) does not hold. The induction equation then has an additional term describing diffusion of the magnetic field by the finite resistivity of the plasma. In general, field lines can then not be labeled in a time-independent way anymore, but in practice this can often be overcome so that one can still meaningfully talk about lines `diffusing across' the fluid (see section \ref{diffusion}). 

Astrophysical conditions are, with some exceptions, close enough to perfect conductivity that intuition based on ideal MHD is applicable as a first step in most cases. The opposite limit of low conductivity is more amenable to applied mathematical analysis but rarely relevant in astrophysics.

\subsection{Field amplification by fluid flows}
\label{stretch}
\begin{figure}
\hfil{\includegraphics[width=0.7\hsize]{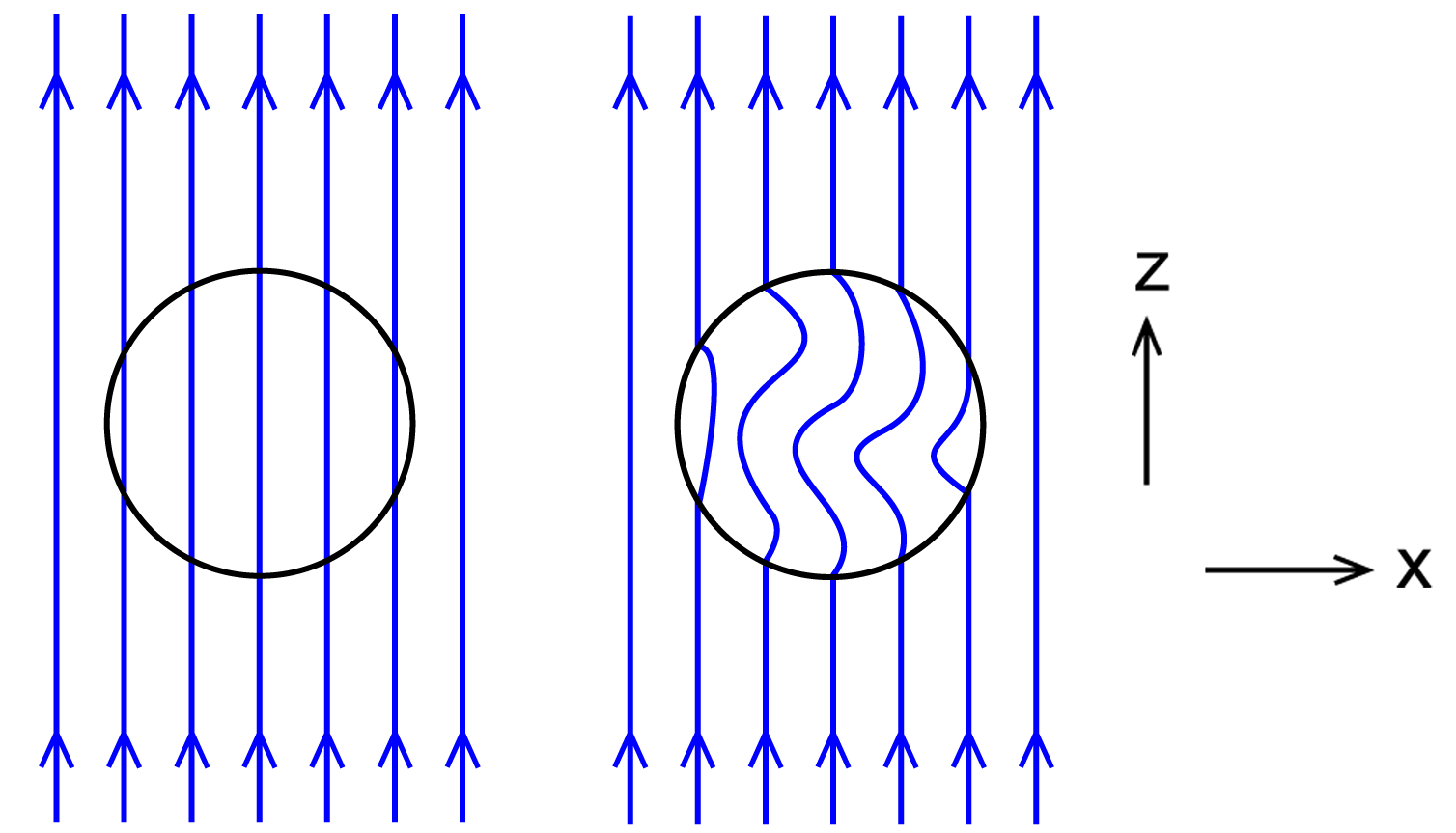}}\hfil
\caption{\label{tangle} \small Field amplification by a complex fluid flow}
\end{figure}

In the absence of diffusion, magnetic fields embedded in a fluid have a strong {\em memory effect}\,: their instantaneous strength and shape reflects the {\em history} of the fluid motions acting on it. 
One  consequence of this is that fluid motions tend to rapidly increase the strength of an initially present magnetic field. This is illustrated in Fig.\,\ref{tangle}. The fluid is assumed to have a constant density, i.e.\ it is incompressible. By the continuity equation (\ref{conti}) the velocity field is then divergence-free, $\bs{\nb}\kern-1pt\cdot{\bf v}=0$. Assume that there is an initially uniform field in the $z$-direction, ${\bf B}=B\,{\bf\hat z}$.  

The field is deformed by fluid motions which we assume to be confined to a volume of constant size. The velocities vanish at the boundary of this sphere. The field lines can then be labeled by their positions at the boundary of the sphere, for example a field line entering at $x_1,-z_1$ and exiting at $x_1,z_1$ can be numbered $1$. Define the length $L$ of the field line as its path length from entry point to exit; the initial length of line number 1 is $L=2z_1$. 

Since the field lines are initially straight, any flow in the sphere will increase their length $L$. Let $\delta$ be the average distance of field line number 1 from its nearest neighbor. As long as the field is frozen-in, the mass enclosed between these field lines is constant, and on account of the constant density of the fluid, its volume $L\delta$ is also constant\footnote{~This argument assumes a 2-dimensional flow, as suggested by Fig.\,\ref{tangle}. Except at special locations in a flow, it also holds in 3 dimensions. To see this requires a bit of visualization of a magnetic field in a shear flow.}. Hence $\delta$ must decrease with increasing $L$ as $\delta\sim 1/L$. By flux conservation (\ref{Phi}), the field strength then increases as $L$. 

A bit more formally, consider the induction equation in the form (\ref{ind1}), and by taking the second term on the right to the left, write it as
\beq {\rd {\bf  B}\over \rd t}=-{\bf B}\,{\bs{\nb}\bf\cdot v}+({\bf B\cdot\kern -2pt\bs{\nb}}){\bf v}.\label{ind2}\eeq
It then describes the rate of change of $\bf B$ in a frame comoving with the fluid. Like the mass density, the field strength can change by compression or expansion of the volume (first term). This change is modified, however, by the second term which is intrinsically magnetohydrodynamic. Under the assumed incompressibility the first term vanishes, and the induction equation reduces to
\beq {\rd {\bf  B}\over \rd t}=({\bf B\cdot\kern-2pt\bs{\nb}}){\bf v}.\qquad(\rho={\rm cst.})\label{ind3}\eeq

\begin{figure}
\href{run:/Fig1.4.mp4}{\hspace{10cm}}\hspace{-10cm}
\centerline{\includegraphics[width=0.7\hsize]{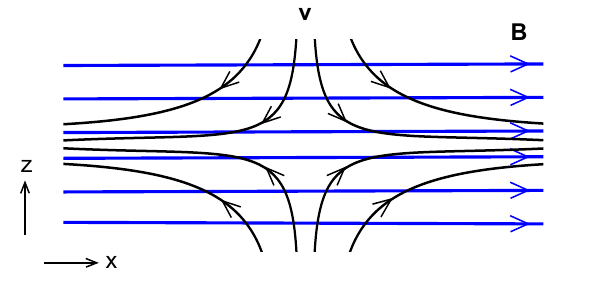}}\hfil
\caption{\label{figstretch}\small Amplification of a uniform field (blue) by a converging-diverging flow (black). 
\href{run:Fig1.4.mp4}{\can Animation} shows  the flow of randomly spaced (black) and regularly spaced (red) fluid particles.
}
\end{figure}

To visualize what this means, consider some simple examples. 
In the first example we consider the effect of amplification of a magnetic field by a flow which converges on the field lines and diverges along them (`stretching'). Assume an initially uniform magnetic field in the $x$-direction (Fig.\,\ref{figstretch}) in a stationary, divergence-free velocity field:
\beq {\bf B}(t=0)=B_0\,{\bf\hat x},\qquad v_x=ax,\qquad v_z=-az,\qquad v_y=0\label{strflow}.\eeq
At t=0, eq.\ (\ref{ind3}) yields:
\beq {\rd {\bf B}\over\rd t}{\vert}_{t=0}={\pa{\bf B}\over\pa t}\vert_{t=0}=B_0\,a\,{\bf\hat x}\label{bstretch}.\eeq
The first equality holds since at $t=0$ the field is uniform, so that ${(\bf v\cdot\kern -2pt\bs{\nb}) B}=0$.
Eq.\ (\ref{bstretch}) shows that the field stays in the $x$-direction and remains uniform. The assumptions made at $t=0$ therefore continue to hold, and (\ref{bstretch}) remains valid at arbitrary time $t$ and can be integrated with the result:
\beq {\bf B}(t)=B_0\, e^{a t}{\,\bf\hat x}.\label{stretchexp}\eeq
Stretching by a flow like (\ref{strflow}) thus increases $B$ exponentially for as long as the stationary velocity field is present. This velocity field is rather artificial, however. If the flow instead results from a force that stretches the left and right sides apart at a constant velocity, the rate of divergence along the field decreases with time, and the field strength increases only linearly (exercise\,: verify this for yourself, using mass conservation and taking density constant).

When the flow $\bf v$ is perpendicular to ${\bf B}$ everywhere, the term $({\bf B\cdot\kern-2pt\bs{\nb}}){\bf v}$ describes an effect that is conceptually rather different (see Fig.\,\ref{shear}). Take the magnetic field initially uniform in the $z$-direction, the flow in the $x$-direction, constant in time with a piece-wise linear shear in $z$:

\begin{figure}
\hfil{\includegraphics[width=0.4\hsize]{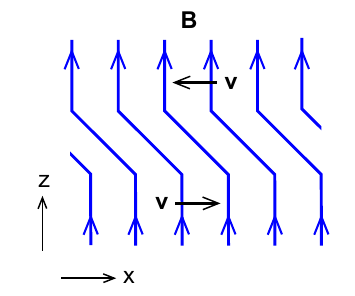}}\hfil
\caption{\label{shear} \small Field amplification by a shear flow perpendicular to $\bf B$}
\end{figure}

\beq {\bf B}(t=0)=B_0\,{\bf\hat z}, \qquad v_y=v_z=0\quad (t=0),\label{shz}\eeq
and
\beq  v_x=-v_0\quad(z<-1),\quad v_x=v_0 z\quad(-1<z<1),\quad v_x=v_0\quad(z>1).\eeq
The induction equation (\ref{indn}) then yields (analogous to {\cpr problem \ref{pr.0}}):
\beq B_z(t)=B_0={\rm~ cst},\eeq
\beq B_x=B_0v_0 t\quad (-1<z<1),\quad B_x=0\quad (z<-1,z>1).\eeq
In the shear zone $-1<z<1$ the $x$-component of ${\bf B}$ grows linearly with time. One could call this kind of field amplification by a flow perpendicular to it `shear amplification'.  

Shear zones of arbitrary horizontal extent like the last example are somewhat artificial, but the effect takes place in essentially the same way in a rotating shear flow (such as a differentially rotating star for example). In this case, the field amplification can be described as a process of `winding up' of field lines. A simple case is described by {\cpr problem \ref{pr.rotstar}}. 

In these examples the flow field was assumed to be given, with the magnetic field responding passively following the induction equation. As the field strength increases, magnetic forces will eventually become important, and determine the further evolution of the field in a much more complex manner. Evolution of a magnetic field as described above, the so-called {\em kinematic} approximation, can hold only for a limited period of time in an initially sufficiently weak field (see also {sect.~\ref{tilde}}).

\section{Magnetic force and magnetic stress}\label{magfor}

\subsection{Magnetic pressure and curvature force}
\label{magpres}
The Lorentz force is perpendicular to  $\bf B$. Along the magnetic field, the fluid motion is therefore subject only to the normal hydrodynamic forces. This makes the mechanics of a magnetized fluid extremely {\em anisotropic}.  

To get a better feel for magnetic forces, one can write the Lorentz force ({\ref{FL}}) in alternative forms. Using the vector identities ({sect. \ref{idents}}), 
\beq 
{\bf F}_{\rm L}= {1\over 4\pi}({\bf\bs{\nb}\times B}){\bf\times B}=-{1\over 8\pi}\bs{\nb}B^2 +{1\over 4\pi}({\bf B\ncd\bs{\nb}}){\bf B}. \label{prescurv}
\eeq
The first term on the right is the gradient of what is called the {\em magnetic pressure} $B^2/8\pi$. The second term describes a force due to the variation of magnetic field strength in the direction of the field. It is often called the magnetic {\em curvature force}. 

These names are a bit misleading. Even in a field in which the Lorentz force vanishes, the magnetic pressure gradient (\ref{prescurv}) is generally nonzero, while the `curvature' term can also be present in places where the field lines are straight. To show the role of curvature of the field lines more accurately, write the magnetic field as 
\beq {\bf B}=B\,{\bf s}, \eeq
where ${\bf s}$ is the unit vector in the direction of $\bf B$. The Lorentz force then becomes
\beq 
{\bf F}_{\rm L}=-{1\over 8\pi}\bs{\nb}B^2 +{1\over 8\pi}{\bf s}~{\bf s}\ncd\nb B^2 +{B^2\over 4\pi}{\bf s}\ncd\bs{\nb}{\bf s}. \label{prescurv1}
\eeq
Combining the first two terms formally, we can write this as
\beq 
{\bf F}_{\rm L}=-{1\over 8\pi}\bs{\nb}_\perp B^2+{B^2\over 4\pi}{\bf s}\ncd\bs{\nb}{\bf s}, \label{prescurv2}
\eeq
where $\bs{\nb}_\perp$ is the projection of the gradient operator on a plane perpendicular to $\bf B$. The first term, perpendicular to the field lines, now describes the action of magnetic pressure more accurately. The second term, also perpendicular to $\bf B$ contains the effects of field line curvature. Its magnitude is 
\beq 
\vert {B^2\over 4\pi}{\bf s}\ncd\bs{\nb}{\bf s}\vert={B^2\over 4\pi R_{\rm c}},
\eeq
where 
\beq
R_{\rm c}=1/\vert {\bf s}\ncd\bs{\nb}{\bf s}\vert 
\eeq
is the radius of curvature of the path ${\bf s}$. 
[{\cpr problem \ref{pr.7}}\,: magnetic forces in a $1/r^2$ field.]

As an example of magnetic curvature forces, consider an axisymmetric azimuthally directed field, ${\bf B}=B\,{\bf\hat\varphi}$, in cylindrical coordinates ($\varpi,\varphi,z$). The strength $B$ is then a function of $\varpi$ and $z$ only. 
The unit vector in the azimuthal direction $\bs{\hat\varphi}$ has the property $\bs{\hat\varphi}\cdot\bs{\nb}\bs{\hat\varphi}=-\bs{\hat\varpi}/\varpi$, so that
\beq {1\over 4\pi}({\bf B\cdot\bs{\nb}}){\bf B}=-{1\over 4\pi}{B^2\over \varpi}\bs{\hat\varpi}.\eeq
The radius of curvature of the field line is thus the cylindrical radius $\varpi$. The curvature force is directed inward, toward the center of curvature. For an azimuthal field like this, it is often also referred to as the {\em hoop stress} (like the stress in hoops keeping a barrel together). [{\cpr problem \ref{pr.5}}\,: magnetic forces in a $1/\varpi$ field.]
 
\subsection{Magnetic stress tensor}
\label{mstress}
The most useful alternative form of the Lorentz force is in terms of the {\em magnetic stress tensor}. Writing the vector operators in terms of the permutation symbol $\epsilon$ ({sect.~\ref{idents}}), one has
\bea 
[({\bs{\nb}\kern-1pt\bf\times B}){\bf\times B}]_i=\epsilon_{ijk}\epsilon_{jlm}{\pa B_m\over\pa x_l}B_k \cr
= (\delta_{kl}\delta_{im}-\delta_{km}\delta_{il}){\pa B_m\over\pa x_l}B_k\cr
={\pa\over \pa x_k}(B_iB_k-{1\over 2}B^2\delta_{ik}),\label{flm}
\eea
where the summing convention over repeated indices is used and in the last line ${\bs{\nb}\bf\ecd B}=0$ has been used. Define the {\em magnetic stress tensor} ${\bf M}$ by its components:
\beq M_{ij}={1\over 8\pi}B^2\delta_{ij}-{1\over 4\pi}B_iB_j. \label{M}\eeq
Eq.\ (\ref{flm}) then shows that the force per unit volume exerted by the magnetic field is minus the divergence of this tensor:
\beq {1\over 4\pi}({\bf \bs{\nb}\kern -1pt\times B}){\bf\times B}=-{\bs{\nb}\bf\cdot M}.\label{divm}\eeq

The magnetic stress tensor thus plays a role analogous to the fluid pressure in ordinary fluid mechanics (explaining the minus sign introduced in its definition), except that it is a rank\,-2 tensor instead of a scalar. This is much like the stress tensor in the theory of elasticity. 
If $V$ is a volume bounded by a closed surface $S$, (\ref{divm}) yields by the divergence theorem
\beq 
\int_V {1\over 4\pi}({\bs{\nb}\kern-1pt\bf\times B}){\bf\times B}~\rd V=\oint_S -{\bf n\cdot M}~\rd S,\label{gauss}
\eeq
where ${\bf n}$ is the outward normal to the surface $S$. This shows how the net Lorentz force acting on a volume V of fluid can be written as an integral of a {\em magnetic stress vector} acting on its surface, the integrand of the right  in (\ref{gauss}). If, instead, we are interested in the forces ${\bf F}_S$ exerted {\em by} the field in the volume $V${\em on} its surroundings, a minus sign is to be added,
\beq {\bf F}_S={\bf n\cdot M}={1\over 8\pi}B^2{\bf n}-{1\over 4\pi}{\bf B}B_n, \label{stress}\eeq
where $B_n={\bf B\cdot n}$ is the component of ${\bf B}$ along the outward normal ${\bf n}$ to the surface of the volume. The vector ${\bf F}_{\rm S}$ is a surface force (per unit  area), not to be confused with the Lorentz force vector which is a volume force.

The stress in a magnetic field differs strongly from that in a fluid under pressure. Unlike normal fluid pressure, magnetic stress does not act perpendicular to a surface but  is  a vector at some angle to the normal to the surface, just as in the case of a sheared elastic medium or viscous fluid ({\cpr problem \ref{sforce}}).

It is often useful to visualize magnetic forces in terms of the surface stress vector (\ref{stress}), evaluated on surfaces of suitable orientation. The sign of this surface force depends on the direction of the normal taken on the surface. The ambiguity caused by this is resolved by deciding on the  volume of interest to which the surface element would belong (for an example see {\ref{disksupport}}).

\subsection{Properties of the magnetic stress. Pressure and tension} \label{stresprop}
\begin{figure}
\hfil{\includegraphics[width=0.4\hsize]{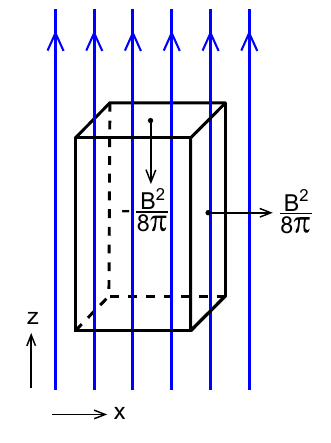}}\hfil
\caption{\label{boxstress} \small Force vectors exerted by magnetic stress on the surfaces of a rectangular box uniform of magnetic field}
\end{figure}

To get some idea of the behavior of magnetic stresses, take the simple case of a uniform magnetic field, in the $z$-direction say, and evaluate the forces exerted by a volume of this magnetic field on its surroundings, see Fig.\,\ref{boxstress}. The force ${\bf F}_S$ which the box exerts on a surface parallel to ${\bf B}$, for example the surface perpendicular to the $x$-axis on the right side of the box (\ref{stress} with $\bf{n}=\hat{\bf x}$), is ${\bf F}_{\rm right}={\bf\hat x\cdot M}$. The components are
\beq F_{{\rm right},x}={1\over 8\pi}B^2-{1\over 4\pi}B_xB_z={1\over 8\pi}B^2, \qquad F_{{\rm right},z}=F_{{\rm right},y}=0.\eeq
Only the magnetic pressure term contributes on this surface. The magnetic field exerts a force in the positive $x$-direction, away from the volume. The stress exerted by the magnetic field at the top surface of the box has the components
\beq 
F_{{\rm top},z}={1\over 8\pi}B^2-{1\over 4\pi}B_zB_z=-{1\over 8\pi}B^2,\qquad F_{{\rm top},x}=F_{{\rm top},y}=0,
\eeq
i.e.\ the stress vector is also perpendicular to the top surface. It is of equal magnitude to that of the magnetic pressure exerted at the vertical surfaces, but of opposite sign. 

On its own, the magnetic pressure would make the volume of magnetic field expand in the perpendicular directions $x$ and $y$. But in the direction {\em along} a magnetic field line the volume would {\em contract}. Along the field lines the magnetic stress thus acts like a negative pressure, as in a stretched elastic wire. As in the theory of elasticity, this negative stress is referred to as the {\em tension} along the magnetic field lines. 

A magnetic field in a conducting fluid thus acts somewhat like a deformable, elastic medium. Unlike a usual elastic medium, however,  it is always under compression in two directions (perpendicular to the field) and under tension in the third (along the field lines), irrespective of the deformation. Also unlike elastic wires, magnetic field lines have no `ends' and cannot be broken. As a consequence, the contraction of the box in Fig.\,\ref{boxstress} under magnetic stress does not happen in practice, since the tension at its top and bottom surfaces is balanced by the tension in the magnetic lines continuing above and below the box. The effects of tension in a magnetic field manifest themselves more indirectly, through the {\em curvature} of field lines (see {eq.\ \ref{prescurv2} ff}). For an example, see {sect.~\ref{disksupport}}.

Summarizing, the stress tensor plays a role analogous to a scalar pressure like the gas pressure, but unlike gas pressure is extremely anisotropic. The first term in (\ref{M}) acts in the same way as a hydrodynamic pressure, but it is never alone. Approximating the effect of a magnetic field by a scalar pressure term is rarely useful (see also {sect. \ref{Bcompress}}).
\bigskip

\subsection{Boundaries between regions of different field strength}
\label{boundequi}

\begin{figure}[t]
\hfil{\includegraphics[width=0.5\hsize]{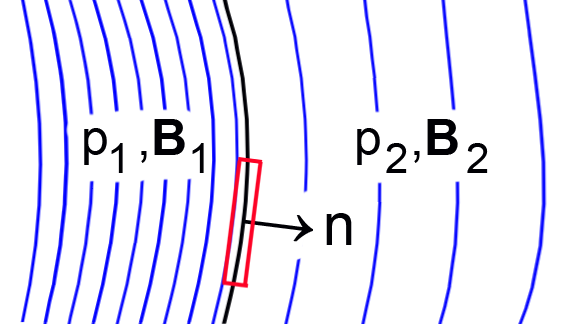}}\hfil
\caption{\small Boundary between two regions of different field strength, with the pill box (red) used for applying Gauss's theorem.}\label{bound} 
\end{figure}

Let $S$ be a surface in the field configuration, everywhere parallel to the magnetic field lines but otherwise of arbitrary shape (a so-called `magnetic surface'). Introduce a `total' stress tensor ${\bf M}_{\rm t}$ by
combining gas pressure with the magnetic pressure term:
\beq {\bf M}_{\rm{t},ij}=(p+B^2/8\pi)\delta_{ij}-B_iB_j/4\pi. \label{stress1}\eeq
The equation of motion is then
\beq \rho{\rm{d}\bf{v}\over\rm{d}t}=-{\rm div}\,{\bf M}_{\rm t} +\rho\bf{g}.\label{eqmob}\eeq

Integrate this equation over the volume of a `pill box' $V$ of infinitesimal thickness $\epsilon$ which includes a unit of surface area of $S$ (Fig.\,\ref{bound}). The left hand side and the gravity term do not contribute to the integral since the volume of the box vanishes for $\epsilon\rightarrow 0$, hence
\beq\int_{\rm pill}{\rm div}\,{\bf M}_{\rm t}~{\rm d} V =0.\eeq
 The stress (\ref{stress1}) is perpendicular to $S$ on the surfaces of the box that are parallel to $\bf B$. Applying Gauss's theorem to the box  then yields, in the limit $\epsilon\rightarrow 0$,
\beq p_1+{B_1^2\over 8\pi}=p_2+{B_2^2\over 8\pi}~.\label{presbal}\eeq
Equilibrium at the boundary between the two regions 1 and 2 is thus governed by the total pressure $p+B^2/8\pi$ only. The curvature force does {\em not} enter in the balance between regions of different field strength. This may sound contrary to intuition. The role of the curvature force near a boundary is more indirect. See sect.\  \ref{curvb}. 

The configuration need not be static for (\ref{presbal}) to apply, but singular accelerations at the surface $S$ must be excluded. The analysis can be extended to include the possibility of a sudden change of velocity across $S$ (so the contribution from the left hand side of \ref{eqmob} does not vanish any more), and $S$ can then also be taken at an arbitrary angle to the magnetic field. This leads to the theory of MHD shock waves.

\subsection{Magnetic buoyancy}
\label{buoy}
Consider a magnetic flux bundle embedded in a nonmagnetic plasma, in pressure equilibrium with it according to (\ref{presbal}). With the external field $B_{\rm e}=0$, the pressure $p_{\rm i}$ inside the bundle is thus lower than $p_{\rm e}$ outside. Assume that the plasma has an equation of state $p={\cal R}\rho T$, where $\cal R$ (erg g$^{-1}$ K$^{-1}$) is the gas constant and $T$ the temperature, and assume that thermal diffusion has equalized the temperature inside the bundle to the external temperature $T$. The reduced pressure then means that the density $\rho_{\rm i}$ is reduced:
\beq \delta\rho\equiv\rho_{\rm i}-\rho_{\rm e}=-{B^2\over 8\pi{\cal R} T}, \eeq
or
\beq \delta\rho/\rho_{\rm e}=-{1\over 2}({v_{\rm Ae}\over c_{\rm i}})^2, \label{delrb}\eeq
where  $c_{\rm i}$ the  isothermal sound speed $c_{\rm i}=(p/\rho)^{1/2}$ and  $v_{\rm Ae}\equiv B/(4\pi\rho_{\rm e})^{1/2}$ is a notional Alfv\'en speed, based on the {\em internal} field strength and the {\em external} density. 
In the presence of an acceleration of gravity $\bf g$, the reduced density causes a {\em buoyancy force} ${\bf F}_{\rm b}$ against the direction of gravity. Per unit volume:
\beq  {\bf F}_{\rm b}={\bf g}\,\delta\rho. \eeq
This causes a tendency for magnetic fields in stars to drift outwards (see {\cpr problems \ref{cbuoy}, \ref{drift}} for the expected speed of this process).

\section{Strong fields and weak fields, plasma-{$\bs{\beta}$}}
\label{tilde}
In the examples so far we have looked separately at the effect of a given velocity field on a magnetic field, and at the magnetic forces on their own. Before including both, it us useful to classify physical  parameter regimes by considering the relative importance of the terms in the equation of motion. Ignoring viscosity and external forces like gravity, the equation of motion ({\ref{eqm}}) is
\beq 
\rho {\rd {\bf v}\over\rd t}=-\bs{\nb}p+{1\over 4\pi}({\bs{\nb}\kern-1pt\bf\times B}){\bf\times B}~.\label{eqm0}
\eeq
A systematic procedure for estimating the relative magnitude of the terms is to decide on a length scale $l$ and a time scale $\tau$ that are characteristic for the problem at hand, as well as characteristic values $v_0$ for velocity and $B_0$ for field strength. Since the sound speed is generally a relevant quantity, we assume a compressible medium, for simplicity with an equation of state as before, $p={\cal R}\rho T$, where $\cal R$ is the gas constant and $T$ the temperature, which we take to be constant here (`isothermal' equation of state). Define dimensionless variables and denote them with a tilde $\tilde\ $:
\beq 
t=\tau\tilde t,\qquad \bs{\nb}=\bs{\tilde\nb}/l, \qquad {\bf v}=v_0 {\bf \tilde v},\qquad {\bf B}=B_0{\bf \tilde B}.
\eeq
With the {\em isothermal sound speed} $c_{\rm i}$,
\beq c_{\rm i}^2=p/\rho={\cal R}T,\eeq
the equation of motion becomes, after multiplication by $l/\rho$,
\beq 
v_0{l\over\tau} {\rd\over\rd\tilde t}\tilde{\bf v}=-c_{\rm i}^2\bs{\tilde\nb}\ln\rho+\va^2(\bs{\tilde\nb}{\bf\times\tilde B}){\bf\times \tilde B},\label{diml0}
\eeq
where $\va$ is the {\em Alfv\'en speed}:
\beq
\va={B_0\over\sqrt {4\pi\rho}}.
\eeq
A characteristic velocity for things happening on the time scale $\tau$ over the length of interest $l$ is $v_0=l/\tau$. Dividing (\ref{diml0}) by $c_{\rm i}^2$ then yields a dimensionless form of the equation of motion:
\beq 
{\cal M}^2 {\rd\over\rd\tilde t}\tilde{\bf v}=-\bs{\tilde\nb}\ln\rho+{2\over\beta}(\bs{\tilde\nb}{\bf\times\tilde B}){\bf\times \tilde B},\label{dimless}
\eeq
where ${\cal M}$ is the {\em Mach number} of the flow,
\beq {\cal M}=v_0/c_{\rm i},\eeq
and $\beta$ is the so-called\footnote{~The plasma-$\beta$ has become standard usage, thanks to the early plasma physics literature. Attempts to introduce its inverse as a more logical measure of the influence of a magnetic field have not been very successful.} {\em plasma}-$\beta$:
\beq \beta={c_{\rm i}^2\over \va^2}={8\pi p\over B_0^2},\label{beta}\eeq
the ratio of gas pressure to magnetic pressure.
Since we have assumed that $l,\tau,v_0,B_0$ are representative values, the tilded quantities in (\ref{dimless}) are all of order unity. The relative importance of the 3 terms in the equation of motion is thus determined by the values of the two dimensionless parameters ${\cal M}$ and $\beta$. 

Assume first the case ${\cal M}\ll 1$: a highly {\em subsonic} flow, so the left hand side of (\ref{dimless}) can be ignored. The character of the problem is then decided by the value of $\beta$. The physics of {\em high-$\beta$} and {\em low-$\beta$} environments is very different. 

If $\beta\gg 1$, i.e.\ if the gas pressure is much larger than the magnetic energy density, the second term on the right is small, hence the first must also be small, $\bs{\tilde\nb}\ln\rho\ll 1$. That is to say, the changes in density produced by the magnetic forces are small. In the absence of other forces causing density gradients, a constant-density approximation is therefore often useful in high-$\beta$ environments.

If $\beta\ll 1$, on the other hand, the second term is large, but since logarithms are not very large, it cannot be balanced by the first term. We conclude that in a low-$\beta$, low-${\cal M}$ plasma the magnetic forces must be small\,: we must have
\beq 
(\bs{\tilde\nb}{\bf\times\tilde B}){\bf\times\tilde B}\sim {\cal O}(\beta)\ll 1.
\eeq
In the limit $\beta\rightarrow 0$ this can be satisfied in two ways. Either the current vanishes, ${\bs{\nb}\bf\kern-1pt\times  B}=0$, or it is parallel to ${\bf B}$. In the first case, the field is called a {\em potential field}. It has a scalar potential $\phi_{\rm m}$ such that ${\bf B}=-\bs{\nb}\phi_{\rm m}$, and with ${\bs{\nb}\bf\kern-1pt\cdot B}=0$, it is governed by the Laplace equation
\beq \nb^2\phi_{\rm m}=0.\eeq

Applications could be the magnetic field in the atmospheres of stars, for example. For reasons that are less immediately evident at this point,  the currents in such an atmosphere are  small enough for a potential field to be a useful first approximation, depending on the physical question of interest. The more general second case,
\beq 
(\bs{\nb}{\bf\times B}){\bf\times  B}=0
\eeq
describes {\em force-free fields} ({sect.~\ref{fff}}). 
Like potential fields, they require a nearby `anchoring' surface ({sect.~\ref{boundrole}}). They are also restricted to environments such as the tenuous atmospheres of stars like the Sun, magnetic A stars, pulsars, and of accretion disks.

If the Mach number ${\cal M}$ is not negligible and $\beta$ is large, the second term on the right of (\ref{dimless}) can be ignored. The balance is then between pressure forces and accelerations of the flow, i.e.\ we have ordinary hydrodynamics with the magnetic field playing only a passive role. We can nevertheless be interested to see how a magnetic field develops under the influence of such a flow. This is the study of the {\em kinematics} of a magnetic field, or equivalently\,: the study of the induction equation for different kinds of specified flows.

 Finally, if $\beta$ is small and the field is not a force-free or potential field, a balance is possible only if ${\cal M}$ is of order $1/\beta\gg 1$. That is, the flows must be supersonic, with velocities $v_0\sim \va$. The magnetic fields in star-forming clouds are believed to be approximately in this regime.

The intermediate case $\beta\approx1$ is sometimes called `equipartition', the `equi' referring in this case to the approximate equality of magnetic and thermal energy densities. The term equipartition is not unique, however; it is also used for cases where the magnetic energy density is comparable with the {\em kinetic} energy density of the flow, i.e.\ when
\beq {1\over 2}\rho v^2\approx {B^2\over 8\pi},\eeq
or equivalently
\beq v\approx \va.\eeq

\section{Force-free fields and potential fields}
\label{ffp}
\subsection{Force-free fields}
\label{fff}

In a force-free field, $\bs{\nb}{\bf\times B}$ is parallel to ${\bf B}$. Hence there is a scalar $\alpha$ such that
\beq \bs{\nb}{\bf\times B}=\alpha{\bf B}.\eeq
Taking the divergence we find:
\beq {\bf B\cdot\kern-1pt\bs{\nb}}\alpha =0,\eeq 
that is, $\alpha$ is constant along field lines. Force-free fields are `twisted'\,: the field lines in the neighborhood  of a field line with a given value of $\alpha$ `wrap around' it, at a pitch proportional to $\alpha$.

The case of a constant $\alpha$ everywhere would be  mathematically interesting, since it leads to a tractable equation. This special case (sometimes called a `linear' force-free field) is of little use, however. Where force-free fields develop in nature, the opposite tends to be the case\,: the scalar $\alpha$ varies strongly between neighboring magnetic surfaces (cf.\ {sects.\ \ref{sheets}, \ref{reco}}).

With magnetic forces vanishing, infinitesimal fluid displacements $\bs\xi$ do not do work against the magnetic field, hence the energy in a force-free magnetic field is an extremum. The opposite is also the case\,: if the magnetic energy of a configuration is a minimum (possibly a local minimum)  the field must be force-free.

To make this a bit more formal imagine a (closed, simply connected) volume $V$ of perfectly conducting fluid in which displacements $\bs\xi$ take place, surrounded by a volume in which the fluid is kept at rest, so that $\bs\xi=0$ there. 
We start with some magnetic field configuration in $V$, not too far from equilibrium\footnote{Relaxation from an arbitrary initial state can involve the formation of current sheets, see sect.\ \ref{tangled}}, and let it relax under the perfectly conducting constraint until an equilibrium is reached. This relaxing can be done for example by adding a source of viscosity so fluid motions are damped out. In doing so the field lines move to different locations, except at their ends, where they are kept in place by the external volume. To be shown is that the minimum energy configuration reached in this way is a force-free field, $(\bs{\nb}{\bf\times B}){\bf \times B}=0$.

Small displacements inside the volume change the magnetic energy $E_{\rm m}$ in it by an amount $\delta E_{\rm m}$, and the magnetic field by an amount $\delta{\bf B}$:
\beq 4\pi~\delta E_{\rm m}=\delta\left[{1\over 2}\int B^2\rd V\right]=\int{\bf B}\cdot \delta{\bf B}~\rd V.\label{dem}\eeq
The velocity of the fluid is related to the displacements $\bs\xi$ by
\beq {\bf v}={\pa\bs\xi/\pa t},\eeq
so that for small displacements the induction equation is equivalent to
\beq \delta{\bf B}={\bs{\nb}\times}({\bs{\xi}\bf\times B}).\label{deltaB}\eeq
({\cpr problem \ref{indint}}).
Using one of the vector identities and the divergence theorem, (\ref{dem}) yields
\beq 4\pi~\delta E_{\rm m}=\int{\bf B\cdot\bs{\nb}\times}({\bs{\xi}\bf\times B})=\eeq
\beq =\oint [({\bs{\xi}\bf\times B})\times{\bf B}]\cdot\rd{\bf S}+\int({\bs{\nb}\kern-1pt\bf\times B})\cdot({\bs{\xi}\bf\times B})~\rd V. \eeq
The surface term vanishes since the external volume is being kept at rest, so that (with \ref{dotcross}):
\beq 4\pi~\delta E_{\rm m}=\int({\bs{\nb}\kern-1pt\bf\times B})\cdot({\bs{\xi}\bf\times B})~\rd V=-\int{\bs{\xi}\cdot}[({\bs{\nb}\kern-1pt\bf\times B}){\bf \times B}]~\rd V.\eeq

If the energy is at a minimum, $\delta E_{\rm m}$ must vanishes to first order for {\em arbitrary} displacements $\bs{\xi}$. This is possible only if the factor in square brackets vanishes everywhere,
\beq ({\bs{\nb}\kern-1pt\bf\times B}){\bf \times B}=0,\eeq
showing that a minimum energy state is indeed force-free. To show that a force-free field in a given volume is a minimum, rather than a maximum, requires examination of the second order variation of $\delta E_{\rm m}$ with $\bs\xi$ (cf.\ \hyperlink{robe}{Roberts 1967}, \hyperlink{kuls}{Kulsrud 2005)}{\footnote{~In addition there is a question of uniqueness of the minimum energy state. See \ \hyperlink{moff}{Moffatt (1985)}.}.  

\subsection{Potential fields}
\label{potf}
The energy of a force-free minimum energy state can be reduced further only by relaxing the constraint of perfect conductivity. Assume that the magnetic field in the {\em external} volume is again kept fixed, for example in a perfectly conducting medium. By allowing magnetic diffusion to take place inside $V$, the lines can `slip with respect to the fluid' in $V$ (cf.\ \ref{diffusion}). 
The only constraint on the magnetic field inside $V$ is now that it is divergence-free.  
We can take this into account by writing the changes in terms of a vector potential,
\beq \delta {\bf B}=\bs{\nb}\kern-1pt\times \delta{\bf A}.\eeq
The variation in energy, $\delta E_{\rm m}$ then is 
\beq 4\pi\delta E_{\rm m}=\int_V{\bf B \cdot\bs{\nb}\kern-1pt\times \delta{\bf A}}~\rd V\eeq
\beq =\int_V\bs{\nb}\kern-1pt\cdot(\delta{\bf A\times B})~\rd V+\int_V\delta{\bf A\cdot\bs{\nb}\kern-1pt\bf\times B}~\rd V\eeq
\beq =\oint_S\delta{\bf A\times B}~\rd{\bf S}+\int_V\delta{\bf A\cdot\bs{\nb}\kern-1pt\bf\times B}~\rd V.\eeq
In order to translate the boundary condition into a condition on $\delta{\bf A}$, a gauge is needed. Eq.~(\ref{deltaB}) shows that in the perfectly  conducting external volume, we can take $\delta{\bf A}$ to be
\beq \delta{\bf A}={\bs{\xi}\bf\times B}. \eeq
In this gauge the condition that field lines are kept in place on the boundary, ${\bs{\xi}\bf\times B}=0$, thus requires setting $\delta{\bf A}=0$ on the boundary. The surface term then vanishes. Since we have no further constraints, $\delta{\bf A}$ is otherwise arbitrary inside $V$, so that the condition of vanishing energy variation is satisfied only when
\beq {\bs{\nb}\kern-1pt\bf\times B}=0 \eeq
inside $V$. That is, the field has a potential $\phi_{\rm m}$, $\bf{B}=-\bs{\nb}\bf\phi_{\rm m}$, governed by the Laplace equation, 
\beq \nb^2\phi_{\rm m}=0.\eeq 

The boundary condition on $\phi_{\rm m}$ is found from ${\bs\nb}\ncd\,{\bf B}=0$. The component of $\bf B$ normal to the surface ${\bf n}$,~ ${\bf n\cdot B}=\pa\phi_{\rm m}/\pa n$, is continuous across it. 
From potential theory it then follows that $\phi_{\rm m}$ is unique. There is only one potential for such Neumann-type boundary conditions, up to an arbitrary constant. This constant only affects the potential, not the magnetic field itself. The minimum energy state of a magnetic field with field lines kept anchored at the boundary is thus a uniquely defined potential field.

\subsubsection{Potential fields as energy source?}\label{poteng}
A consequence of the above is that the magnetic energy density $B^2/8\pi$ of a potential field is not directly available for other purposes. There is no local process that can extract energy from the lowest energy state, the potential field. Its energy can {\em only} be changed or exploited by changes in the boundary conditions.  If the fluid is perfectly conducting, the same applies to a force-free field. 

\subsection{The role of the boundaries in a force-free field}
\label{boundrole}
The external volume that keeps the field fixed in the above is more than a mathematical device. Its presence has a physical significance\,: the magnetic field exerts a stress on the boundary surface, as discussed in {section \ref{mstress}}. The external medium has to be able to take up this stress. 

Irrespective of its internal construction, a magnetic field represents a positive energy density, making it expansive by nature. The internal forces (the divergence of the stress tensor) vanish in a force-free field, but the stress tensor itself does not. It vanishes only where the field itself is zero. At some point there must be something else to take up internal stress, to keep a field together. In the laboratory this is the set of external current carrying coils. In astrophysics, the magnetic stress can be  supported by, for example a stellar interior, a gravitating cloud or an accretion disk. 
The intrinsically expansive nature of magnetic fields can be formalized a bit with an equation for the global balance between various forms of energy, the tensor virial equation (e.g.\ \hyperlink{kuls}{Kulsrud} Chapter 4.6).

Whereas a potential field is determined uniquely by the instantaneous value of the (normal component of) the field on its boundary, the shape of a force-free field configuration also depends on the {\em history} of things happening on its boundary.  Rotating displacements on the boundary wrap the field lines inside the volume around each other. The values of $\alpha$ which measure this wrapping reflect the entire history of fluid displacements on the boundary surface. 

Since the value of $\alpha$ is constant along a field line, it is also the same at its points of entry and exit on the boundary. The consequence is that, unlike in the case of a potential field, the construction of a force-free field \textit{is not possible in terms of a boundary-value problem}. A given force-free field has a unique distribution of $\alpha$ and $B_n$ on its surface. But there is no useful inverse of this fact, because the correspondence of the points of entry and exit of the field lines is not known until the force-free field has been constructed. Force-free fields must be understood in terms of the history of the fluid displacements at their boundary. See {section \ref{lowbet}} for an application.

\subsection{The vanishing force-free field theorem}\label{vanff}
A consequence of the expansive nature of magnetic fields is the following {\em Theorem}\,:
{\em A force-free field which does not exert stress at its boundaries vanishes everywhere inside}. To show this examine the following integral over a closed volume $V$ with surface $S$:
\bea 
a_{ij}\equiv{1\over 4\pi}\int x_i[({\bs{\nb}\bf\times B}){\bf \times B}]_j~\rd V=-\int x_i{\pa M_{jk}\over\pa x_k}~\rd V\nonumber\\
=\int\left [\delta_{ik}M_{jk}-{\pa\over\pa x_k}(x_iM_{jk})\right ]\rd V= \int M_{ij}~\rd V-\oint x_iM_{jk}n_k~\rd S,\label{aij}
\eea
where $n_i$ are again the components of the outward normal to the surface $S$, and $\bf M$ the magnetic stress tensor (\ref{mstress}). For a force-free field, $a_{ij}=0$. Taking the trace of (\ref{aij}) we find
\beq \oint x_iM_{ik} n_k~\rd S=\int M_{ii}~\rd V={3\over 8\pi}\int B^2~\rd V.\eeq
If the stress $M_{ik}$ vanishes everywhere on the surface $S$ (left hand side), it follows that $B=0$ everywhere inside $V$. 

Taking the surface of $V$ to infinity, the theorem also implies that it is not possible to construct a field that is force-free everywhere in an unbounded volume. ({\cpr Problem \ref{magstar}}~: models for magnetic A stars).

Since a force-free field has a higher energy density than a potential field, for given boundary conditions, twisting a magnetic field does not help to `keep it together', contrary to possible expectation. One might think that the `hoop stresses' caused by twisting might help, much like  an elastic band can be used to keep a bundle of sticks together. Instead, the increased magnetic pressure due to the twisting more than compensates the hoop stress. This is illustrated further in the next section.

Summarizing\,: like a fluid under pressure, a magnetic field has an internal stress $\bf M$. The divergence of this stress may vanish, as it does in a force-free field, but there still has to be a boundary capable of taking up the stress exerted on it by the field. If the stress cannot be supported by a boundary, it has to be supported by something else inside the volume, and the field cannot be force-free.

{\cpr Problem \ref{dipenerg}}\,:  stress on the surface of a uniformly magnetized sphere.

\section{Twisted magnetic fields} \label{twist}

\begin{figure}[t]
\hfil{\includegraphics[width=0.3\hsize]{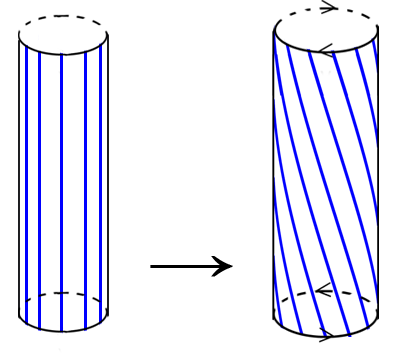}}\hfil
\caption{\small Bundle of field lines embedded in pressure equilibrium in a field-free plasma (a `flux tube'). Twisting it  (right) by applying a torque (arrows) at top and bottom causes its width to expand. The net current along the tube vanishes.}
\label{twistub}
\end{figure}

\subsection{Twisted fluxtubes}
\label{twistube}
Fig.\,\ref{twistub} shows a bundle of field lines embedded in a nonmagnetic plasma, with field strength ${\bf B}_0=B_0\,\hat{\bf z}$ along the $z$-axis. If $p_{\rm i}$ and $p_{\rm e}$ are the internal and external gas pressures, the condition for the tube to be in pressure equilibrium is $p_{\rm i}+B_0^2/8\pi=p_{\rm e}$  (see {\ref{boundequi}}).
We twist a section of the tube by rotating it in opposite directions at top and bottom. Imagine drawing a closed contour around the tube in the field-free plasma. With ({\ref{curr}}) and applying Stokes' theorem to this contour:
\beq  \oint_S {\bf j}\cdot{\rm d}{\bf S}={c\over 4\pi}\int_l {\bf B}\cdot{\rm d}{\bf  l}=0, \eeq
where $l$ is the path along the contour, and $S$ a surface bounded by the contour. Contrary to naive intuition, the net current along the tube therefore vanishes no matter how the tube is twisted (see also {\cpr problem \ref{ctwist}}).

The twisting has produced an azimuthal component $B_\varphi=B_0 \tan \chi$, where $\chi$ is the pitch angle of the twisted field. The magnetic pressure at the boundary of the tube has increased\,: $B^2=B_0^2+B_\varphi^2$, so the tube is not in pressure equilibrium anymore (see {\cpr problem \ref{twistp}}). The additional pressure exerted by $B_\varphi$ causes the tube to expand. 

One's expectation might have been that the tube radius would {\em contract} due to the hoop stress of the added component $B_\varphi$. We see that the opposite is the case. A view commonly encountered is that the twist in Fig.\,\ref{twistub} corresponds to a current flowing along the axis, and that such a current must lead to contraction because parallel currents attract each other. The hoop stress can in fact cause {\em some} of the field configuration in the tube to contract (depending on how the twisting has been applied), but its boundary always expands unless the external pressure is also increased (see {\cpr problem~\ref{ctwist}} to resolve the apparent contradiction). This shows how thinking in terms of currents as in a laboratory setup leads astray.  For more on mistakes made in this context see  Ch.\ 9 in \hyperlink{park1}{Parker (1979)}. For an illustration of the above with the example of MHD jets produced by rotating objects, see {section \ref{jet}}.

\subsection{Magnetic helicity}
\label{helicity}
If $\bf A$ is a vector potential of $\bf B$, the magnetic helicity $H$ of a field configuration is defined as the volume integral
\beq H=\int_V {\bf A}\cdot{\bf B}~ {\rm d}V. \label{hel}\eeq
Its value depends on the arbitrary gauge used for $\bf A$. It becomes a more useful quantity when the condition
\beq {\bf B}\cdot {\bf n}=0 \label{bdotn} \eeq
holds on the surface of the (simply connected) volume $V$. There are then no field lines sticking through its surface, the field is completely `contained' within $V$. In this case $H$ (dimensions G$^2$cm$^4$), has a definite value independent of the gauge. [For proofs of this and related facts, see section 3.5 in \hyperlink{mest}{Mestel (2012)}]. It is a global measure of the degree of twisting of the field configuration. 

\begin{figure}[b]
\hfil\includegraphics[width=0.45\hsize]{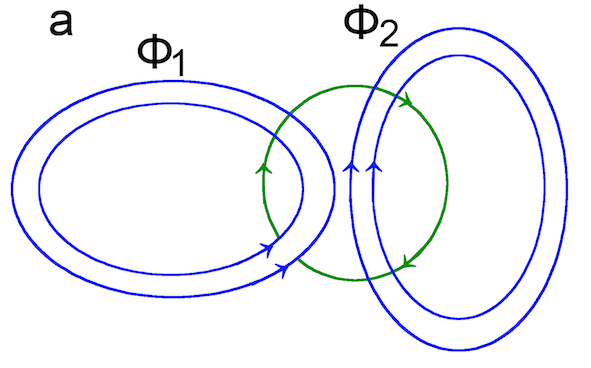}\hfil\includegraphics[width=0.45\hsize]{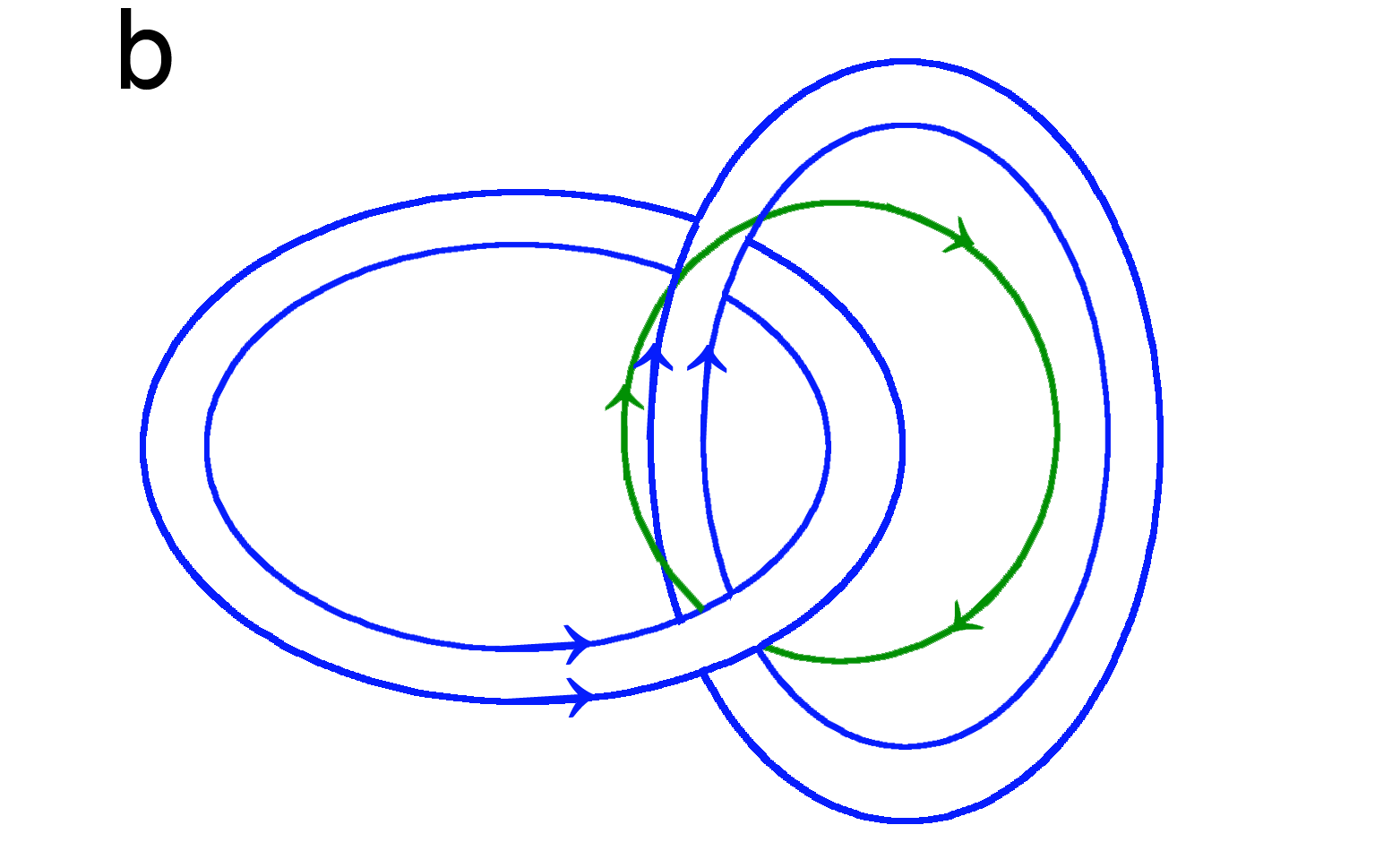}\hfil\caption{\small Helicity of configurations consisting of two untwisted magnetic loops. Green\,: a field line of the vector potential ${\bf A}_1$ of the left loop. In the left panel the loops are  not linked and the helicity of the configuration vanishes. When the loops are linked (b), the helicity does not vanish.}
\label{helicfig}
\end{figure}

Because of the gauge dependence, $H$ does not have locally definable values: there is no `helicity density'. This is not just a computational inconvenience. The physically significant reason is that $H$ is a topological quantity\,: it depends not only on a local degree of twisting, but also on the global `linking' of field lines within the configuration. This is illustrated in Fig.\,\ref{helicfig}, showing two loops of magnetic field, with magnetic fluxes $\Phi_1$ and $\Phi_2$. If they are not linked (panel a), each of the loops can be fit inside its own volume satisfying  (\ref{bdotn}), hence the helicity is a sum over the two loops. If they are untwisted, in the sense that all field lines close on going once around the loop without making turns around the cross section of the tube, the helicity of each vanishes. If the loops are linked, however, (panel b) it can be shown that $H=2\Phi_1\Phi_2$ ({\cpr problem \ref{link}}). For an intuitive picture how this comes about in terms of the definition (\ref{hel}), inspect the location of loop 2 relative to the vector potential of loop 1. In case (b), the field of loop 2 runs in the same direction as the field lines of ${\bf A}_1$, in case (a), much of it runs in the direction opposite to ${\bf A}_1$.
 
The importance of $H$ lies in its approximate conservation property. In ideal MHD, $H$ is an exactly conserved quantity of a magnetic configuration. Within a volume where (\ref{bdotn}) holds, $H$ does not change as long as perfect flux freezing holds everywhere. In practice, this is not a very realistic requirement, since  it may not be possible all the time to find a volume where (\ref{bdotn}) holds. In addition flux freezing is rarely exact over the entire volume. {\em Reconnection} ({sects.\ \ref{sheets}, \ref{reco}}) is bound to take place at some point in time at some place in the volume. This changes  the topology of the field, and consequently its helicity. This can happen even if the size of the reconnecting volume and the energy released in it are tiny. It turns out, however, that $H$  is often conserved in a more approximate sense. In laboratory experiments a helical magnetic field configuration that is not in equilibrium, or unstable, first evolves on the fast time scale of a few Alfv\'en crossing times $L/\va$. During this phase $H$ is approximately conserved. Following this, the magnetic helicity evolves more slowly by reconnection processes. 

Helicity does not behave like magnetic energy. In the presence of reconnection, $H$ can decrease during a relaxation process in which magnetic energy is released, but reconnection can also cause $H$ to {\em increase}. The evolution of magnetic fields on the surface of magnetically active stars (like the Sun) is sometimes described in terms of `helicity ejection'. In view of the global, topological nature of magnetic helicity, this usage of the term helicity is misleading. In a magnetic eruption process from the surface of a star magnetic helicity can decrease even when the thing being ejected is not helical at all. For more on this see {section \ref{recohel}}. 

\hypertarget{curhel}{}There are other ways of characterizing twist, for example the quantity $h=(\bs{\nb}{\bf\times B})\cdot {\bf B}/B^2$ with dimension 1/length, called current helicity. In a force-free field  ({sect.~\ref {fff}}) its value is equal to $\alpha$. In contrast to $H$ which is a property of the configuration as a whole, current helicity is a locally defined quantity. Since it does not have a conservation property, however, its practical usefulness is limited. As in the case of the electrical current (sect.\ \ref{statcur}), it makes no sense to talk about a flux of current helicity or advection of current helicity by a flow, for example.

Ultimately, these facts are all a consequence of the non-local (solenoidal vector-) nature of the magnetic field itself.  The usefulness of analogies with the conservation properties of other fluid quantities is intrinsically limited.

\section{Stream function}\label{stream}
Though in practice all magnetic fields in astrophysics are 3-dimensional in one or the other essential way, 2-dimensional models have played a major role in the development of astrophysical MHD, and their properties and nomenclature have become standard fare. Historically important applications are models for steady jets and stellar winds.

If ($\varpi,\varphi,z$) are cylindrical coordinates as before, an axisymmetric field is independent of the azimuthal coordinate:
\beq \pa{\bf B}/\pa\varphi=0.\eeq
(In numerical simulations axisymmetric models are sometimes called `2.5-dimensional').
Such a field can be decomposed into its {\em poloidal} and {\em toroidal} components
\beq {\bf B}={\bf B}_{\rm p}(\varpi,z)+{\bf B}_{\rm t} (\varpi,z),\eeq
where the poloidal field contains the components in a {\em meridional plane} $\varphi=$ cst.: 
\beq {\bf B}_{\rm p}=(B_\varpi,0,B_z), \eeq
and 
\beq {\bf B}_{\rm t}=B_\varphi\, \bs{\hat\varphi}\eeq 
is the azimuthal field component (the names toroidal and azimuthal are used interchangeably in this context). 
Define the {\em stream function}\footnote{~A stream function can be defined for any axisymmetric solenoidal vector. In (incompressible) fluid flows, it is called the Stokes stream function.}, a scalar $\psi$, by
\beq \psi(\varpi,z)=\int_0^\varpi\varpi B_z\,\rmd\varpi.\eeq
Using ${\rm div}\,{\bf B}=0$, the poloidal field can be written as
\beq 
B_z={1\over\varpi}\pa\psi/\pa\varpi,\qquad B_\varpi=-{1\over\varpi}\pa\psi/\pa z.\label{streamb}
\eeq
From this it follows that $\psi$ is constant along field lines\,: ${\bf B}\cdot\kern-1pt\bs{\nb}\psi=0$. The value of $\psi$ can therefore be used to label a field line (more accurately\,: an axisymmetric magnetic surface). It equals (modulo a factor $2\pi$) the magnetic flux contained within a circle of radius $\varpi$ from the axis. It can also be written in terms of a suitable axisymmetric vector potential $\bf A$ of ${\bf B}$:
\beq \psi=\varpi A_\varphi,\eeq
but is a more useful quantity for 2D configurations\footnote{~Plotting contour lines $\psi={\rm cst.}$, for example, is an elegant way to visualize field lines in 2-D.}. Stream functions can also be defined  more generally, for example in planar symmetry ({\cpr problem \ref{planar}}).

\section{Waves}\label{waves}
A compressible magnetic fluid supports three types of waves. Only one of these resembles the sound wave familiar from ordinary hydrodynamics. It takes some time to develop a feel for the other two. The basic properties of the wave modes of a uniform magnetic fluid also turn up in other MHD problems, such as the various instabilities. Familiarity with these properties is important for the physical understanding of time-dependent problems in general.

The most important properties of the waves are found by considering first the simplest case, a homogeneous magnetic field $\bf B$ in a uniform fluid initially at rest (${\bf v}=0$).
The magnetic field vector defines a preferred direction, but the two directions perpendicular to $\bf B$ are equivalent, so the wave problem is effectively two-dimensional. In Cartesian coordinates ($x,y,z$), take the initial magnetic field  ${\bf B}$ (also called `background' magnetic field) along the $z$-axis,
\beq {\bf B}=B\,{\bf\hat z}, \eeq
where $B$ is a (positive) constant. Then the $x$ and $y$ coordinates are equivalent, and one of them, say $y$, can be ignored by  restricting attention to perturbations $\delta q$ that are independent of $y$:
\beq \pa_y \delta q=0. \eeq

Write the magnetic field as ${\bf B}+\delta{\bf B}$, the density as $\rho+\delta \rho$, the pressure as $p+\delta p$, where $\delta{\bf B}$, $\delta p$ and $\delta\rho$ are small perturbations. Expanding to first order in the small quantities $\delta q$ (of which $\bf v$ is one), the linearized equations of motion and induction are:
\beq 
\rho{\pa{\bf v}\over\pa t}=-\bs{\nb}\delta p+{1\over 4\pi}(\bs{\nb}\kern-1pt\times\delta{\bf B}){\bf\times B}.
\eeq
\beq {\pa\delta{\bf B}\over\pa t}=\bs{\nb}\times({\bf v\times B}).\label{contl}\eeq
The continuity equation becomes
\beq {\pa\delta\rho\over\pa t}+\rho\bs{\nb}\kern-1pt\cdot{\bf v}=0\label{cntw}\eeq
(since the background density $\rho$ is constant). To connect $\delta p$ and $\delta\rho$, assume that the changes are adiabatic:
\beq \delta p=\delta\rho\left({\pa p\over\pa\rho}\right)_{\rm ad}\equiv \cs^2\delta\rho,\label{adia}\eeq
where the derivative is taken at constant entropy, and $\cs$ is the adiabatic sound speed. For an ideal gas with ratio of specific heats $\gamma=c_p/c_v$,
\beq c_{\rm s}^2=\gamma\, p/\rho. \label{cs} \eeq
The components of the equations of motion and induction are 
\beq 
\rho{\pa v_x\over\pa t}=-{\pa\delta p\over\pa x}+{B\over 4\pi}({\pa\delta B_x\over\pa z}-{\pa\delta B_z\over\pa x}),\qquad\rho{\pa v_z\over\pa t}=-{\pa\delta p\over\pa z},  \label{vxz}
\eeq
\beq \rho{\pa v_y\over\pa t}={B\over 4\pi}{\pa\delta B_y\over\pa z},\label{vy}\eeq
\beq 
{\pa\delta B_x\over\pa t}=B{\pa v_x\over\pa z}, \qquad {\pa\delta B_z\over\pa t}=-B{\pa v_x\over\pa x}, \label{indc}\eeq
\beq
 \qquad {\pa\delta B_y\over\pa t}=B{\pa v_y\over\pa z}.\label{by}
 \eeq
The $y$-components  only involve $\delta B_y$ and $v_y$. As a result,  eqs. (\ref{cntw} -- \ref{by}) have solutions in which $v_x=v_z=\delta B_x=\delta B_z=\delta p=\delta\rho=0$, with
 $\delta B_y$ and $v_y$ determined by  (\ref{vy}) and (\ref{by}). These can be combined into the wave equation 
\beq \left({\pa^2\over\pa t^2}-\va^2{\pa^2 \over\pa z^2}\right)(\delta B_y,v_y)=0,\label{vaeq}\eeq
where
\beq \va={B\over \sqrt{4\pi\rho}}\eeq
and the amplitudes are related by
\beq \delta \vert B_y\vert/B=\vert v_y\vert/\va.\label{amplrel}\eeq
These solutions are called {\em Alfv\'en waves} (or `intermediate wave' by some authors). Since they involve only the $y$ and $z$ coordinates, one says that they `propagate in the $y-z$ plane only'.

The second set of solutions involves $v_x,v_z,\delta B_x,\delta B_z,\delta p,\delta\rho$, while  $v_y=\delta B_y=0$. Since the undisturbed medium is homogeneous and time-independent, the perturbations can be decomposed into plane waves, which we represent in the usual way in terms of a complex amplitude. Any physical quantity $q$ thus varies in space and time as
\beq q=q_0\exp[i(\omega t-{\bf k\cdot x})],\label{plane}\eeq
where $\omega$ is the circular frequency and $q_0$ is a (complex) constant. The  direction of the wave vector ${\bf k}$ (taken to be real) is called the `direction of propagation' of the wave. With this representation time and spatial derivatives are replaced by $i\omega$ and $-i{\bf k}$ respectively. The continuity equation yields
\beq \delta\rho-\rho{\bf k\cdot v}/\omega=0.\eeq
With the adiabatic relation (\ref{adia}), this yields the pressure:
\beq \delta p=\cs^2\rho~{\bf k\cdot v}/\omega.\eeq
Substitution in (\ref{vxz}, \ref{indc}) yields a homogenous system of linear algebraic equations:
\beq 
\omega\rho v_x=k_x\rho(k_xv_x+k_zv_z)\cs^2/\omega-{B\over 4\pi}(k_z\delta B_x-k_x\delta B_z)
\eeq
\beq \omega\rho v_z=k_z\rho(k_xv_x+k_zv_z)\cs^2/\omega\eeq
\beq \omega\delta B_x=-k_zBv_x\eeq
\beq \omega\delta B_z=k_xBv_x.\eeq

Writing this as a matrix equation $A_{ij}q_j=0$, where ${\bf q}=(v_x,v_z,\delta B_x,\delta B_z)$, the condition that nontrivial solutions exist is ${\rm det}\,{\bf A}=0$. This yields the {\em dispersion relation}   for the compressive modes (the relation between frequency $\omega$ and wavenumber ${\bf k}$):
\beq 
\omega^4-\omega^2k^2(\cs^2+\va^2)+k_z^2k^2\cs^2\va^2=0.\label{dr0}
\eeq
Let $\theta$ be the angle between ${\bf k}$ and ${\bf B}$:
\beq k_x=k\sin\theta,\qquad k_z=k\cos\theta,\eeq
and set $u\equiv\omega/k$, the phase velocity of the mode. Then (\ref{dr0}) can be written as
\beq u^4-u^2(\cs^2+\va^2)+\cs^2\va^2\cos^2\theta =0.\label{dr}\eeq

The four roots of this equation are real and describe the {\em magnetoacoustic} waves. There are two wave modes, the {\em slow} and the {\em fast} magnetoacoustic modes, or `slow mode' and `fast mode' for short, each in two opposite directions of propagation (sign($\omega/k)=\pm 1$). They are also collectively called the {\em magnetosonic modes}.
\bigskip

\subsection{Properties of the Alfv\'en wave}\label{awavprop}

\begin{figure}[t]
\hfil
\href{run:Fig1.10-Traveling-A-wave.mp4}{\hspace{0.45\hsize}}
\href{run:Fig1.10-Standing-A-wave.mp4}{\hspace{0.45\hsize}}
\hspace{-0.9\hsize}
{\includegraphics[width=0.9\hsize]{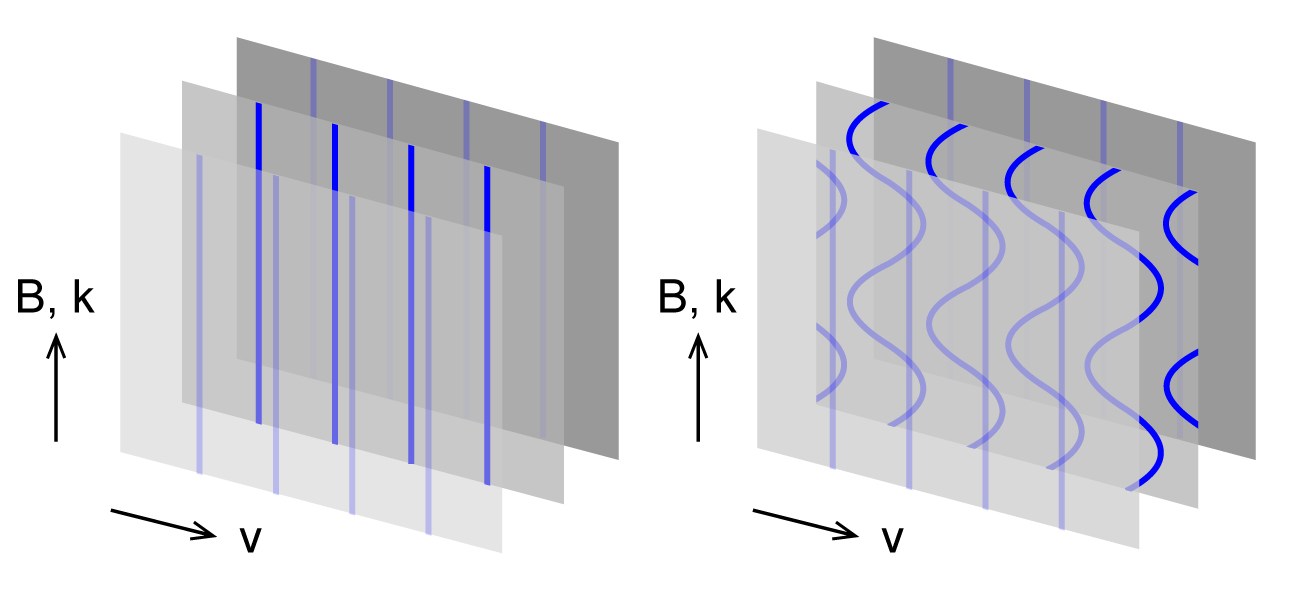}}\hfil
\caption{\small Linear Alfv\'en wave propagating along a plane magnetic surface 
(wave amplitude $\delta B/B$ exaggerated). 
\href{run:Fig1.10-Traveling-A-wave.mp4}{\can Traveling A-wave}~ 
\href{run:Fig1.10-Standing-A-wave.mp4}{\can Standing A-wave}.
}
\label{awave}
\end{figure}

With (\ref{plane}), ({\ref{vaeq}}) becomes:
\beq \omega^2=k_z^2\va^2,\eeq
the dispersion relation of the Alfv\'en wave. Since this only involves $k_z$ one says that Alfv\'en waves `propagate along the magnetic field'\,: their frequency depends only on the wavenumber component along ${\bf B}$. This does not mean that the wave travels along a single field line. In the plane geometry used here, an entire plane $x=$cst. moves back and forth in the $y$-direction (Fig.\,\ref{awave}).  Each plane $x=$ cst. moves independently of the others. 
The direction of propagation of the wave energy is given by the group velocity:
\beq {\pa\omega\over\pa{\bf k}} = \va \bf \hat{z}.\eeq
That is, the energy propagates along field lines, independent of the `direction of propagation' $\bf \hat{k}$ (but again\,: not along a single field line). From 
(\ref{amplrel}) it follows that
\beq {(\delta B)^2\over 8\pi}={1\over 2}v_y^2,\eeq
so there is `equipartition'  between magnetic and kinetic energy in an Alfv\'en wave (as in any harmonic oscillator). Further properties of the Alfv\'en wave are:
{\parindent=-5pt \noindent\par
{\bf -} A linear Alfv\'en wave is {\em incompressive}, i.e.\ $\delta\rho =0$, in contrast with the remaining modes. \noindent\par
{\bf -}  It {\href{run:transversal.mp4}{\can is {\em transversal}}\,: the amplitudes $\delta {\bf B}$ and $\bf v$ are 
perpendicular to the direction of propagation (as well as being perpendicular to $\bf B$). \noindent\par
{\bf -}  It is nondispersive\,: all frequencies propagate at the same speed. \noindent\par
{\bf -}  In an incompressible fluid, it remains linear at arbitrarily high amplitude (e.g.\ \hyperlink{robe}{Roberts} 1967).}
\parindent=12pt

An example of a plane linear Alfv\'en wave traveling on a single magnetic surface is shown in Fig.\,\ref{awave}.  The amplitude of the wave has been exaggerated in this sketch.  The magnetic pressure due to the wave, $(\delta B)^2/8\pi$ excites a compressive wave perpendicular to the plane of the wave (`mode coupling'). The wave will remain linear only when $\delta B/B\ll 1$. 

An Alfv\'en wave can propagate along a magnetic surface of any shape, not necessarily plane. For it to propagate as derived above, however, conditions on this plane have to be uniform (${\bf B},\rho, \cs$ constant). Though it cannot strictly speaking travel along a single field line, it can travel along a (narrow) magnetic surface surrounding a given field line\,: a `torsional' Alfv\'en wave.
\bigskip

\begin{figure}[h]
\hspace{0.2\hsize}\href{run:Torsional-A-wave.mp4}{\hspace{0.42\hsize}} \hspace{-0.62\hsize}
\hfil\includegraphics[width=0.4\hsize]{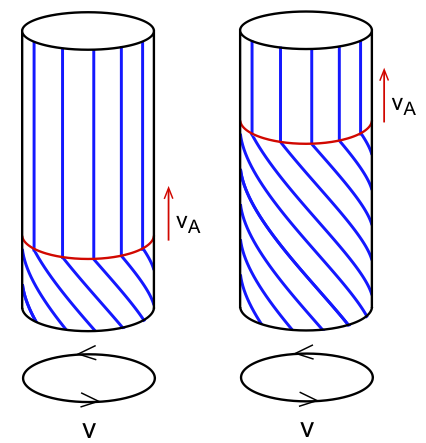}\hfil
\caption{\small\href{run:Torsional-A-wave.mp4}{\can Torsional Alfv\'en wave} propagating along a cylindrical magnetic surface.}
\label{torsA}
\end{figure}

\subsubsection{Torsional Alfv\'en waves}\label{torsa}
\label{torsiA}

Instead of moving an infinite plane $x= $cst., we can also produce a more localized Alfv\'en wave by `rotating a bundle of field lines'. This is illustrated in Fig.\,\ref{torsA}. At time $t=0$, a circular disk perpendicular to the (initially uniform) magnetic field is put into rotation at a constant angular velocity $\Omega$. A wave front moves up at the Alfv\'en speed along the rotating field bundle. This setup corresponds to an Alfv\'en wave of zero frequency\,: apart from the wave front it is time-independent. Of course, an arbitrary superposition of  wave frequencies is also possible. Since the wave is nondispersive, the wave front remains sharp. See {\cpr problems \ref{Awavej}} and {\cpr \ref{cylwave}}.
\bigskip

\subsection{Properties of the magnetoacoustic waves}
\label{ma}
In the second set of waves, the density and pressure perturbations do not vanish, so they share some properties with sound waves. The phase speed $u$ ({\ref{dr}}) depends on the sound speed, the Alfv\'en speed and the angle $\theta$ between the wave vector and the magnetic field. Introducing a dimensionless phase speed $\tilde u$,
\beq u=\tilde u (\cs\va)^{1/2},\label{utild}\eeq
 (\ref{dr}) becomes
 \beq \tilde u^4-({\cs\over\va}+{\va\over\cs})\tilde u^2+\cos^2\theta =0.\label{drtild}\eeq
 This shows that  the properties of the wave can be characterized by just two parameters\,: the angle of propagation $\theta$ and the ratio of sound speed to Alfv\'en speed, $\cs/\va=(\gamma\beta/2)^{1/2}$ (cf. {eqs.\ \ref{beta}, \ref{cs}}). The phase speed does not depend on frequency\,:  like the Alfv\'en wave, the waves are nondispersive. Since they are anisotropic, however, the phase speed is not the same as the group speed.

 The solutions of eq.~(\ref{dr}) are

 \beq 
 u^2=({\omega\over k})^2={1\over 2}(\cs^2+\va^2)[1\pm(1-4\cos^2\theta/b^2)^{1/2}],\label{roots}
 \eeq
where
\beq b={\cs\over \va}+{\va\over \cs}\ge 2.\eeq
  
\begin{figure}[t]
\hfil{\includegraphics[width=0.9\hsize]{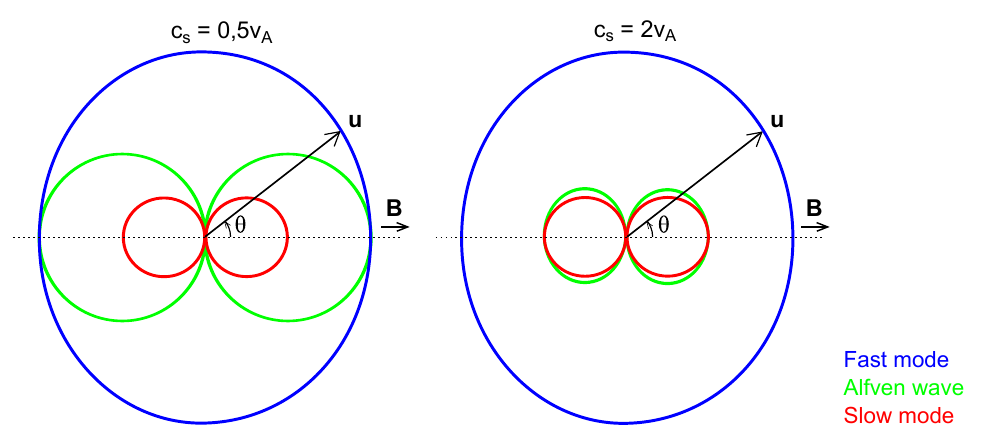}}\hfil
\caption{\small Propagation diagram for the MHD waves in a uniform medium, showing the phase speed (length of the vector $\bf u$) as a function of the angle $\theta$ of the wave vector with respect to the direction of the magnetic field.}
\label{phasediag}
\end{figure}

Hence $u^2$ is positive, and the wave frequencies real as expected (for real wave numbers $k$).
The $+$($-$) sign corresponds to the fast (slow) mode. The limiting forms for $b\gg 1$ (either because $\va\gg\cs$ or $\cs\gg\va$)  are of interest. In these limits the fast mode speed is
\beq 
u_{\rm f}\rightarrow (\cs^2+\va^2)^{1/2},\qquad (\cs\gg\va {\ \rm or\ } \va\gg \cs)\label{ufast}
\eeq
i.e.\ $u_{\rm f}$ is the largest of $\va$ and $\cs$, and is independent of $\theta$\,: the fast mode propagates isotropically in these limiting cases. 
For $\va\gg\cs$ or $\cs\gg\va$ the slow mode speed becomes:
\beq 
u_{\rm s}^2\rightarrow {\cs^2\va^2\over\cs^2+\va^2}\cos^2\theta, \qquad (\cs\gg\va {\ \rm or\ } \va\gg \cs)\label{uslow}
\eeq
i.e.\ $u_{\rm s}$ is smaller than both $\va$ and $\cs$ in these limiting cases, and its angular dependence  $\cos^2\theta$ is the same as that of the Alfv\'en wave.

\begin{figure}[t]
\hspace{0.07\hsize}\href{run:Fig1.13.mp4}{\hspace{0.8\hsize}}\hspace{-0.87\hsize}
\hfil\includegraphics[width=0.8\hsize]{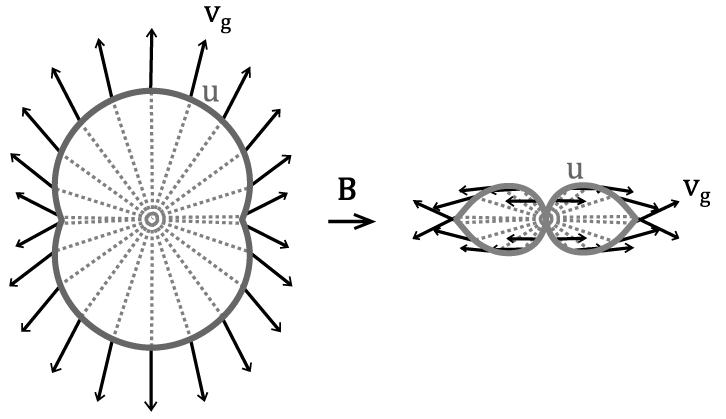}\hfil
\caption{\small Propagation diagrams of the fast mode (left) and slow mode (right) for  $\va=\cs$. Curves (u) show the phase speed as in Fig.\,\ref{phasediag}. Arrows\,: direction of the group speed ${\bf v}_{\rm g}$ as a function of the direction (dotted) of the phase speed. 
\hfill\break 
\href{run:Fig1.13.mp4}{\can Propagation diagrams with  $v_{\rm A}/c_{\rm s}$ varying from 0.5 to 2.}
}
\label{slowpg}
\end{figure}

Another interesting limiting case is $\cos^2\theta\ll1$, that is, for wave vectors nearly perpendicular to $\bf B$. Expressions (\ref{ufast}) and (\ref{uslow}) hold in this limit as well, but they now apply for arbitrary $\va$, $\cs$. In this context, (\ref{ufast}) is called the {\em fast magnetosonic speed} while the quantity
\beq c_{\rm c}={\cs\va\over(\cs^2+\va^2)^{1/2}} \label{cusp}\eeq
is called the {\em cusp speed}. 

The nomenclature used in describing wave propagation can be a bit confusing in the case of the Alfv\'en and slow waves. Because of their anisotropy,  the wave vector $\bf k$ is not the direction in which the energy of the wave flows. The case $\cos^2\theta\ll1$ is called `propagation perpendicular to $\bf B$', though the wave energy actually propagates {\em along} $\bf B$  in the Alfv\'en wave and (approximately) also in the slow mode.  {\em Propagation diagrams}, shown in {Fig.~\ref{phasediag}} can help visualization. These polar diagrams show the absolute value of the phase speed,  as a function of the angle $\theta$ of $\bf k$ with respect to $\bf B$. 

The curves for the Alfv\'en wave are circles. The slow mode is neither a circle nor an ellipse, though the difference becomes noticeable only when $\va$ and $\cs$ are nearly equal. This is shown in {Fig.~\ref{slowpg}}, the propagation diagrams for $\va/\cs=1$. It also shows the direction of the group speed, as a function of the angle $\theta$ of the wave vector. At $\va/\cs=1$ the group speed shows the largest variations in direction, both in the fast and in the slow mode. At other values of $\va/\cs$ the direction of the group speed of the slow mode is close to $\bf B$ for all directions of the wave vector, like in the Alfv\'en wave, while the group speed of the fast mode is nearly longitudinal, as in a sound wave. 

The symmetry between $\cs$ and $\va$ suggested by ({\ref{drtild}}) and the propagation diagrams in {Figs.~\ref{phasediag}} and {\ref{slowpg}} is a bit deceptive. The behavior of the waves is still rather different for $\beta>1$ and $\beta<1$. 
For $\beta>1$ the fast mode behaves like a sound wave, modified a bit by the additional restoring force due to magnetic pressure (Fig.\,\ref{fluiddisp}, top left). At low $\beta$ (top right), it propagates at the Alfv\'en speed, with fluid displacements nearly perpendicular to $\bf B$ as in the Alfv\'en wave, but in contrast with the Alfv\'en wave nearly isotropically. 

\begin{figure}[t]
\center{\includegraphics[width=0.95\hsize]{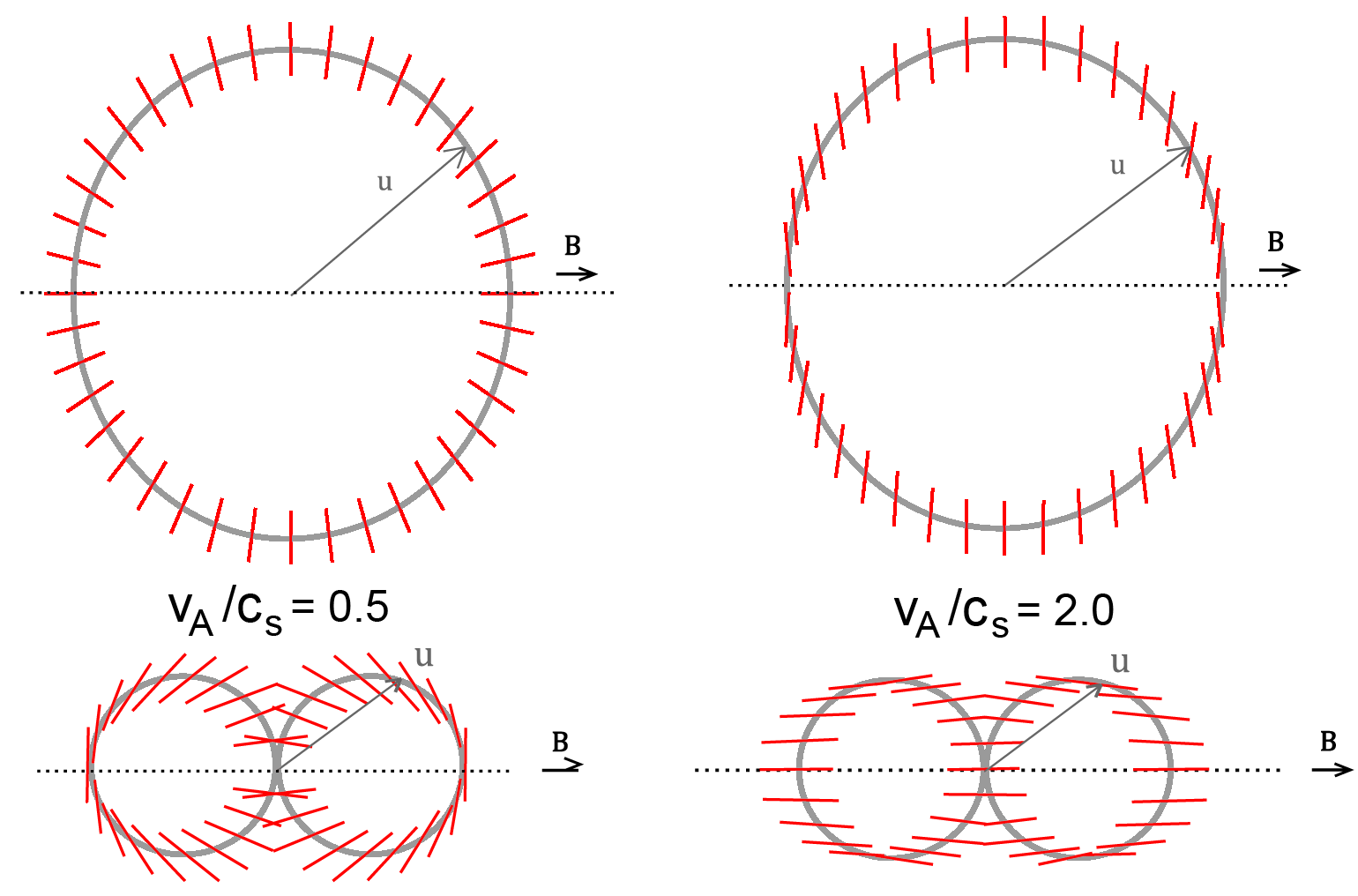}\hfil}
\caption{\small Directions of fluid displacement (red) in the fast mode (top) and  slow mode (bottom) at low (left) and high (right) $\va/\cs$. Grey\,: angular dependence of the phase speed as in Fig.\,\ref{phasediag}}
\label{fluiddisp}
\end{figure}

The slow mode has the most complex properties of the MHD waves. 
At low $\beta$ (high $\va/\cs$) it behaves like a 1-dimensional sound wave `guided by the field'. The fluid executes a `sloshing' motion along field lines (Fig.~\ref{fluiddisp}, lower right). Sound waves propagating along neighboring field lines behave independently of each other in this limit. The dominance of the magnetic field prevents the wave's pressure fluctuations from causing displacements perpendicular to the field (in other words, the nonlinear coupling to other waves is weak). 

In the high-$\beta$ (low $\va/\cs$) limit, the propagation diagram of the slow wave approaches that of the Alfv\'en wave.  The fluid displacements ({Fig.~\ref{fluiddisp}}, lower left) become perpendicular to the wave vector. The flow is now in the $x-z$ plane instead of the $y$-direction. At high $\beta$, the slow mode can be loosely regarded (in the limit $\beta\rightarrow \infty$\,: exactly) as a second {\em polarization state} of the Alfv\'en wave (polarization here in the `wave' sense, not the polarization of a medium in an electric field as described in {sect.~\ref{polariz}}). 

See the flow patterns and field line shapes of standing slow mode waves for these parameter combinations\,: 
\href{run:Slowmode-vA0.5cs-theta0.1.mp4}{\can ($\va/\cs=0.5$, $\theta=0.1$)}, 
\href{run:Slowmode-vA0.5cs-theta0.4.mp4}{\can ($\va/\cs=0.5$, $\theta=0.4$)},
\href{run:Slowmode-vA2cs-theta0.1.mp4}{\can ($\va/\cs=2$, $\theta=0.1$)}, and 
\href{run:Slowmode-vA2cs-theta0.4.mp4}{\can ($\va/\cs=2$, $\theta=0.4$)} ($\theta$ in radians).

\subsubsection{Summary of the magnetoacoustic properties}
The somewhat complex behavior of the modes is memorized most easily in terms of the asymptotic limits $\va \gg \cs$ and $\va \ll \cs$\,:\\
- The energy of the fast mode propagates roughly isotropically; for $\va \gg \cs$ with the speed of an Alfv\'en wave and displacements perpendicular to the field ({Fig.~\ref{fluiddisp}}, top right), for $\va \ll \cs$ with the speed and fluid displacements of a sound wave (top left).\\
- The energy of the slow mode propagates roughly along field lines; for $\va \gg \cs$ at the sound speed with displacements along the field  (Fig.~\ref{fluiddisp}, bottom right), for $\va \ll \cs$ at the Alfv\'en speed and  with displacements that vary from perpendicular to $\bf B$ to parallel, depending on the wave vector (bottom left).

\subsubsection{Waves in inhomogeneous fields}
The waves as presented in this section are of course rarely found in their pure forms. When the Alfv\'en speed or the sound speed or both vary in space, the behavior of MHD waves is richer. The subject of such inhomogeneous MHD waves has enjoyed extensive applied-mathematical development. Important concepts in this context are `resonant absorption' and `linear mode conversion'. 

\section{Poynting flux in MHD}\label{poy}
The Poynting flux $\bf S$ of electromagnetic energy:
\beq {\bf S}={c\over 4\pi}{\bf E}\times{\bf B}, \eeq
is usually thought of in connection with electromagnetic waves in vacuum. It is in fact defined quite generally, and has a nice MHD-specific interpretation.
With the MHD expression for the electric field, ${\bf E}=-{\bf v\times B}/c$, we have
\beq {\bf S}={1\over 4\pi}{\bf B}\times({\bf v\times B}).\eeq
$\bf S$ thus vanishes in flows parallel to $\bf B$. Writing out the cross-products, and denoting by ${\bf v}_\perp$ the components of $\bf v$ in the plane perpendicular to $\bf B$:
\beq {\bf S}={\bf v_\perp}{B^2\over 4\pi}. \label{poymhd}\eeq 

An MHD flow does not have to be a wave of some kind for the notion of a Poynting flux to apply. It also applies in other time dependent flows, and even in steady flows  (the MHD flow in steady magnetically accelerated jets from accretion disks or black holes for example). See {\cpr problem \ref{PoyA}}.

Expression (\ref{poymhd}) suggests an interpretation in terms of magnetic energy being carried with the fluid flow. But the magnetic energy density is $B^2/8\pi$. To see where the missing factor 2 comes from, consider the following analogy from ordinary hydrodynamics. In adiabatic flow, i.e.\ in the absence of dissipative or energy loss processes so that entropy is constant in a frame comoving with the flow, the equation of motion can be written as\footnote{~For the following see also Landau \& Lifshitz \S5}
\beq {\rmd {\bf v}\over\rmd t}=-{\bs \nb} w ~~+({\rm additional~ forces})/\rho,\eeq
where $w$ is the heat function or {\em enthalpy} of the fluid,
\beq w=(e+p)/\rho,\eeq
$e$ is the internal (thermal) energy per unit volume, and $p$ the gas pressure. At constant entropy, $w$ satifies ${\bs \nb} w={\bs \nb}p/\rho$. Conservation of energy in a flow can be expressed in terms of the {\em Bernoulli integral}. In the absence of additional forces, it is given by:
\beq E={1\over 2}v^2+w.\eeq
In a steady adiabatic flow, $E$ can be shown to be constant along a flow line (the path taken by a fluid element). 

\begin{figure}[t]
\center{\includegraphics[width=0.8\hsize]{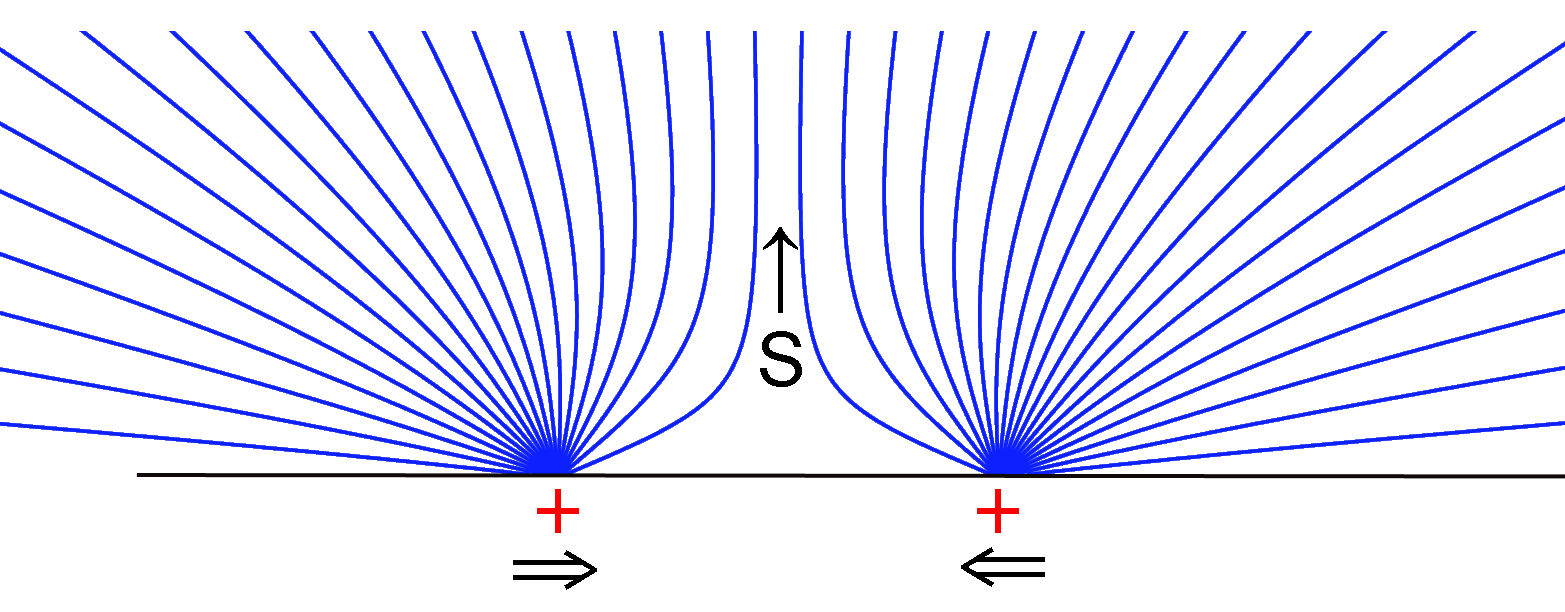}\\
\includegraphics[width=0.8\hsize]{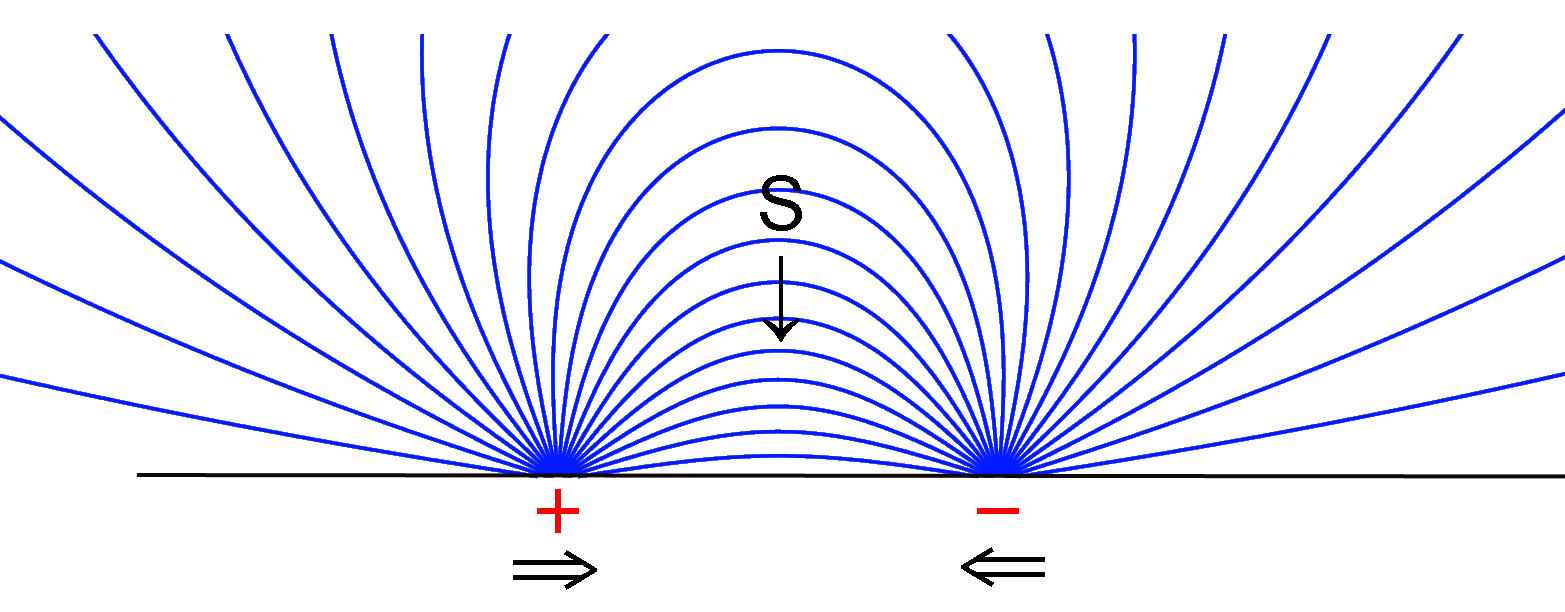}}
\caption{\small Field lines extending from patches of magnetic flux on a surface. When the patches are displaced with respect to each other (horizontal arrows), the sign of the Poynting flux $S$ depends on the relative polarities of the patches.}
\label{pf}
\end{figure}

The above shows that the thermodynamic measure of thermal energy in hydrodynamic flows is the enthalpy, not just the internal energy $e$. What is the physical meaning of the additional pressure term in the energy balance of a flow? In addition to the internal energy carried by the flow, the $p\,\rmd V$ work done at the source of the flow must be accounted for in an energy balance\,: this adds the additional $p$. The analogous measure of magnetic energy in MHD flows would be:
\beq w_{\rm m}\equiv e_{\rm m}+p_{\rm m},\label{menth}\eeq
where $e_{\rm m}$ and $p_{\rm m}$ are the magnetic energy density and magnetic pressure. Both are equal to $B^2/8\pi$, hence $w_{\rm m}=B^2/4\pi$. The Poynting flux in MHD, (\ref{poymhd}) can thus be interpreted as the  flux of a {\em magnetic enthalpy} in the plane perpendicular to $\bf B$. Note, however, that it cannot be simply added to the hydrodynamic enthalpy (in a Bernoulli integral, for example), since the two flow in different directions.

A Poynting flux can be evaluated from a solution of the MHD equations. It can be useful for the interpretation of results, but intuitions about Poynting flux can also lead astray. As an example, imagine a force-free field configuration as in Fig.\ \ref{pf}. Two patches of magnetic flux with field lines extending into the volume above the surface are brought together by a slow displacement at the boundary (horizontal arrows). If the patches have the same polarity (sign of $\bf B\cdot n$), they repell each other, and the displacement requires energy input to be provided at the surface. The Poynting flux is upward (into the volume). If they are of opposite polarity, the Poynting flux is {\it downward}\,: the displacement taps energy  from the field configuration. 

\section{Magnetic diffusion}
\label{diffusion}
Assume (without justification for the moment) that in the fluid frame the current is proportional to the electric field:
\beq {\bf j}^\prime=\sigma_{\rm c} {\bf E}^\prime,\label{ohm} \eeq
that is, a linear `Ohm's law' applies, with electrical conductivity $\sigma_{\rm c}$ (not to be confused with $\sigma$ the charge density).
In the non-relativistic limit (where ${\bf B}={\bf B}^\prime$, ${\bf j}={\bf j}^\prime=c/4\pi\,\bs{\nb}\times{\bf B}$): 
\beq {{\bf j}\over\sigma_{\rm c}}={\bf E}^\prime=({\bf E}+{{\bf v}\over c}\times{\bf B}),\label{ohm1}\eeq
and the induction equation ({\ref{inM}}) becomes
\beq 
{\pa{\bf B}\over \pa t}=\bs{\nb}\times[{\bf v}\times{\bf B}-\eta\bs{\nb}\times{\bf B}],\label{inddif}
\eeq
where 
\beq \eta={c^2\over 4\pi \sigma_{\rm c}}, \eeq
with dimensions cm$^2$/s is called the {\em magnetic diffusivity}.
Eq.\ (\ref{inddif}) has the character of a diffusion equation. If $\eta$ is constant in space :
\beq 
\pa_t{\bf B}=\bs{\nb}\times({\bf v}\times{\bf B})+\eta{\nabla}^2{\bf B}.\qquad ( \eta={\rm cst.})\label{inddifr}
\eeq
(using {\rm div}\,${\bf B}=0$). The second term on the right has the same form as in diffusion of heat (by thermal conductivity), or momentum (by viscosity).  Eq.\ (\ref{inddif}) is a parabolic  equation, which implies that disturbances can propagate at arbitrarily high speeds.  It is therefore not valid relativistically; there is no simple relativistic generalization.

In the presence of diffusion ($\eta\ne 0$) Alfv\'en's theorem  ({\ref{fluxc}}) no longer applies. The flux of magnetic field lines through a loop comoving with the fluid (as in Fig.\,\ref{loop}) is no longer constant in time. Individual field lines cannot be identified anymore with the fluid elements `attached' to them, which was so useful for visualizing time dependent magnetic fields.  

\begin{figure}[t]
\hfil{\includegraphics[width=0.4\hsize]{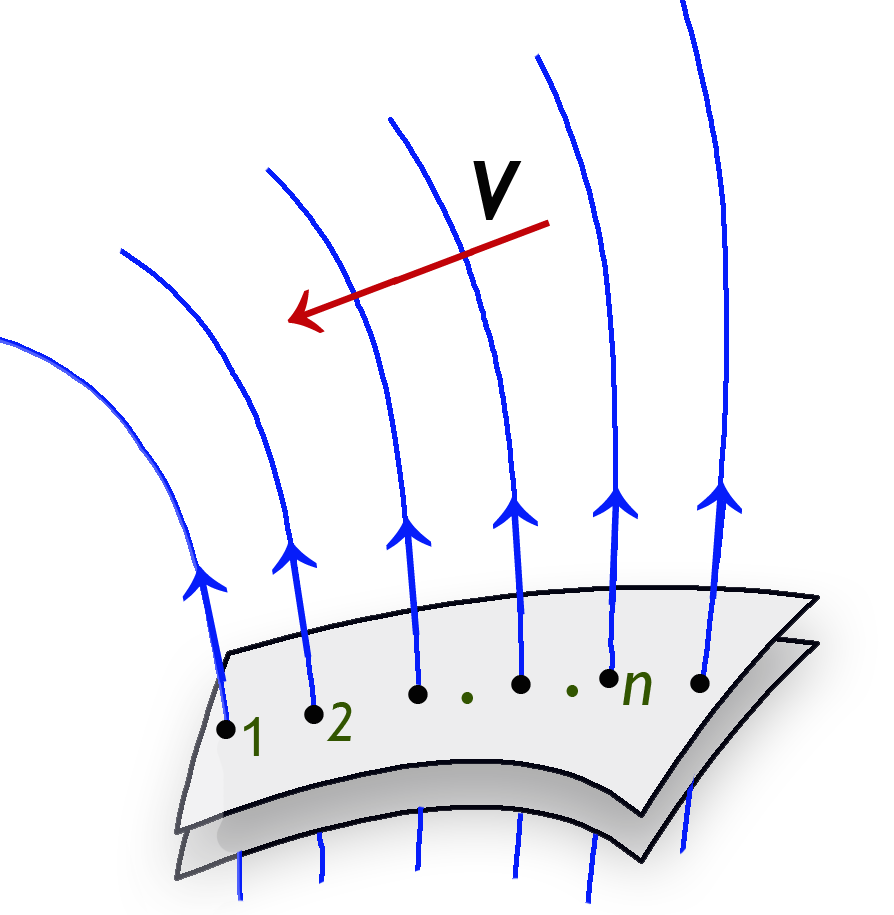}}\hfil
\caption{\small Flow of fluid `across field lines' when magnetic diffusion is present. Though not carried with the flow, the field lines can still be labeled in a time-independent way if there is a surface (shaded) where diffusion is sufficiently slow for the field lines to be `frozen-in' on the time scales of interest.}
\label{anchor}
\end{figure}

Even when diffusion is important, however, it is often possible to label field lines as if they had an individual, time independent identity.  Sufficient for this is that somewhere in the volume considered there is a region where field lines can be associated with fluid elements in a time-independent manner. This could for example be a solid highly conducting surface, or a fluid region of  high conductivity. In such a region field lines can again be labeled as before. Thanks to div$\,{\bf B}=0$, this label can then be extended along the entire length of each field line. We can then speak of a speed at which field lines diffuse across a fluid (if the fluid is more or less at rest), or how fast a fluid flows across the field (if the field is more or less stationary). The latter case is illustrated in Fig.~\ref{anchor}. 

The importance of diffusion, i.e.\ the influence of the second term in (\ref{inddif}) can be quantified with a dimensional analysis, in the same way as was done for the equation of motion in {sect.~\ref{tilde}}. If $L$ and $V$ are characteristic values for the length scale and velocity of the problem, this yields a nondimensional form of (\ref{inddifr}):

\beq 
{\pa{\bf\tilde B}\over\pa{\tilde t}}=\bs{\tilde\nb}\times({\bf \tilde v}\times{\bf \tilde B})+{1\over {\rm R_m}}{\tilde\nabla}^2{\bf \tilde B}, \label{inddift}
\eeq
where ${\rm R_m}$ is the {\em magnetic Reynolds number}\,:
\beq {\rm R_m}=L\,V/\eta.\eeq

If ${\rm R_m}$ is large, the most common case in astrophysics ({\cpr exercise \ref{Rm}}), the diffusion term can be ignored and ideal MHD used except in subvolumes  where small length scales develop, such as in current sheets.

The form of the diffusion term in (\ref{inddif}) is appropriate for processes that can be represented by an `Ohm's law'. Among other things, this assumes that the mean velocities of all components of the fluid (the neutrals and the two charge carriers) are the same to sufficient accuracy. This is not the case if the mean velocity of the electrons due to the current they carry, or if  the velocity of the  neutral component relative to the charged components is significant. In these cases additional terms appear in (\ref{inddif}), corresponding to {\em Hall drift} and {\em ambipolar diffusion}, respectively ({see \ref{hall-am}}). 

\section{Current sheets}
\label{sheets}

When field lines of different directions get close together, the finite conductivity of the plasma has to be taken into account. Field lines are no longer tied to the fluid, and magnetic energy is released. Normally the topology of the field lines also changes\,: they {\em reconnect}. 

A prototypical case is shown in Fig.\,\ref{csheet}\,: a current sheet separates magnetic field lines running  in opposite directions. The sheet runs perpendicular to the plane of the drawing. The close-ups illustrate that significant differences can exist in the internal structure of the sheet. In a pressure supported sheet (top right) the field strength vanishes in the middle. Equilibrium of the structure is provided by a varying fluid pressure $p$, such that $p+ B^2/8\pi=\rm{cst}$ across the sheet. The current is perpendicular to the magnetic field (and to the plane of the drawing).

\begin{figure}[h]
\hfil{\includegraphics[width=0.7\hsize]{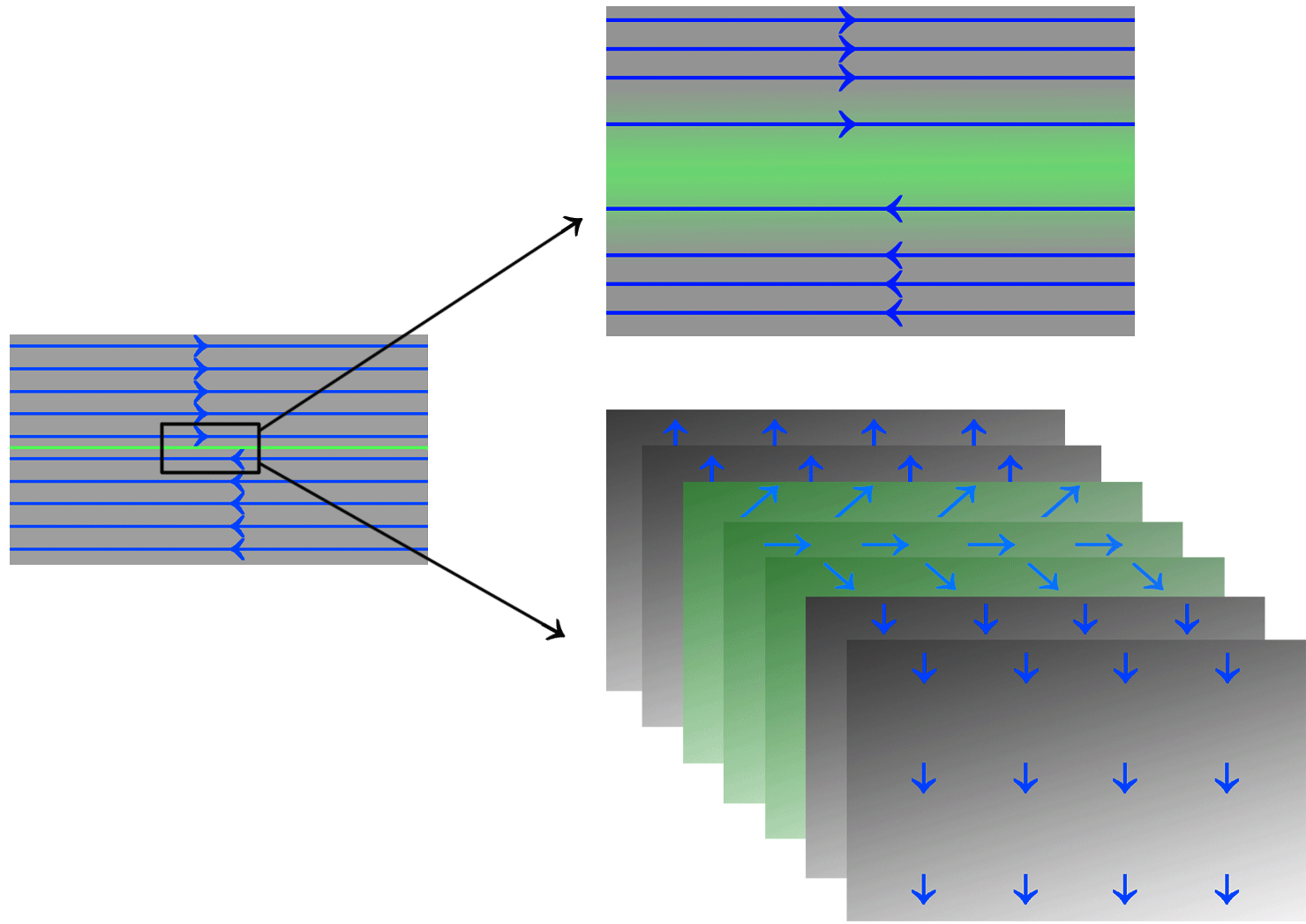}}\hfil
\caption{\small The two types of current sheets at a 180$^\circ$ change in direction of the field lines (blue). Top right\,:  a two-dimensional sheet perpendicular to the plane of the drawing.  $B$ vanishes in the middle, the current is perpendicular to $\bf B$ and to the plane. Current density indicated in green. Bottom right\,:  a force-free sheet. The current is in the plane of the sheet, parallel to {\bf B}, and $B$ is constant across the layer.}
\label{csheet}
\end{figure}

In a {\em force-free current sheet} on the other hand (botton right), fluid pressure is constant (more generally\,: small, $\beta\ll1$). The absolute value $B$ of the field strength is constant across the sheet, and the current is parallel to the field lines instead of perpendicular. Instead of the field strength vanishing in the middle, the direction of the field lines rotates over 180$^\circ$. In this case the space between the opposing field lines is filled with other field lines, instead of  with fluid. The mechanism of reconnection is correspondingly different in the two cases ({sect.~\ref{reco}}).

\chapter{Supplementary}

\label{epicyc}

\section{Alfv\'en's theorem}
\label{der.1a}
This somewhat intuitive derivation follows \hyperlink{ferr}{Ferraro \& Plumpton}. See also \hyperlink{kuls}{Kulsrud}.
Consider,  as in {section \ref{motion}}, a surface $S(t)$ bounded by a loop of fluid elements $C(t)$, which changes in shape and position with time. Denote the path length along $C$ by $s$, the unit vector along $C$ by $\bf \hat{s}$. The flux through the loop is
\beq \Phi(t)=\int_S{\bf B}({\bf r},t)\cdot \rd{\bf S}, \eeq
where ${\bf S}=S{\bf n}$, with {\bf n} a unit vector normal to $S$, and $\rd \bf S$ the surface element of $\bf S$. To be shown is that $\Phi$ is constant in time if the magnetic field satisfies the ideal MHD induction equation $\pa_t\,{\bf B}=\bs{\nb}\times({\bf v\times B})$. 

\begin{figure}
\hfil{\includegraphics[width=0.45\hsize]{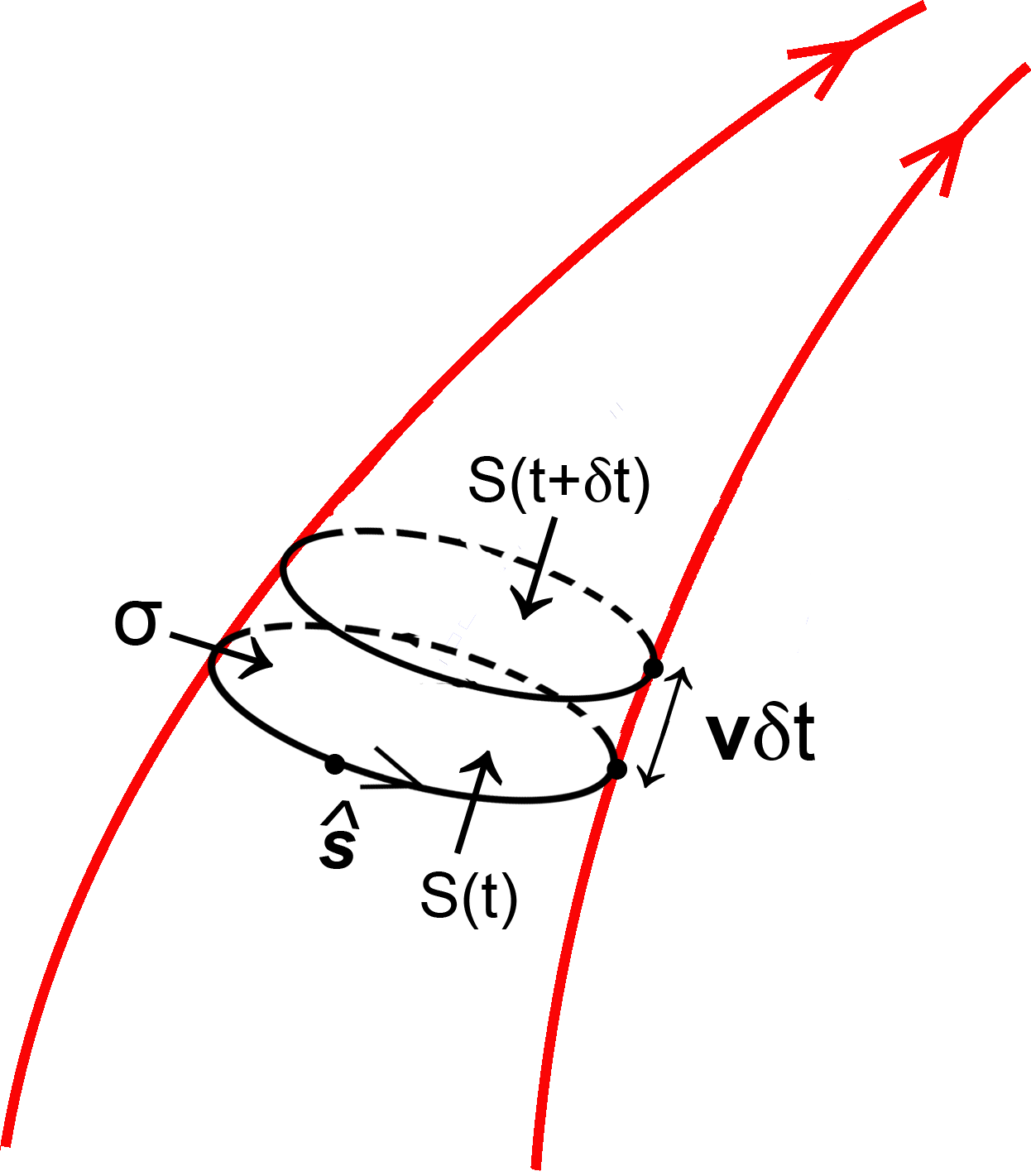}}\hfil
\caption{\label{APF}\small Flux loops for heuristic derivation of Alfv\'en's theorem. Red\,: flow lines}
\end{figure}

Fig.\,\ref{APF} shows the loop at two instants in time separated by an infinitesimal interval $\delta t$. We are interested in the difference $\delta\Phi$ in the flux through the loop between these two instants :
\beq 
\delta\Phi=\int_{S(t+\delta t)}{\bf B}({\bf r},t+\delta t)\cdot \rd{\bf S}-\int_S{\bf B}({\bf r},t)\cdot \rd{\bf S}.
\eeq
We write this as
\beq \delta\Phi= I_1+I_2,\label {i1i2}\eeq
with
\beq 
I_1=\int_{S(t+\delta t)}{\bf B}({\bf r},t+\delta t)\cdot \rd{\bf S}-\int_{S(t)}{\bf B}({\bf r},t+\delta t)\cdot \rd{\bf S},
\eeq
\beq 
I_2=\int_{S(t)}{\bf B}({\bf r},t+\delta t)\cdot \rd{\bf S}-\int_{S(t)}{\bf B}({\bf r},t)\cdot \rd{\bf S}.
\eeq
In this way we have divided the change in $\Phi$ into a part ($I_1$) that is due to the change in position of the loop, and a contribution $I_2$ due to the change of ${\bf B}$ in time at a fixed location. Hence
\beq I_2=\delta t\int_S{\pa{\bf B}\over\pa t}\cdot \rd{\bf S}.\label{i2} \eeq

We now apply the divergence theorem to the volume $V$ defined by the surfaces $S(t+\delta t)$, $S(t)$, and the side surface $\sigma$ traced out by the loop in the course of its displacement from $t$ to $t+{\delta} t$. With $\bs{\nb}\cdot {\bf B}=0$ and using Gauss's theorem, the total flux through these surfaces is
\beq 
\int_{S(t+{\delta} t)}{\bf B\cdot}\, \rd {\bf S}-\int_{S(t)}{\bf B\cdot}\, \rd{\bf S}+\int_\sigma{\bf B\cdot}\,\rd{\bf S}=\int_V{\bs{\nb}\bf\cdot B}~\rd\tau=0,
\eeq
where $\rd\tau$ is the volume element of $V$.The minus sign in the second term appears because we want the flux $\Phi$ to have the same sign at the two instants $t$ and $t+\delta t$. With the sign of $\Phi$ chosen to correspond to the outward normal on the upper surface in Fig.\,\ref{APF}, and the outward normal being required for application of the theorem, a change of sign is thus needed for  the lower surface, represented by the second term.

To evaluate the flux through the side surface, we divide it into infinitesimal surface elements spanned by the vectors $\rd{\bf s}$ and ${\bf v}\delta t$, where $s$ is the arc length along the loop. The sign of $\rd{\bf s}$ is chosen such that the vector $\rd{\bf S}=\rd{\bf s}\times{\bf v}\delta t$ has the direction of the outward normal on $\sigma$. With these definitions we have
\beq 
\int_\sigma{\bf B\cdot}\rd{\bf S}=\oint_C{\bf B\cdot}({\rd\bf s\times v})\delta t=\delta t\oint_C{\bf v\times B}\cdot\rd{\bf s}.
\eeq
Applying Stokes' theorem to the closed contour $C$:
\beq
\int_\sigma{\bf B\cdot}\rd{\bf S}=\delta t\int_S\bs{\nb}\bf\times({\bf v \times B})\cdot\rd{\bf S}.
\eeq
Hence $I_1$ becomes
\beq 
I_1=\int_{S(t+\delta t)}{\bf B}\cdot \rd\,{\bf S}-\int_{S(t)}{\bf B}\cdot \rd\,{\bf S}=-\delta t\int_S\bs{\nb}\bf\times({\bf v \times B})\cdot\rd{\bf S},
\eeq
and with (\ref{i1i2}, \ref{i2}):
\beq
\delta\Phi=\delta t\int_S[{\pa{\bf B}\over\pa t}-\bs{\nb}\bf\times({\bf v \times B})]\cdot\rd{\bf S}.
\eeq
If the induction equation is satisfied, the integrand vanishes and we have shown that
\beq {{\rd\Phi}\over\rd t}=0.\eeq

\section{Conditions for `flux freezing'}
\label{freeze}
The `freezing' of field lines to the fluid described by ({\ref{fluxc}}) was a consequence of the expression for the electric field, 
\beq\mathbf{E}=-\mathbf{v\times B}/c,\label{ee1}\eeq
from which the MHD induction equation followed directly. In deriving this we have made the assumption of perfect conductivity. Flux freezing is often described intuitively in a rather different way, however, namely in terms of  charged particles being bound to the field by their orbital motion of  around magnetic field lines. Collisions between particles would make them jump to orbits on neighboring field lines. This `orbit' picture of tight coupling applies if the {\em cyclotron frequency} $\omega_B$ of the charges (also called gyro frequency of Larmor frequency) is much larger than  the  collision rate $\omega_{\rm c}$ . Such a plasma is said to be {\em collisionless} or  strongly {\em magnetized}. 

In this strongly magnetized limit,  the conductivity in a direction perpendicular to the magnetic field would therefore {\em vanish}, since the charges would just circle around, instead of jumping between field lines. This is the opposite of the infinite conductivity assumed in deriving (\ref{ee1}). Because of this it is sometimes argued that an MHD approximation is not valid in a collisionless plasma. Eq.\ (\ref{ee1}) still applies in this limit, however, for a different reason. If an electric field is present and collisions negligible, particles  drift across the magnetic field with a velocity, independent of their charge, of $\mathbf{v}=-c\,\mathbf{ E\times B}/B^2$ (the so-called `E cross B drift', {\cpr problem \ref{pr.2}}). 
In a frame comoving with this velocity,  the electric field vanishes. If the mean velocities of the different particle species in the plasma are also representative of the velocity $\bf v$ of the fluid as a whole, the observation of an electric field component perpendicular to the magnetic field just means that we are not observing in the rest frame of the fluid. Translating to this frame the electric field disappears again, just as in the case of perfect conductivity.

If there is a significant difference in the mean velocity between electrons and ions, or if the velocity of the charge-neutral component of the plasma relative to the charge carriers is important, we are in an intermediate regime between the limits of perfect conductivity and perfect magnetization. The MHD induction equation must then be extended to include Hall drift and/or ambipolar diffusion terms, see {section \ref{hall-am}}. 

\section{Magnetic surfaces, Euler potentials}
\label{Euler}

An explicit {\em labeling} of field lines with numbers that are constant along field lines, is sometimes useful. In 3 dimensions such a label requires 2 scalars, counting in two directions perpendicular to the field lines. Let $\chi({\bf r},t)$ be a scalar function, taken sufficiently smooth, with the property that ${\bf B}\cdot\bs{\nb}\chi=0$. Isosurfaces $\chi=$ cst.\ are called {\em magnetic surfaces}\,: continuous surfaces that are everywhere parallel to the magnetic field.  The surfaces can be labeled by their value of $\chi$. With another scalar $\xi$ define a second, intersecting, set of magnetic surfaces, i.e.\ such that $\bs{\nb}\chi$ is nowhere parallel to $\bs{\nb}\xi$. Each field line can then be labeled with its values of $\chi$ and $\xi$. Their values can be chosen such that
\beq {\bf B}=\bs{\nb}\chi\times\bs{\nb}\xi ,\label{eulr}\eeq
and are then called {\em Euler potentials}.  [Exercise\,: show that they satisfy div\,${\bf B}=0$.]  They are not unique\,: for a given field ${\bf B}({\bf r})$ there exists substantial freedom of choice for $\chi$ and $\xi$. 

In general, the labeling is possible only in a limited volume. Field lines on a torus, for example, generally wrap around it on `irrational' paths\,: without ever returning to the same point. To label field lines on such a surface, a cut has to be made across it, where the labels are then discontinuous.  For more about Euler potentials, see \hyperlink{ster}{Stern} (1970).

\section{Reconnection}
\label{reco} 

\begin{figure}[ht]
\hfil\includegraphics[width=0.7\hsize]{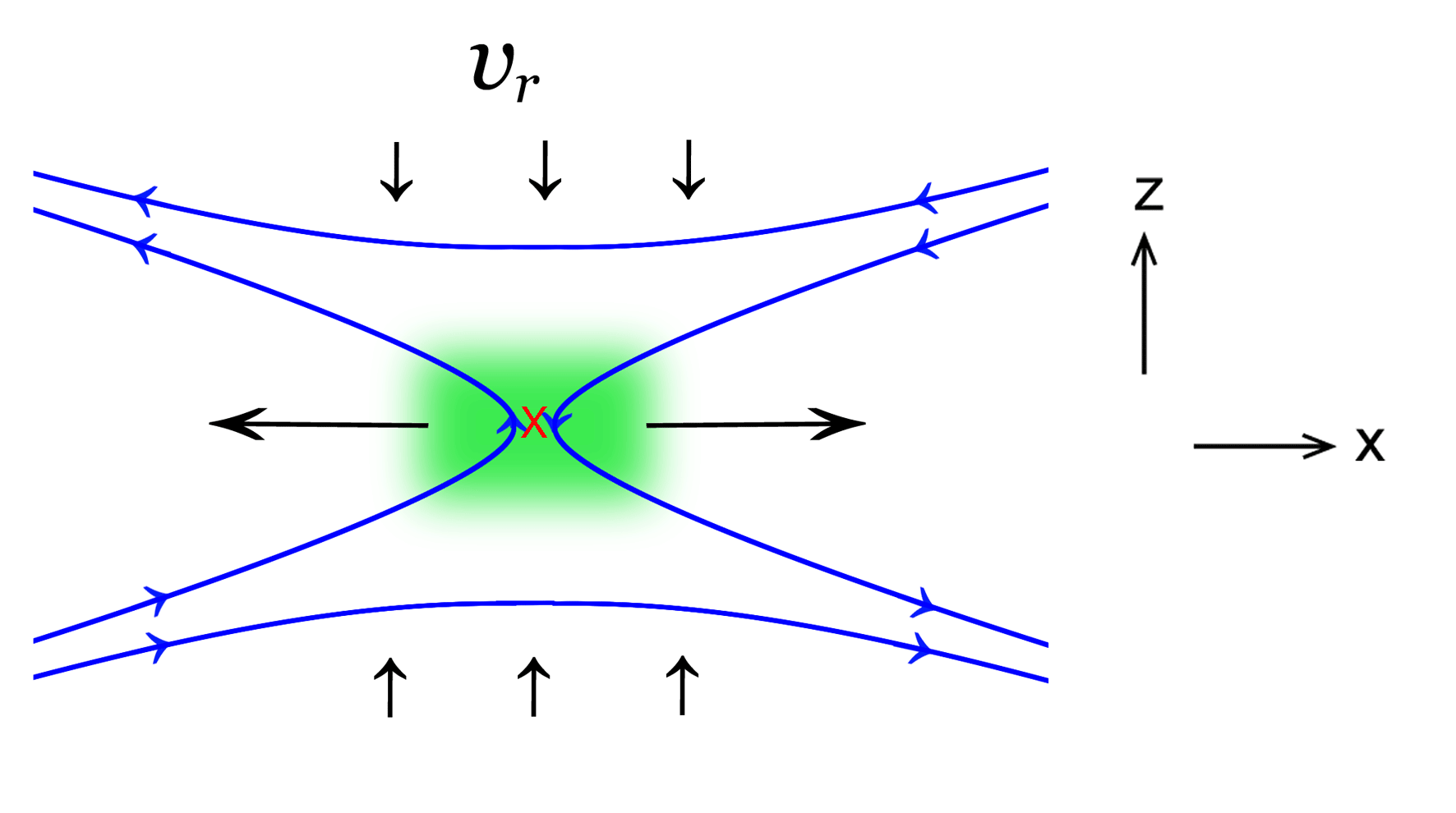}\hfil
\caption{\label{uzden} \small Inflow and outflow (black arrows) near a reconnection point {\color{red} $\times$} (current density in green).}
\end{figure}

By reconnection we mean the change of topology of a (part of) a field configuration, when neighboring field lines running in different directions `touch' and exchange the directions of their paths. Since there are open issues in theories of reconnection, the subject is not included in chapter 1. 

\subsection{Reconnection in a pressure supported current sheet}
\label{pressreco}
The simplest and most widely studied case is that of a {\em pressure supported} current sheet, as in {Fig.\,\ref{csheet}} (upper right panel). It is a model for reconnection in solar flares of the classical `two ribbon' type (cf.\ Priest 1982 Ch.\ 10) and the geotail of the Earth's magnetosphere. The field lines on the two sides of the sheet are antiparallel, the field strength vanishes in the middle of the sheet, and the configuration is independent of the $y-$coordinate (perpendicular to the plane). One envisages that finite diffusion allows this sheet to change into a configuration like Fig.\,\ref{uzden}. There is to be a `reconnection point' {\textsf X} (actually a line  perpendicular to the plane) of some extent still to be determined. Because of the symmetry assumed, there could in principle be several such points along the sheet, but in less symmetrical practice the process will typically start at one point which then preempts reconnection in its immediate neighborhood. The fluid, with the field lines frozen into it, flows towards {\textsf X} (top and bottom arrows). Inside {\textsf X} the field lines from above and below are not frozen-in any more and exchange directions. The field lines newly connected in this way are strongly bent. The tension force pulls them away from {\textsf X} (horizontal arrows). This causes more fluid carrying field lines with it to flow towards {\textsf X}, resulting in a quasi-stationary reconnection flow. 

A logical definition of the rate of reconnection at an {\textsf X} point is the number of field lines reconnected, i.e.\ a magnetic flux  per unit time and per unit length in the direction perpendicular to the plane, $\dot\Phi$ (dimensions G cm s$^{-1}$). This can be written in terms of the velocity $v_{\rm r}$ at which field lines flow into the reconnection region (vertical arrows in Fig.\,\ref{uzden})\,: 
\beq \dot\Phi=B_{\rm i} v_{\rm r},\eeq
where $B_{\rm i}$ is the field strength in the inflow region. The task of a reconnection theory is then to provide a number for $v_{\rm r}$.  Each of the reconnecting field lines carries an amount of fluid with it that has to change direction by $90^\circ$ from vertical inflow  to horizontal outflow. If $d$ is the thickness of the reconnection region, $D$ its extent in the $x-$direction, and $\rho$ the density, this amount is $D/d\,\cdot\,\rho/B$, per unit of reconnected magnetic flux. The inertia of the mass to be accelerated out of the reconnection region by the tension forces thus increases proportionally to $D$. The inflow speed $v_{\rm r}$ and the reconnection rate decrease accordingly:  $v_r\sim 1/D$.

The sketch in Fig.\,\ref{uzden} assumes $D\approx d$ and a correspondingly high reconnection rate. This is in line with indirect evidence such as observations of solar flares and with reconnection in the plasma laboratory. A problem has been that such high rates are not found  in the results of numerical MHD simulations that include only an Ohmic diffusivity. Instead of forming a small reconnection area, the reconnection region tends to spread out into a layer of significant width $D\gg d$ (a `Sweet-Parker' configuration).  For more on this topic see \hyperlink{kuls}{Kulsrud} Ch.\ 14.5.

Much better agreement is obtained in MHD simulations of ion-electron plasmas if Hall drift ({sect.~\ref{hall-am}}) is included [for a detailed description see \hyperlink{uzde}{ Uzdensky \& Kulsrud} (2006)].  Assuming  that the plasma is sufficiently highly ionized, ion-neutral collisions are frequent and ambipolar drift can be ignored. At the high current density in the reconnection region, however,  the Hall term cannot be ignored; it becomes the dominant term in the induction equation ({\ref{indha}}). Inside the reconnection region the electron fluid  is effectively decoupled from the ion fluid, and the evolution of the magnetic field is understood from the form ({\ref{inde}}) of the induction equation: the velocity determining the evolution of the magnetic field is that of the electron fluid.  This greatly increases the speed of the process, since the field lines can now reconnect without having to deflect the heavy ion fluid through the reconnection region. 

The coupling of the field to the ions becomes important again at some distance away from the reconnection region. The curvature force of the reconnected field lines at this distance drives the ion flow outward, as in classical reconnection models without a Hall term in the induction equation. Reconnection becomes effectively `point-like' \break($d\approx D$, a so-called `Petschek' configuration) and fast (inflow at a significant fraction of the Alfv\'en speed). 

In a positron-electron plasma, the Hall effect is absent (cf.\ {sect.~\ref{hall-am}}). Reconnection then depends on more details of the plasma physics.  If  the finite inertia of the charge carriers is included, reconnection is again found to assume the Petschek configuration (cf.\ \hyperlink{jaro}{Jaroschek et al.}\ 2004). 

These results provide a (partial) justification for the popular astrophysical assumption that reconnection proceeds roughly at `a tenth of the Alfv\'en speed'. Slightly more precisely:  a tenth of the Alfv\'en speed in the inflow region, with $v_{\rm A}$ evaluated from the component of $\bf B$ in the plane of reconnection (the plane of Fig.\,\ref{uzden}).

\subsection{Reconnection in tangled fields}
\label{tangled}
In MHD magnetic fields have a strong memory: their configuration depends on the history of the flows acting on them. Flows can be envisaged in which nontrivial topologies develop, such as for example a 3-stranded braiding pattern.~\hyperlink{park2}{Parker (1972)} raised the question of the mechanical equilibrium of such `tangled' configurations, introducing the idea of {\em topological reconnection}. Imagine  a nontrivial field configuration, constructed so as to be smoothly varying in space but not  in mechanical equilibrium, and let it relax to equilibrium. Parker argued that the degrees of freedom of  magnetic equilibria relaxed from nonequilibrium fields generally are insufficient to allow for them to be smooth functions of space: they must contain tangential singularities i.e. current sheets. This idea has gained increasing acceptance, especially since it has become testable with numerical simulations. As an example, consider the relaxation to equilibrium of a random, but smoothly varying initial field configuration in a box with periodic boundary conditions in all 3 directions. Such fields do indeed develop current sheets; the distribution of the sheets in space is  fractal, the time scale on which they develop is an Alfv\'en crossing time. On a somewhat longer time scale the sheets reconnect,  leaving a smooth equilibrium (\hyperlink{brai}{Braithwaite 2015}). 

The periodic boundary conditions in experiments  such as these allow the fluid displacements  much freedom. In fields anchored at boundaries relaxation is more restricted. This happens in low-$\beta$ environments such as the solar corona.

\subsection{Reconnection at low $\beta$}
\label{lowbet}
A low-$\beta$ field is close to a force-free configuration (unless very dynamic, $v\sim\va\gg\cs$, {see \ref{tilde}}).  If a current sheet develops,  it will be closer to the force-free variety (lower right panel in Fig.\ \ref{csheet}) than the pressure supported current sheet discussed above and assumed in most reconnection studies. Since force-free fields are so intimately connected with their boundaries ({sect.~\ref{boundrole}}), reconnection in this case is much more strongly determined by the boundaries than in a pressure supported sheet. 

\begin{figure}[t]
\hfil{\includegraphics[width=0.5\hsize]{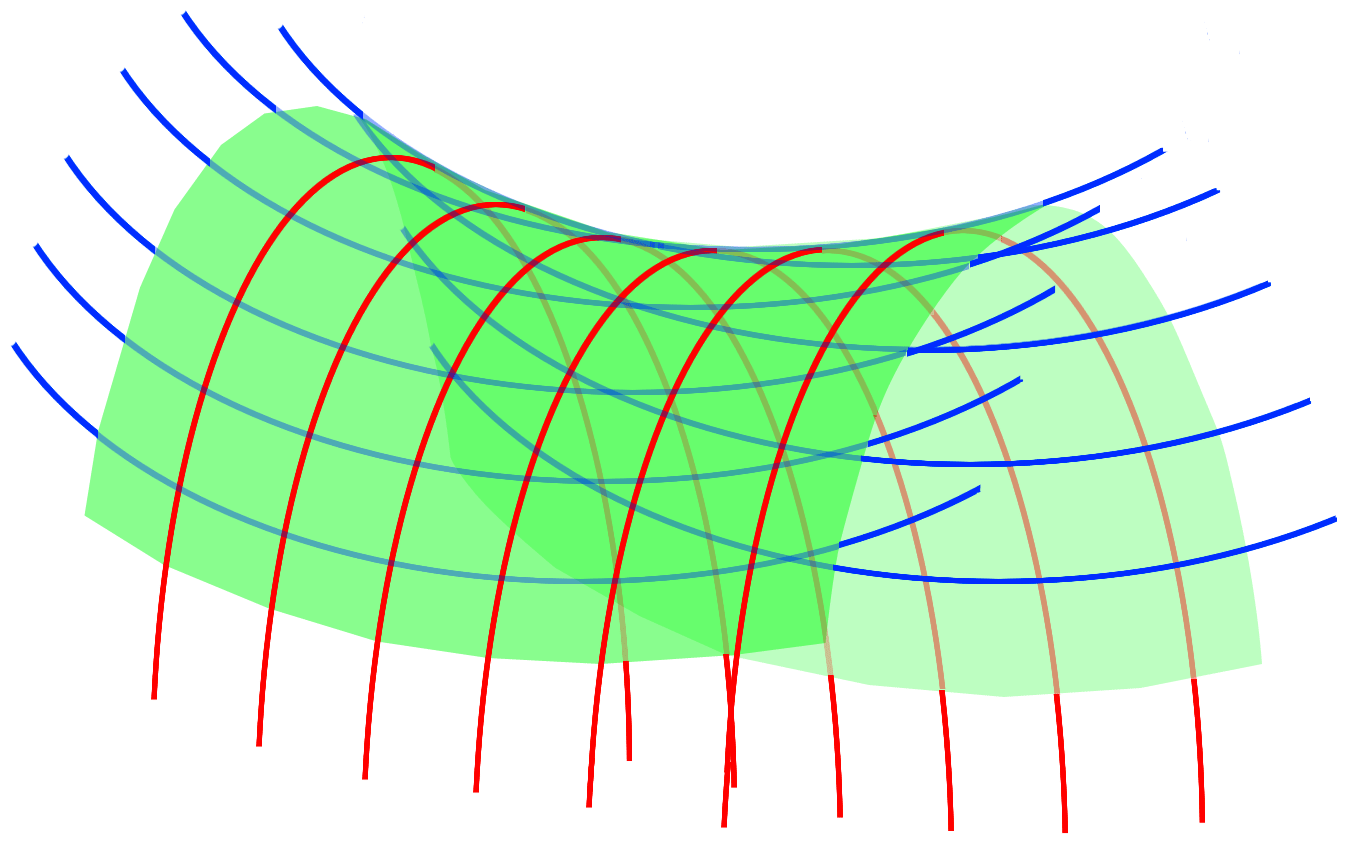}}\hfil{\includegraphics[width=0.4\hsize]{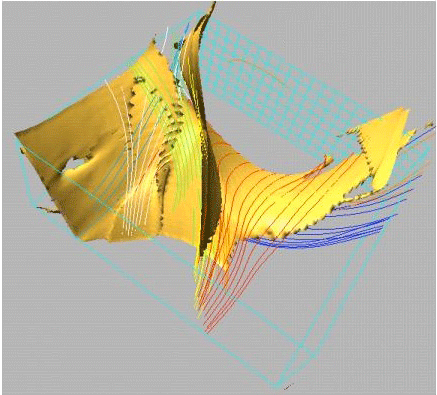}}\hfil
\caption{\label{nordl} \small Left (schematic)\,:  Field lines of different directions wrapping over (red) and under (blue) a force-free current sheet  (green). Right\,: complex current sheet (yellow) and field lines wrapping around it in a low-$\beta$ numerical simulation (detail, from \protect\hyperlink{gals}{ Galsgaard \& Nordlund} 1996).}
\end{figure}

A well-studied application  is magnetic heating of the solar corona. The field in the low-density, but highly conducting atmosphere is anchored below the Sun's visible surface, in the high-density plasma of the convective envelope. The fluid motions in the envelope displace the footpoints of the magnetic field lines extending into the atmosphere. The convective displacements at the two ends of a magnetic loop are uncorrelated, and the field gets wrapped into a tangled configuration. The displacements are slow compared with the time scale on which the force free configuration settles to equilibrium: the Alfv\'en travel time along the field. The wrapping takes therefore place as a series of  force-free configurations.  
Any sufficiently nontrivial flow field imposed at one or both ends of a bundle of field lines causes  {\em chaotic mixing} (\url{http://en.wikipedia.org/wiki/Chaotic_mixing}). 
Suppose the imposed displacements at the boundary are smooth in space and in time. The direction of the field lines then varies smoothly across the volume. As the wrapping process proceeds, however, the distance over which their direction varies decreases  exponentially with time. A fractal set of force-free current sheets develops in the volume. When the length scale across the sheets decreases below the limit where ideal MHD applies, reconnection and energy dissipation set in. Because of the exponential nature of chaotic mixing, this happens after a finite amount of time.

This chaotic mixing process is a kinematic contribution to the rapid development of currents sheets (i.e.\ describable by the induction equation alone). Concurrent with it there is a {\em dynamic} contribution, resulting from the Lorentz forces that develop in a tangled flux bundle. As emphasized in Parker's topological view (\ref{tangled}), Lorentz forces locally to lead to configurations as sketched in the left panel of Fig.\,\ref{nordl}.  The tension in the magnetic field lines wrapping around the two sides of a current sheet squeezes it from both sides, much like neighboring links in a chain press on each other at their surface of contact. 

In a force-free sheet, the volume between the two directions is filled with field lines of intermediate direction (lower right panel in {Fig.\,\ref{csheet}}). For reconnection to proceed, these have to somehow `get out of the way' to make room for new field lines entering the reconnection region. 
 Being part of a force free field (\ref{fff}), however, these intermediate field lines are tied to boundaries. This is unlike  reconnection in the pressure supported sheet of Fig.~\ref{uzden}, where the fluid leaving the reconnection region does not have such a restriction. Close to a boundary this `line tying' restriction is strong; the small length scales  that form are just those developing  on the nearby boundary due to the chaotic mixing effect of the displacements imposed there. Further from the boundaries, field  lines can move out of the way more freely, and magnetic tension forces cause current sheets become thinner more rapidly 
(\hyperlink{vanb}{van Ballegooijen} 1985). 
With increasing distance from the boundary, this process converges to Parker's topological reconnection picture.

Numerical simulations of a such a low-$\beta$ field tangled by displacements at a boundary  (\hyperlink{gals}{Galsgaard \& Nordlund} 1996) show how the magnetic configuration develops a fractal distribution of current sheets (Fig.\,\ref{nordl}). It takes a finite number of uncorrelated displacements before the sheets are thin enough for dissipation to set in.  When these displacements are halted in the simulation, dissipation drops quickly by broadening of the current sheets: the wrapping pattern of entire field configuration `freezes'. When the displacements  are then resumed, dissipation immediately springs into action again. This illustrates how in a low-$\beta$ field reconnection is directly governed by what happens on its boundaries.

\subsection{Energy conversion in reconnection}

In the reconnection region (green in Fig.\ \ref{uzden}) finite resistivity or plasma processes involving electric fields convert magnetic energy into heat or energetic particle distributions. This process accounts only for  a small part of the energy released, however. The magnetic free energy that is released in the reconnection process derives from a much larger volume than the reconnection region.  The tension in the strongly bent post-reconnection field lines  in Fig.\ \ref{uzden} causes them to snap to the sides. In this way most of the magnetic energy of the pre-reconnection configuration is converted into bulk kinetic energy rather than to particle heating at the reconnection point itself. This kinetic energy may be dissipated further downstream, for example through shock waves, but this is unrelated to the plasma physics happening in the reconnection region itself. If the filling factor of the reconnection volume in the field configuration is small, plasma processes  happening in it (such as the acceleration of fast particles)  are energetically inconsequential compared with the total energy release. 

A useful analogy is the failure of a steel cable under tension. Almost all energy released derives from the elastic stretching of the wires. It appears as kinetic energy of the ends snapping away. The longer the cable, the larger the amount of energy released. A minor amount of energy, independent of the length of the wire, is dissipated  locally at the point where the wire breaks through plastic deformation.

\subsection{Energy storage and dissipation}

For a given field configuration (the pressure-supported current sheet, for example) the question addressed above concerned the speed at which field lines would reconnect, expressed as a velocity of inflow. One can also wonder how much magnetic energy is converted, and into which forms. A classical topic in this connection is magnetic heating of the solar corona. Magnetic field lines connect  different parts of the surface of the Sun through the corona. Convective flows displace these footpoints randomly, if they are sufficiently far apart. Twisting of  the field configuration by the displacements stores energy in the configuration, and reconnection as discussed in sect.\ \ref{lowbet} releases (some of) it. The `coronal heating' question to be answered is how much energy is dissipated for given  displacements at the surface. 

The answer depends on these displacements, but also on the speed of reconnection, in a somewhat non-intuitive way. In the limit of  very efficient reconnection, the field stays at the lowest energy state determined by its boundary conditions. The magnetic energy that can be released in reconnection consequently  {\it vanishes} in this limit. If reconnection is not perfectly efficient, on the other hand, the amount of magnetic free energy stored in the configuration, and released in reconnection is finite and {\it increases} with {\it decreasing} speed of reconnection. This complicates estimates of the magnetic heating rate. The question requires simulations such as mentioned in sect.\ \ref{lowbet}. 

Not all displacements between footpoints store energy into a field configuration. An example to the contrary is shown in Fig.\ \ref{pf}. Displacements that bring footpoints of opposite polarity closer together actually {\it extract} energy from the configuration. The Poynting flux in this case is downward, into the convection zone. This illustrates  the limited usefulness of Poynting flux arguments in situations like a magnetic stellar atmosphere.  In a region of mixed polarities the average Poynting flux has contributions of either sign. It  is not directly related to the rate of magnetic dissipation in the atmosphere, which can be larger than the net Poynting flux.

\section{Charged clouds}
\label{clouds}
While ({\ref{dsigma}}) shows that charge is conserved in MHD, one might ask what happens when a charge density is present in the initial conditions. For example, can one still use MHD in a star or a conducting gas cloud that has a net charge?  

In MHD, as in classical electrostatics, a net  charge in a volume of conducting fluid object appears only on its surface. If a charge were initially distributed through a star, its volume would quickly get polarized by small displacements between the positive and negative charge carriers, canceling the charge density in the inside and leaving the net charge as a surface layer on its boundary. The time scale for this to happen is governed by the plasma frequency or the light travel time across the volume, which are usually very fast compared with the MHD time scales of interest (see {sect.~\ref{applic}}). Alternatively, this can be seen as a setting a minimum on the time scale of the process under consideration, for the MHD approximation to apply.

Physical charge densities, i.e.\  net imbalance between plusses and minuses as measured in the rest frame of the fluid, thus appear only on boundaries between the conducting fluid and non-conducting parts of the system. In frames other than the rest frame of the fluid, charge densities in a conducting fluid appear only as frame transformation quantities, see next section.

\section{The charge density in MHD}
\label{chargedd}

In the limit $v\ll c$, charge densities in MHD are  `small' in some sense, and are ignored because not needed for solving the equations. It is instructive nevertheless  to check that the procedure followed for deriving the MHD approximation has not introduced an inconsistency with Maxwell's equations. In addition, a closer look at charge densities is important for interpreting their role in relativistic MHD.

To test if the third of Maxwell's equations ({\ref{chargeM}}) is satisfied (which was not used in deriving the MHD equations), take the divergence  of the electric field under the assumption of perfect conductivity, eq.\  ({\ref{ee}}). Using the formula for the divergence of a cross product:
\beq 
4\pi\sigma={\bf\bs{\nb}\cdot E}=-\bs{\nb}\cdot({\bf v\times B})/c={\bf B}\cdot({\bs{\nb}\bf\times v})/c-{\bf v}\cdot({\bs{\nb}\bf\times B})/c.\label{ghc}
\eeq
This is valid for arbitrary $v/c$, since the expression for $\bf E$ is relativistically correct in ideal MHD (but not when diffusion terms are added, because of the unbounded signal speeds they imply). This does not, in general, vanish. Hence we have the disconcerting conclusion that a charge density appears in a theory which uses charge neutrality as a starting assumption. 

The key here is the distinction between charge and charge density. To see how charge densities appear consider first an {\em irrotational flow} (i.e.\ ${\bs{\nb}\bf\times v}=0$), so that the first term vanishes. The second term describes a charge density that appears in a flow $\bf v$ in the direction of the current density $\mathbf{j}$. It can be understood as the consequence of a Lorentz transformation between the rest frame of the plasma and the frame in which (\ref{ghc}) is applied. The relativistic current is the four-vector $J=(\mathbf{j},c\sigma)$. If the primed quantities ${\bf j}^\prime,\sigma^\prime$ are measured in the rest frame of the plasma, and unprimed quantities in the frame in which the plasma velocity is $\bf v$, a Lorentz transformation yields [with $\beta=v/c,\gamma=(1-\beta^2)^{-1/2}$]:
\bea j &=\gamma(j^\prime-\beta c\sigma^\prime),\\
        c\sigma &=\gamma(c\sigma^\prime-\beta j^\prime).\label{transsigma}
\eea
If the charge density $\sigma^\prime$ in the plasma frame vanishes, we have $j=\gamma j^\prime$, and
\beq \sigma=-\beta\gamma j^\prime/c =-\beta j/c. \eeq
With the current ${\bf j}=c\bs{\nb}{\bf\times B}/4\pi$ this agrees with the second term  in (\ref{ghc}). [For an interesting paradox about such charge densities, see problem {\cpr problem \ref{curwire}}.]

Thus, while individual charges are Lorentz invariant, the charge density is not because it is a quantity  per unit of volume, and volumes are subject to Lorentz contraction. In the presence of a current, it can be given an arbitrary (positive or negative) value $\vert\sigma\vert < j/c$ by a change of reference frame\footnote{~Since  $j^2-c^2\sigma^2$ is Lorentz invariant, a sufficient condition for this to be possible is that $c^2\sigma^2< j^2$. This is unlikely to be violated in a medium of any significant conductivity. In the opposite case $c^2\sigma^2> j^2$, there is a frame in which the current vanishes but a charge density is present. The absence of a current in the presence of the electric field implies that the medium is an electric insulator.}.

This formal result can be made more intuitive by taking a microscopic view of the plasma, consisting of charge carriers of opposite sign with densities $n^+,n^-$, which we assume to be equal in the rest frame of the plasma. In the presence of a current, they move in opposite directions. A Lorentz transformation with the velocity $v$ then has a different effect on the two charge densities, since  one is flowing in the same direction as $\bf v$, the other in the opposite. The Lorentz contraction of the two carrier densities differs, so they are not the same anymore in the observer's frame.
 
To see what the first term in (\ref{ghc}) means, consider the plasma in an inertial frame locally comoving at some point in space. The electric field, as well as the charge density due to the second term in  (\ref{ghc}) then vanish at this point. Unless ${\bs{\nb}\bf\times v}=0$, however, the electric field does not vanish away from this point. It varies approximately linearly with distance, implying the presence of a charge density. In a frame rotating locally and instantaneously with rotation vector $\bs{\Omega}=\bs{\nb}\bf\times v$, this charge density vanishes again. By observing in  frames rotating at different rates, charge densities of either sign appear\footnote{~Rotating reference frames are relativistically problematic. Here we need only a locally corotating frame in a region of infinitesimal extent, involving arbitrarily small velocities.}. 

Charge densities thus disappear in a locally comoving, locally corotating frame of reference\,: the frame in which the plasma is `most at rest'. An observer in this frame sees neither a charge density nor an electric field, and senses no electrical force. Since this frame is in general a different one at each point in the flow, it is not a practically useful frame for calculations. It illustrates, however, that  charge densities in MHD can be regarded as frame transformation quantities. This is the case in general, not only in the limit $v/c\ll 1$. Whether relativistic or not, flows do not create charges. 

What remains to be checked is if the charge density (\ref{ghc}) needs to be taken into account in the   non-relativistic equation of motion, in frames other than the fluid frame, since we have ignored this possibility in deriving the MHD equations. The electric field, of order $v/c$, must be kept in order to arrive at the MHD equations. The force density $\sim {\bf E}\sigma$ is one order in $v/c$ higher, hence can be consistently ignored in all reference frames, in the nonrelativistic limit.

Relativistic MHD is beyond the scope of this text. For a practical formulation and its use in numerical applications see \hyperlink{komm}{Komissarov} (1999). 

\section{Applicability limits of MHD}
\label{applic}
A number of different considerations lead to limits of applicability.
Start with the condition that a charge density is absent (in the frame in which the plasma is `most at rest', cf.\ {sect.~\ref{chargedd}}). In the example of a charged cloud ({sect.~\ref{clouds}}) this requires that the time scale on which a charge density initially present in the volume is canceled by the formation of a surface charge, is short compared with the MHD time scales of interest. This time scale $\tau_{\rm e}$ is determined by the most mobile charge carriers (i.e.\ electrons, usually). In a volume of size $L$, equate the electrostatic energy of the charge density  with the kinetic energy of the charges, of mass $m$, when accelerated in the corresponding electric field. This yields
\beq \tau_{\rm e}^2\approx m/(4\pi n e^2)=1/\omega_{\rm p}^2, \eeq
where $\omega_{\rm p}$ is called the {\em plasma frequency}. This is independent of the size $L$ of the volume.  

In this electrostatic adjustment process the charges that were assumed to be present initially do not move to the boundary themselves; this would take longer. Instead, their presence is canceled by {\em polarization}\,: small  contractions or expansions of the electron fluid relative to the ions. The time of this process is given by  $1/\omega_{\rm p}$. For the MHD approximation to hold,  the shortest time scale of interest in  the MHD process studied must be longer than this. 

The charge density ({\ref{ghc}}) also sets a minimum plasma density that must be present for MHD to be applicable\,:  the density of charges implied by the current cannot exceed the total density  of charged particles $n$, i.e. $\sigma/e<n$. If the length scale on which $\bf B$ varies in a direction perpendicular to itself is  $L_B$, the second term in ({eq.~\ref{ghc}}) thus requires that $ n> vB/(4\pi  L_B e c)$.
This is not the most relevant limit involving $L_B$, however. The condition that the current density can be carried by particles moving below the speed of light is stricter by a factor $c/v$. With $j=(c/4\pi)\, B/L_{\rm B}= v n e$ :
\beq n > {B\over 4\pi e L_B}\label{limB}.\eeq
If the first term in (\ref{ghc}) dominates, and $L_v$ is a characteristic value of $v/\vert\bs{\nb}\times\bf v\vert$, $n$ has to satisfy the additional condition
\beq n> vB/(4\pi  L_v e c),\eeq
or, writing $\Omega\equiv v/L_v$:
\beq n> n_{\rm GJ}\equiv {\Omega B\over 4\pi e c}.\label{limv}\eeq
By analogy with the pulsar application where it was first derived, this is called the {\em Goldreich-Julian density}. If the charged particle density $n$ is pre-determined, (\ref{limB}) and (\ref{limv}) can alternatively be seen as setting minima on the macroscopic length scale $L_B$ and time scale $L_v/v$ for MHD to be applicable.

The moving charges carrying the electrical current also carry momentum. A change of the current in response to a changing field requires a force acting on these charges. When should one worry about the inertia of the charges responding to this force? As a model, take the shearing field configuration of {Fig.\,\ref{shear}}. A velocity of amplitude $v$ changes direction over a length $L$ perpendicular to it. The current density $j$, carried by particles of mass $m_{\rm c}$, charge $e$ and number density $n_{\rm c}$ moving at a velocity $v_{\rm c}$ is $j=e n_{\rm c}v_{\rm c}$. The momentum $p_ j$ carried by these charges is $p_j=j m_{\rm c}/e$. The current in this example increases linearly with time at the shear rate, $\tau_v^{-1}=v/L$. The force (per unit volume) needed to achieve this increase is $\pa_t p_{\rm j}=(j/\tau_v)(m_{\rm c}/e)$. To see how important this is, compare it with the Lorentz force acting on the fluid as a whole, $F_{\rm L}=j B/c$. The ratio is:
\beq {\pa_t p_{\rm j}\over F_{\rm L}}={m_{\rm c}c\over e B \tau_v}= {1\over\omega_B\tau_v},\eeq
where $\omega_B=e B/(m_{\rm c}c)$ is the cyclotron frequency  of the charges, $\approx 3\,10^6 B$  s$^{-1}$ for electrons. The condition that the inertia of  a current carried by electrons can be neglected is thus
\beq \tau_v\gg 3\,10^{-7} B,\eeq
usually a mild condition except for  time scales deep inside microscopic plasma processes (see also {sect.~\ref{origj}} below).

Plasma processes of various kind can set in already at densities higher than (\ref{limv}), leading to an
effective increase of the electrical resistivity (`anomalous resistivity'). 
Even when the configuration is well within the limits for ideal MHD to apply on global scales, its evolution can lead to the formation of current sheets\,: regions where the direction of the field lines changes so rapidly across magnetic surfaces that field lines are locally not `frozen-in' anymore.  Reconnection of field lines takes place there, with consequences for the field configuration as a whole ({sects.\ \ref{sheets}, \ref{reco}}).

\subsubsection{Collisionless plasmas}\label{colless}
In many astrophysical fluids the collision frequency $\omega_{\rm c}$ is much lower than the 
(electron) cyclotron frequency $\omega_B$\,: the plasma is called `collisionless', or `strongly magnetized'.  This differs from the Ohmic conductivity picture that figures somewhat implicitly in Chapter 1. This might lead one to conclude that nearly collisionless situations are in the domain of plasma physics, not MHD. In the limit $\omega_{\rm c}\ll\omega_B$, however, the charge carriers are again tied to the field lines (though for a different reason than under the assumption of infinite conductivity, see {sect.~\ref{freeze}}). As a result, the MHD equations again apply on length scales where local fluid properties like pressure, density and velocity can be meaningfully defined. (Hall drift and ambipolar diffusion terms in the induction equation may become relevant, however, see {section \ref{hall-am}}). 

For the scales of structures observed in the universe this is almost always the case. In the solar wind, for example, the Coulomb interaction length can be of the order of an astronomical unit, but the observed structure of its flows and magnetic fields agrees closely with expectations from MHD theory. Another example are the structures seen in jets and radio lobes of  active galactic nuclei, where Coulomb interaction times can be of the order of the age of the universe. Yet their morphology, with ages of order $10^6$ years, closely resembles structures expected from normal compressible MHD fluids.  For more on this see \hyperlink{park3}{Parker (2007}, Chs.\ 1 and 8). 

\section{The microscopic view of currents in MHD}
\label{origj}
With the consistency of MHD established, the subject of currents has become of marginal significance (`just take the curl of your field configuration'). This argument does not always suffice to dispel concerns  how a current can come about without a battery driving it. What sets the particles in motion that carry the current?

The origin of the current of course lies in the fluid flow. The current in resistive MHD results from the flow across field lines (as in {Fig.~\ref{anchor}}). In the limit of infinite conductivity of the plasma, the flow speed associated with this current becomes infinitesimally small, but conceptually is still present. 
Since flow across stationary field lines only occurs at finite resistivity, a microscopic view of the source of currents requires a model for the resistivity. 

As a simple model which allows for an adjustable resistivity, assume a plasma with a small degree of ionization, and consider first the case when the collision frequency $\omega_{\rm c}$ is large compared with the cyclotron frequency $\omega_B$ of the charge carriers.  Assume a magnetic field anchored in an object that is stationary in the frame of reference. The finite resistivity  of the fluid allows it to flow across $\bf B$  with a velocity $\bf v$ (cf.\ {sect.~\ref{diffusion}}). Between collisions, charges traveling with the flow are displaced from the direction of the mean flow by their orbits in the field $\bf B$. Charges of opposite sign are displaced in opposite directions, i.e.\ a current results, perpendicular to $\bf v$ and $\bf B$. In the opposite case of high magnetization, the charges orbit around the field lines with infrequent collisions. The collision probability is enhanced during the part of their orbits where they are traveling opposite to $\bf v$. Such collisions reduce the velocity of the orbits and at the same time shift their guiding centers, again in opposite directions for the $+$'s and $-$'s. 

For a given velocity, the current produced in this way is proportional to the density of charges participating in the collisions, i.e.\ the degree of ionization in the present example. Conversely, for a given value of the current, the incoming flow speed decreases with increasing degree of ionization. The current distorts the magnetic field; the associated magnetic force is balanced by the momentum transfer in the collisions.  In the limit of vanishing resistivity, the component of the flow speed perpendicular to the field vanishes, in a stationary field configuration. In the general nonstationary case, the magnetic field changes by being `dragged with the flow' as described by the MHD induction equation.

A current needs a finite time to be set up, or to change direction. 
The discussion in {section \ref{applic}} shows that the time scale on which this happens is related to the orbital time scale of the charges in the magnetic field. For further discussion of the above in the context of a practical application, see {section \ref{eMHD}}.

\section{Hall drift and ambipolar diffusion}
\label{hall-am}
The induction equation in the form ({\ref{inddif}}) includes only an `Ohm's law' resistivity. For it to be relevant as  the dominant deviation from ideal MHD,  differences between the mean velocities of neutrals, ions and electrons must be unimportant. When such differences do become relevant, Hall drift and/or ambipolar diffusion have to be taken into account. Let $n,\,n_{\rm e},\,n_{\rm i}$ be the number densities of neutrals, electrons and ions, and $\rho$, $\rho_{\rm i}$ the mass densities of  neutrals and ions, respectively. The MHD induction equation then has the form
\beq 
{\pa{\bf B}\over\pa t}=\bs{\nb}\times[({\bf v}+{\bf v}_{\rm H}+{\bf v}_{\rm a})\times{\bf B}-\eta\bs{\nb}\times{\bf B}],\label{indha}
\eeq
where ${\bf v}$ is the velocity of the neutral fluid component and
\beq {\bf v}_{\rm H}\equiv{\bf v}_{\rm e}-{\bf v}_{\rm i}= -{c\over 4\pi}{\bs{\nb}\times{\bf B}\over e\,n_{\rm e}}=-{{\bf j}\over e\, n_{\rm e}},\label{hall} \eeq
the {\em Hall drift}, is the velocity ${\bf v}_{\rm e}$ of the electrons relative to the velocity ${\bf v}_{\rm i}$ of the ions. The term
\beq 
{\bf v}_{\rm a}\equiv{\bf v}_{\rm i}-{\bf v}= {1\over 4\pi}{(\bs{\nb}\times{\bf B})\times{\bf B}\over\gamma\rho\rho_{\rm i}}={{\bf F}_{\rm L}\over\gamma\rho\rho_{\rm i}}\label{ambi},
\eeq
is the {\em ambipolar drift velocity}, where ${\bf F}_{\rm L}$ is the Lorentz force, and $\gamma$ a `friction coefficient' (of order $3\,10^{13}$ cm$^{3}$ s$^{-1}$g$^{-1}$ for astrophysical mixtures). For a derivation of these expressions and the assumptions involved see e.g.\ \hyperlink{balb}{Balbus} (2009). 

The main assumption made is apparent in the form of the expressions on the right in (\ref{hall}) and (\ref{ambi}). The second equality in (\ref{hall}) shows that the Hall drift is just the mean velocity of the electrons associated with the current they carry. This involves the approximation that the contribution of the ions to the current can be neglected, which requires that  the ratio $m_{\rm e}/m_{\rm i}$ of electron to ion mass is neglected. 

The Lorentz force ${\bf F}_{\rm L}$ acts directly only on the charged components, since the neutrals do not sense the magnetic field. Transmission of the force from the ions to the neutral component involves momentum transfer through collisions, and associated with this there is a velocity difference ${\bf v}_{\rm  a}$ between neutrals and ions. The collision rate is proportional to both the density of the neutrals and the density of the ions. This explains the  appearance of their densities in the denominator of  the ambipolar drift velocity (\ref{ambi}). Both ${\bf v}_{\rm H}$ and ${\bf v}_{\rm a}$ vanish at large electron density $n_{\rm e}$ ($=n_{\rm i}$, for singly ionized atoms). 

Eq.\ (\ref{indha}) is the appropriate form of the induction equation when Hall and ambipolar drift can be regarded as small flows relative to the neutrals. If the plasma is nearly fully ionized, on the other hand, the ions dominate the mass density $\rho$. In this case it is more useful to write (\ref{indha}) in terms of the mean velocity of the ions ${\bf v}_{\rm i}={\bf v} + {\bf v}_{\rm a}$ instead of the neutrals.  With the first equality in (\ref{ambi}), this yields
\beq 
{\pa{\bf B}\over\pa t}=\bs{\nb}\times[({\bf v}_{\rm i}+{\bf v}_{\rm H})\times{\bf B}-\eta\bs{\nb}\times{\bf B}].\label{indh}
\eeq
The ambipolar process is then just the drift of the inconsequential neutral component relative to the ions. With the first equality in (\ref{hall}) we can write this also in terms of the velocity ${\bf v}_{\rm e}$ of the electron fluid alone:
\beq 
{\pa{\bf B}\over\pa t}=\bs{\nb}\times[{\bf v}_{\rm e}\times{\bf B}-\eta\bs{\nb}\times{\bf B}].\label{inde}
\eeq
This shows that in an ion-electron plasma the magnetic field is strictly speaking not `frozen-in'  in the plasma as a whole, but in its electron component. Eq.\ (\ref{inde}) is not a useful form for use in the full MHD problem, however, since the equation of motion involves the velocity of the heavy component, not the electrons (in the limit $m_{\rm e}/m_{\rm i}\rightarrow 0$). Instead, (\ref{indha}) or (\ref{indh}) are needed, depending on the degree of ionization. 

The Hall drift in the above applies to an ion-electron plasma. Its existence depends on the mass difference between the positive and negative charge carriers. In a {\em pair plasma}, consisting of equal numbers of positrons and electrons, the situation is symmetric, the current is carried equally by the positive and negative charges, and the Hall drift vanishes. Hall drift probably plays a crucial role in the reconnection of field lines in ion-electron plasmas ({sect.~\ref{reco}});  in pure pair plasmas reconnection is likely to function differently. This is relevant for the magnetohydrodynamics of pulsar winds as observed in a young pulsar like the Crab, which are believed to be consist of nearly pure pair plasmas. 

The ambipolar drift velocity can be decomposed as
\beq {\bf v}_{\rm a}={\bf v}_{\rm so}+{\bf v}_{\rm ir}, \eeq
where ${\bf v}_{\rm so}$ and ${\bf v}_{\rm ir}$ are its solenoidal (${\rm div}\,{\bf v}_{\rm so}=0$) and irrotational ($\bs{\nb}{\bf\times v}_{\rm ir}=0$) components. The two have different consequences. The irrotational flow component is the `divergent/convergent' one\,: it causes a pile-up of the charged relative to the neutral fluid (as opposed to a rotational motion, which in principle can proceed in a stationary state). Pressure balance of the fluid as a whole implies that a gradient in pressure develops between the two, by which  ${\bf v}_{\rm ir}$ eventually stalls, leaving only ${\bf v}_{\rm so}$ (\hyperlink{gold2}{Goldreich and Reisenegger} 1992). See {\cpr problem \ref{ambs}}.

A somewhat implicit assumption in the above is that the density of the charge carriers relative to that of the neutrals is determined instantaneously, compared with the MHD time scales of interest. This is rarely a concern, since ionization/recombination rates are usually very frequent compared with the MHD time scale. It becomes important, however, for the magnetohydrodynamics of a neutron star interior. Relaxation of the  ratio of charged (proton+electron) to neutron density, mediated  by slow weak interaction processes, plays a role in this case (\hyperlink{gold2}{Goldreich and Reisenegger} 1992).

\section{Curvature force at a boundary}\label{curvb}
The curvature force does not enter in the condition for equilibrium between regions of different field strength. It plays the intuitively expected role more indirectly.
This is illustrated by the example in Fig.\ \ref{curveb},  showing the shape of a potential field (section \ref{fff}) wrapping around a perfectly conducting plate (such that ${\bf B}=0$ on its surface). The balance between the  tension force and the magnetic pressure gradient causes the field lines to pile up towards the edge of the plate, increasing the field strength there. The curvature force makes itself felt through this effect on the field strength, the balance of forces at the boundary itself involves only the magnetic  pressure $B^2/8\pi$ (see also {\cpr problem \ref{edge}}).

\begin{figure}[h]
\hfil{\includegraphics[width=0.6\hsize]{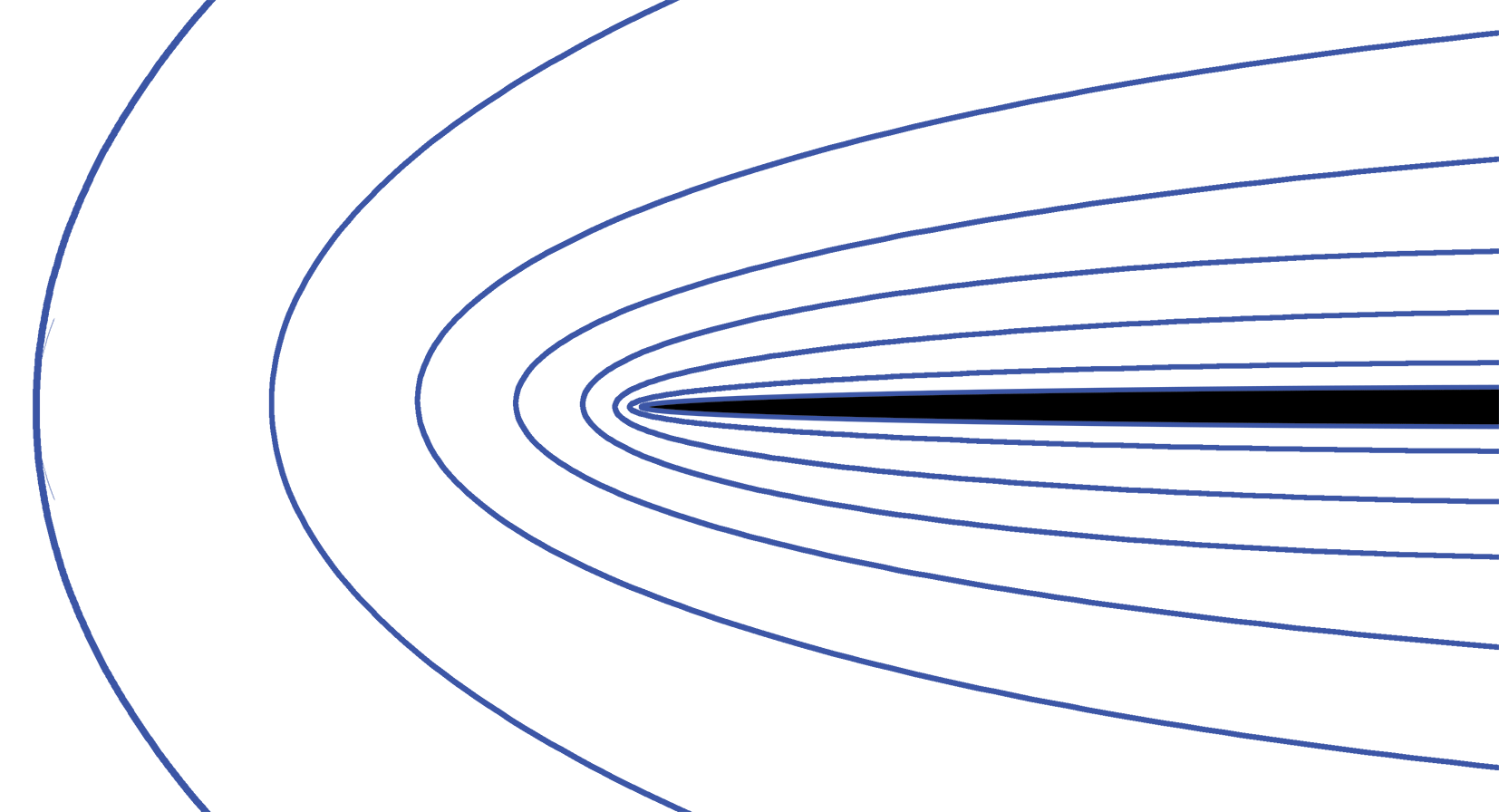}}\hfil
\caption{\label{curveb} \small Wrapping (schematic) of field lines around a thin field-free plate (black). Separation between neighboring field lines shown is inversely proportional to field strength. The curvature force causes the lines to pile up at its edge.}
\end{figure}

\section{Surface stress\,: example}
\label{disksupport}

As an example for evaluating forces by using the stress tensor, consider support against gravity by a magnetic field [Fig.\,\ref{disk}, a model for the (partial) support of an accretion disk by a magnetic field  embedded in it]. The acceleration of gravity acts towards the left, ${\bf g}=-g{\bf\hat x}$. Mass is present in a layer of thickness $2h$ around the midplane $z=0$. Let $L=\vert{\bf B}\vert/\pa_x\vert{\bf B}\vert$ be a characteristic length on which $\bf B$ varies along the plane, and use a unit of length such that $L\approx {\cal O}(1)$. We take the layer to be thin, $h\ll 1$ in this unit. Outside the layer ($\vert z\vert>h$) the field is assumed independent of the $y$-coordinate, and symmetric about the midplane, $B_z(-h)=B_z(h)$, $B_x(-h)=-B_x(h)$. Inside the layer the field can be of arbitrary shape.

The question to answer is how large the field strength must be to support the disk significantly against gravity. We do this by calculating the magnetic stresses over a flat box of unit surface area, $-h<z<h$, $0<x<1$, $0<y<1$. The magnetic force on the mass in this box is then given by the right hand side of ({\ref{gauss}}). It has contributions from the top, bottom, and side surfaces, and the outward normal is to be taken on these surfaces. The top and bottom surfaces have normals ${\bf n}=(0,0,1)$, $(0,0,-1)$, respectively and  yield contributions
\bea F_{x,{\rm top}}= &M_{xz}(h)&=  B_x(h)B_z(h)/4\pi,\\
F_{x,{\rm bot}}=&-M_{xz}(-h)&=-B_x(-h)B_z(-h)/4\pi=B_x(h)B_z(h)/4\pi.\eea
The contributions from the sides are smaller by a factor $h$ on account of the assumed thinness of the layer, hence will be ignored. Summing up, the magnetic force on the disk, per unit surface area in the $x-$direction is
\beq {\bf F}_{\rm m}= {\bf\hat x}\, B_x B_z/2\pi\label{diskforce}\eeq
(where $B_x$ is evaluated at the top surface of the layer).
The force of gravity ${\bf F}_{\rm g}$ is found from the mass in the box. Let
\beq\Sigma=\int_{-h}^h\rho~\rd z \eeq
be the {\em surface mass density} of the layer. The force per unit surface area of the disk is then
\beq {\bf F}_{\rm g}=-{\bf\hat x}\,g\Sigma.\eeq
For the magnetic field to contribute significantly to support (compared with the rotational support due orbital motion), we must have
\beq {B_xB_z}\sim 2\pi\Sigma g,\eeq
implying also that $B_xB_z >0 ~ (z=h)$, i.e.\ the field is bent outward, as sketched in Fig.\,\ref{disk}.

\begin{figure}[t]
\hfil{\includegraphics[width=0.7\hsize]{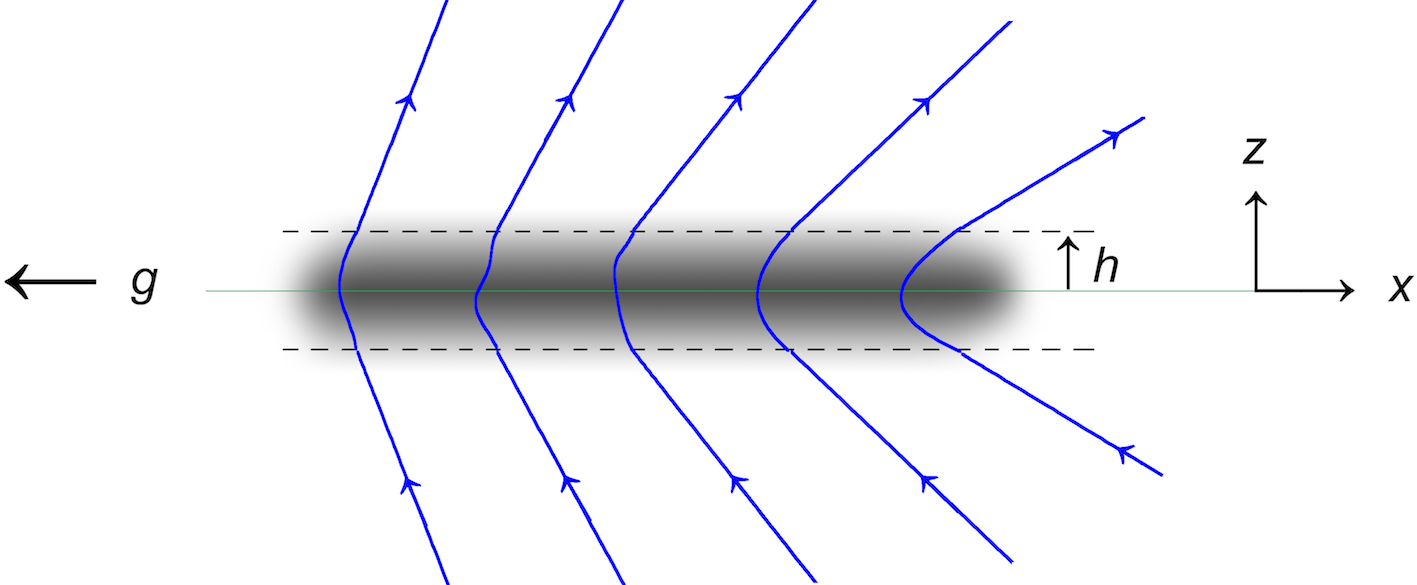}}\hfil
\caption{\label{disk} \small A thin accretion disk with magnetic support against the gravity of a central mass (located to the left).}
\end{figure}

The example shows how by evaluating the stress on its surface, we have been able to circumvent the details of the magnetic forces inside the box. Of course this gives a useful result only if the magnetic fields on the top and bottom surfaces are known. In the applications we have in mind, approximate vacuum conditions hold outside the layer, and $\bf{B}$ is a {\em potential field} there, ${\bs{\nb}\kern-1pt\bf\times B}=0$ ({sect.~\ref{potf}}). It is determined only by the distribution of field lines crossing the disk, i.e.\ the vertical component $B_z (x,z=h)$. With $B_z$ given as a function of $x$, the magnetic potential $\phi_{\rm m}(x,z)$ above and below the disk can be found, and from this  $B_x=-\pa_x\phi_{\rm m}$, and the force (\ref{diskforce}).

Rotating the figure by $-90^\circ$, it can also be seen as the sketch for a {\em quiescent prominence}\,: a sheet of cool gas floating in the solar atmosphere, supported against gravity by a magnetic field (see Ch.\ 11 in \hyperlink{prie}{Priest} 2014). 

\section{`Compressibility' of a magnetic field}
\label{Bcompress}

Suppose the volume of the box of uniform field in {Fig.\,\ref{boxstress}} is changed by expansions in $x$, $y$, and $z$, and we ask the question how this changes the magnetic pressure in it. Let the expansion take place uniformly, i.e.\ by displacements $\Delta x\sim x$, $\Delta y\sim y$, $\Delta z\sim z$. The magnetic field then changes in magnitude, but remains uniform ({\cpr problem \ref{pr.6}}). At a given point in the volume, orient the axes such that $z$ is parallel to $\bf B$. Consider an expansion in $x$ and $y$ only, i.e.\ $\Delta z=0$. Under such a change the volume $V$ of the box is proportional to its cross sectional area $A$ in the ($x,y$)-plane. By conservation of magnetic flux, the number of field lines crossing $A$ remains unchanged, hence $B\sim 1/A\sim 1/V$. The magnetic pressure thus varies as $ B^2/8\pi\sim V^{-2}$ for expansions perpendicular to $\bf B$. 

On adiabatic expansion, the pressure of a gas with a ratio of specific heats $c_p/c_v=\gamma$ varies as $p\sim V^{-\gamma}$. We can therefore say that, for expansions perpendicular to the field, the magnetic pressure varies like that of a gas with $\gamma=2$; somewhat stiffer than a fully ionized ideal gas ($\gamma=5/3$). The tension in the field, however, makes  the analogy with a compressible gas inapplicable for expansions {\em along} ${\bf B}$.

Consider next the case when the box contains a magnetic field that is of unspecified configuration, except it is known or arbitrarily assumed that the 3 components are of the same average mean square strength. That is, if $<>$ indicates an average over the volume,
\beq <B_x^2>=<B_y^2>=<B_z^2>.\eeq
Expand the volume uniformly in the $x$-direction by a small amount, such that the new position $x_1$ of everything in the box is related to its old position $x_0$ by $x_1=(1+\epsilon)x$. Since the induction equation is linear in $\bf B$, the effect of the expansion can be done separately on the 3 components of $\bf B$.  $B_x$ remains unchanged, $B_{1x}(x_1)=B_{0x}(x_0)$, but the  $y$- and $z$-components are expanded in the perpendicular direction, and are reduced by the factor $1+\epsilon$. To order $\epsilon$ the  magnetic pressure thus changes by the factor
\beq B_1^2(x_1,y,z)/B_0^2(x_0,y,z)=[1+2/(1+\epsilon)^2]/3=1-{4\over 3}\epsilon.\label{43}\eeq
The same factor applies for small expansions simultaneously in several directions ({\cpr problem \ref{pr.6}a}). When the expansion is {\em isotropic and uniform} in all directions, (\ref{43}) even holds for fields of different strength in the 3 directions ({\cpr problem \ref{pr.6}b}). 

This is the basis for the statement sometimes encountered that a `tangled' magnetic field behaves like fluid with $\gamma=4/3$, analogous to the pressure of a photon field. The statement is nevertheless of very limited use. Volume changes in a fluid are rarely even remotely isotropic (in the incompressible limit\,: never). Where they are not,  the strengths of the field components rapidly become unequal on distortions of the volume, after which the factor (\ref{43}) does not apply anymore. Examples were given in {section \ref{stretch}}.

\begin{figure}[t]
\hfil\includegraphics[width=0.85\hsize]{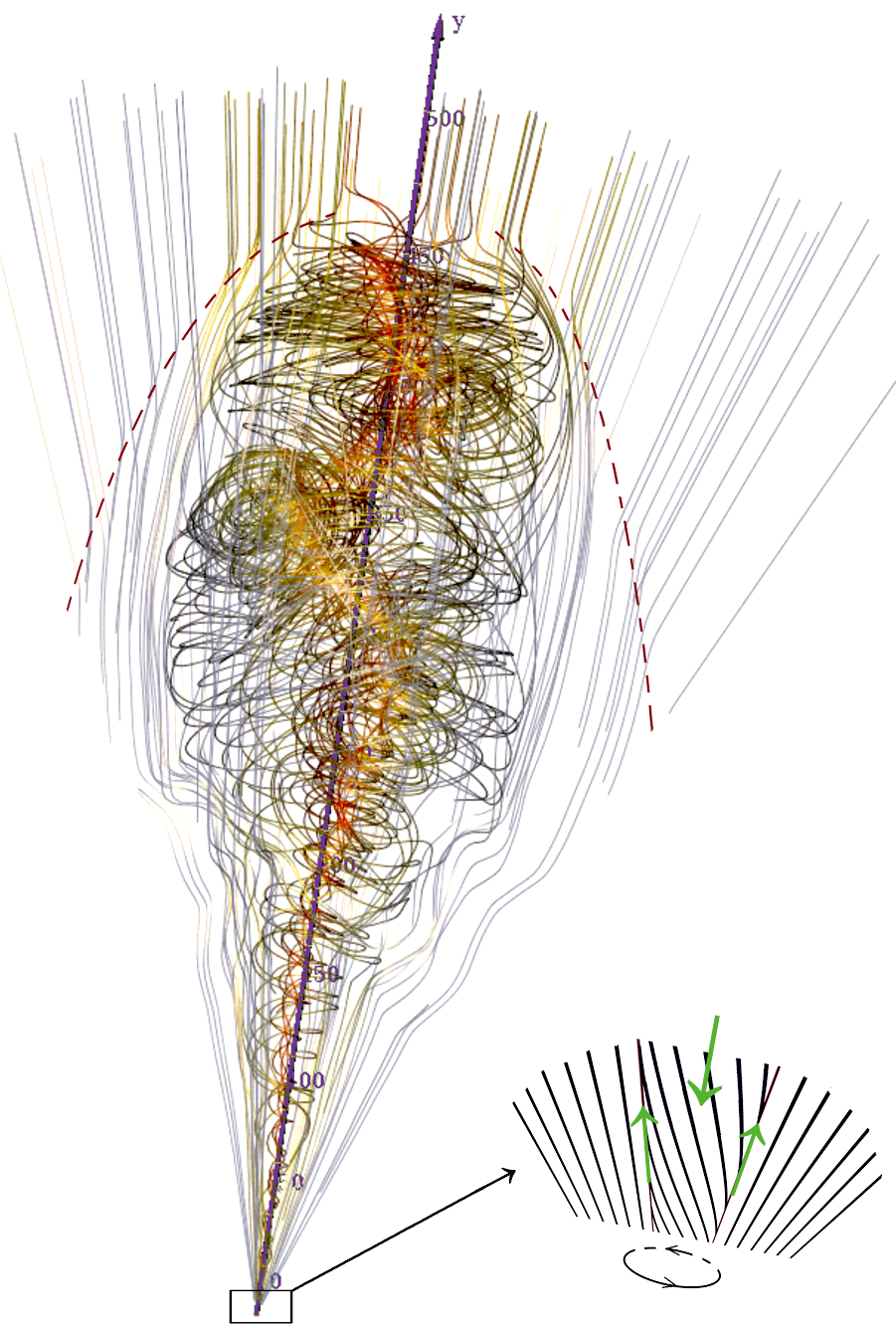}\hfil
\caption{\small Jet from a 3D MHD simulation by \protect\hyperlink{moll}{Moll (2009)}. 
Rotation applied near the origin (inset) of an initially radial magnetic field  (thin lines) produces an outflow with an increasingly twisted field that starts buckling under the pressure of the azimuthal field component. Ahead and surrounding the jet a fast mode shock (dashed) propagates into the surrounding magnetic field. The volume current in the interior of  the jet is compensated by the surface current along the interface with the surrounding magnetic field (green arrows: poloidal component of the current). The pressure of the azimuthal field component causes the jet to expand into the surroundings.}  
\label{molljet}
\end{figure}

\section{Twisted magnetic fields\,: jets}
\label{jet}

An application where twisted magnetic fields play an essential role is the production of fast outflows from magnetic fields anchored in rotating objects (stars, accretion disks, black holes). 

Consider a rotating disk containing an ordered magnetic field which accelerates an outflow of mass from the surface of the disk. At some distance the field becomes highly twisted ({Fig.~\ref{molljet}}). The pressure $B_\varphi^2/8\pi$ contributed by the azimuthal (twist) component of this field causes it to expand into the pre-existing field configuration (cf.\ {sect.~\ref{twistube}}), and produces a shock wave around and ahead of the jet. 
Near the base of the jet the twisted configuration is stabilized by the surrounding field. As the twist angle increases with distance from the source, while the pressure of the external field declines, the field configuration of the jet starts buckling under the pressure of its azimuthal component. Since the poloidal field strength ($\sim 1/r^2$) declines more rapidly with distance $r$ than the azimuthal component $B_\varphi$ ($\sim 1/r$), this happens inevitably at some finite distance from the source.

The inset in Fig.\,\ref{molljet} shows how the `closing of currents' in a jet is not an issue. The current is not needed for understanding a magnetorotationally powered jet in the first place, but if one insists it can be computed from the field configuration. It closes through the head of the jet and along its boundary with the environment (compare with {section \ref{twistube}} and {\cpr problems \ref{twistp}, \ref{ctwist}}). 

The interest in magnetically powered jets has given rise to a thriving subculture in the literature fueled by misunderstanding of the role of currents. Currents are confused with the jet itself, the direction of the currents confused with the direction of the twist in the magnetic field, and unnecessary elaborations made on the closing of currents. In  this tradition currents are correctly understood as due to the rotation at the base of a jet, but are then incorrectly dissociated from this process and reinterpreted as the source of the toroidal field,  thereby mixing up cause and effect. These misunderstandings can be avoided with a grasp of the basics of MHD.
\clearpage\noindent

\section{Magnetic helicity and reconnection}
\label{recohel}

Since a field of zero energy also has zero magnetic helicity (sect.\ \ref{helicity}), a field configuration with a finite helicity cannot decay completely, however far out of equilibrium or unstable. At least not as long as flux freezing holds, since helicity is conserved only in ideal MHD. In practice, relaxation of the configuration can quickly lead to the formation of  length scales that are small enough for reconnection to become effective (current sheets), after which energy and helicity can both decline (at a slower rate). 

Reconnection does not necessarily cause the helicity of a configuration to decrease, however. Helicity can also be created (quasi out of nothing) by a change of topology.  Fig.\,\ref{helic} shows an example of two disconnected planar magnetic  loops, embedded in a field-free environment. They have the same amount magnetic flux, both are untwisted and have zero helicity. Assume that they are brought together by a fluid flow until they touch. (This can be achieved with negligible input of energy). Reconnection between the loops {\em releases} magnetic energy, but the resulting single loop now has a finite helicity. It is contained in the shape of its path ({\em writhe}, \url{https://en.wikipedia.org/wiki/Writhe}). Folding this loop out onto a plane, it is seen that this amount of helicity corresponds to a twist of one turn around the plane loop. [The unfolding step requires input of energy, but does not change helicity]. 

\begin{figure}[t]
\hfil\includegraphics[height=0.15\hsize]{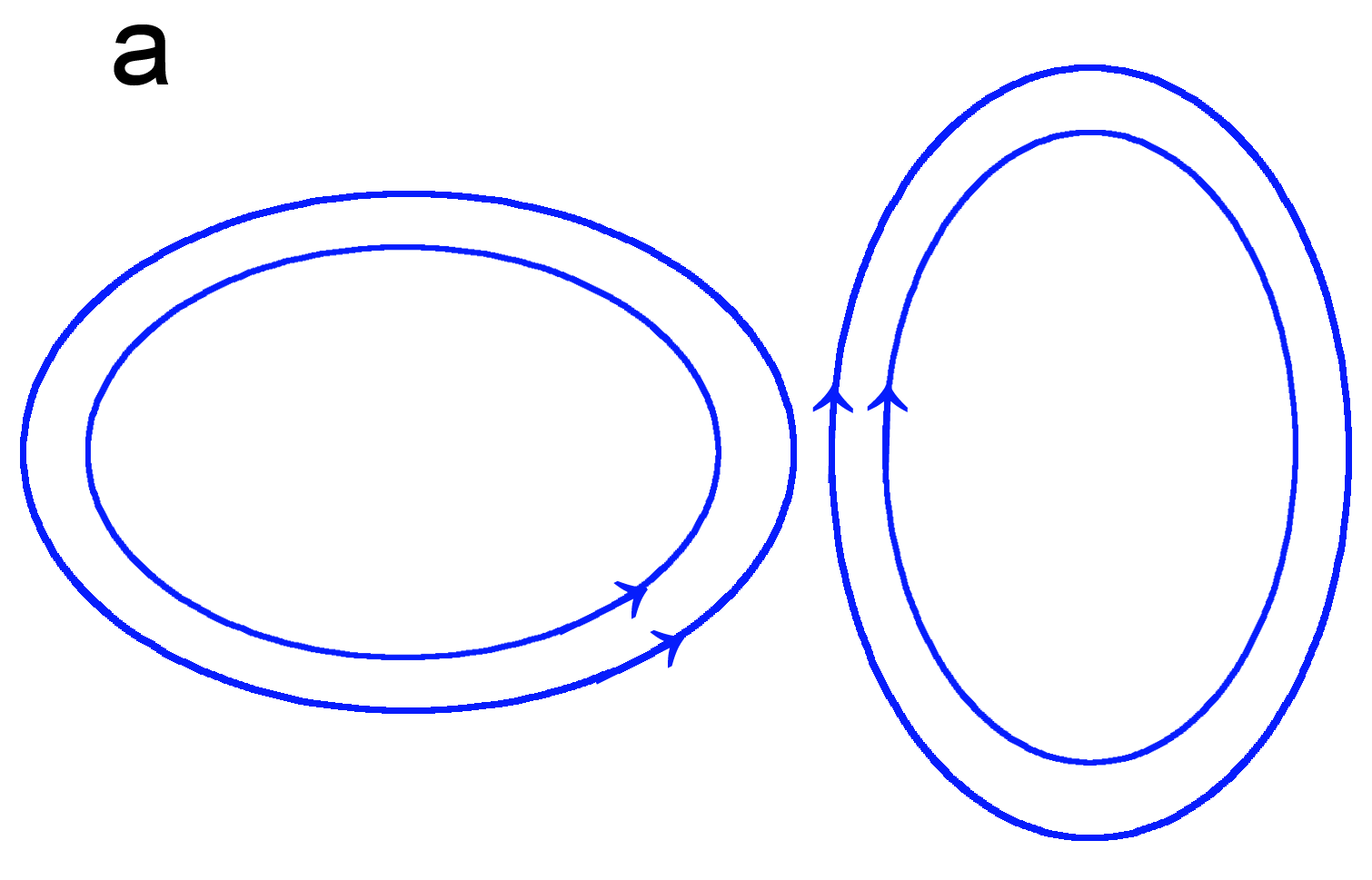}\hfil\includegraphics[height=0.15\hsize]{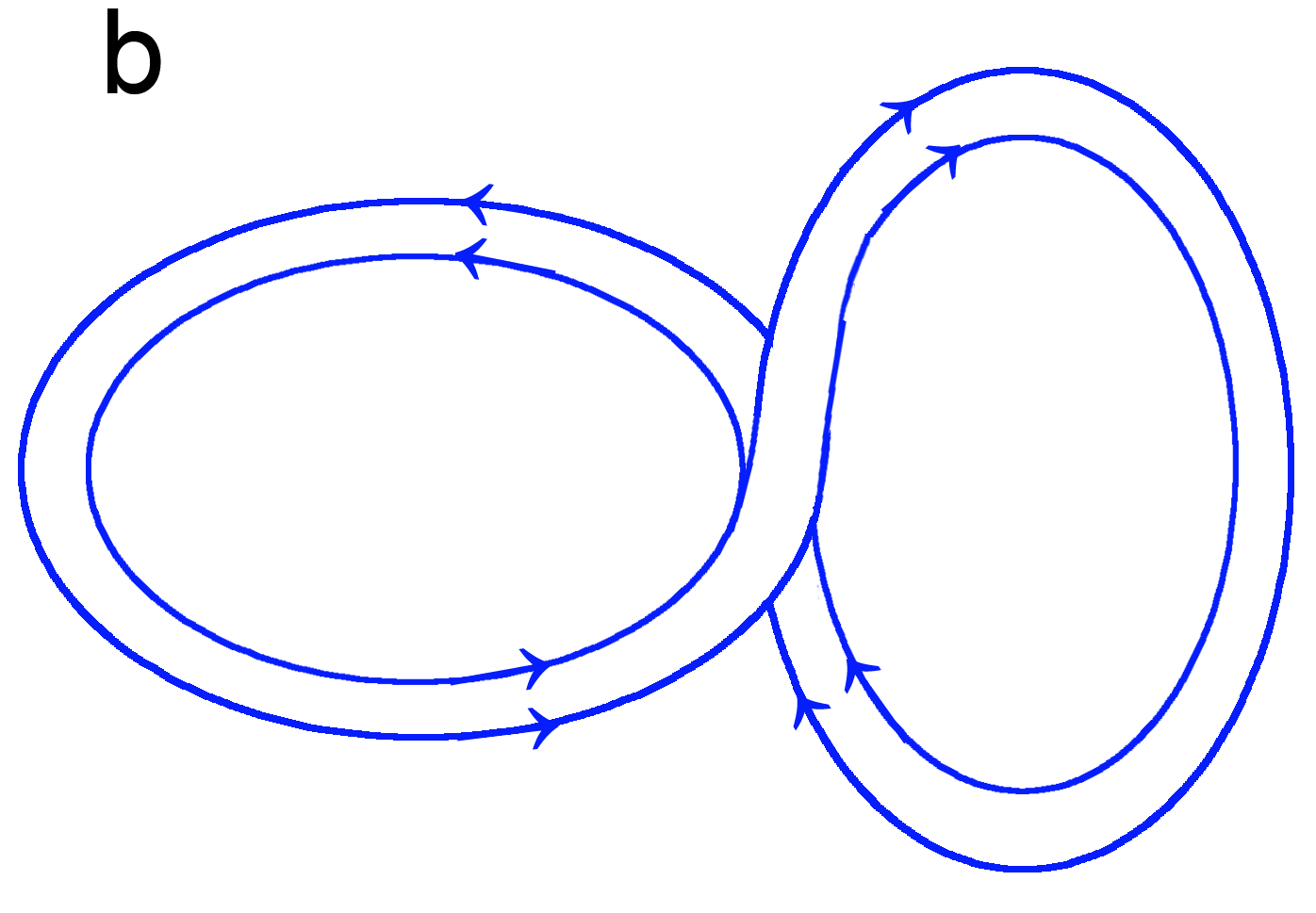}\hfil\includegraphics[height=0.15\hsize]{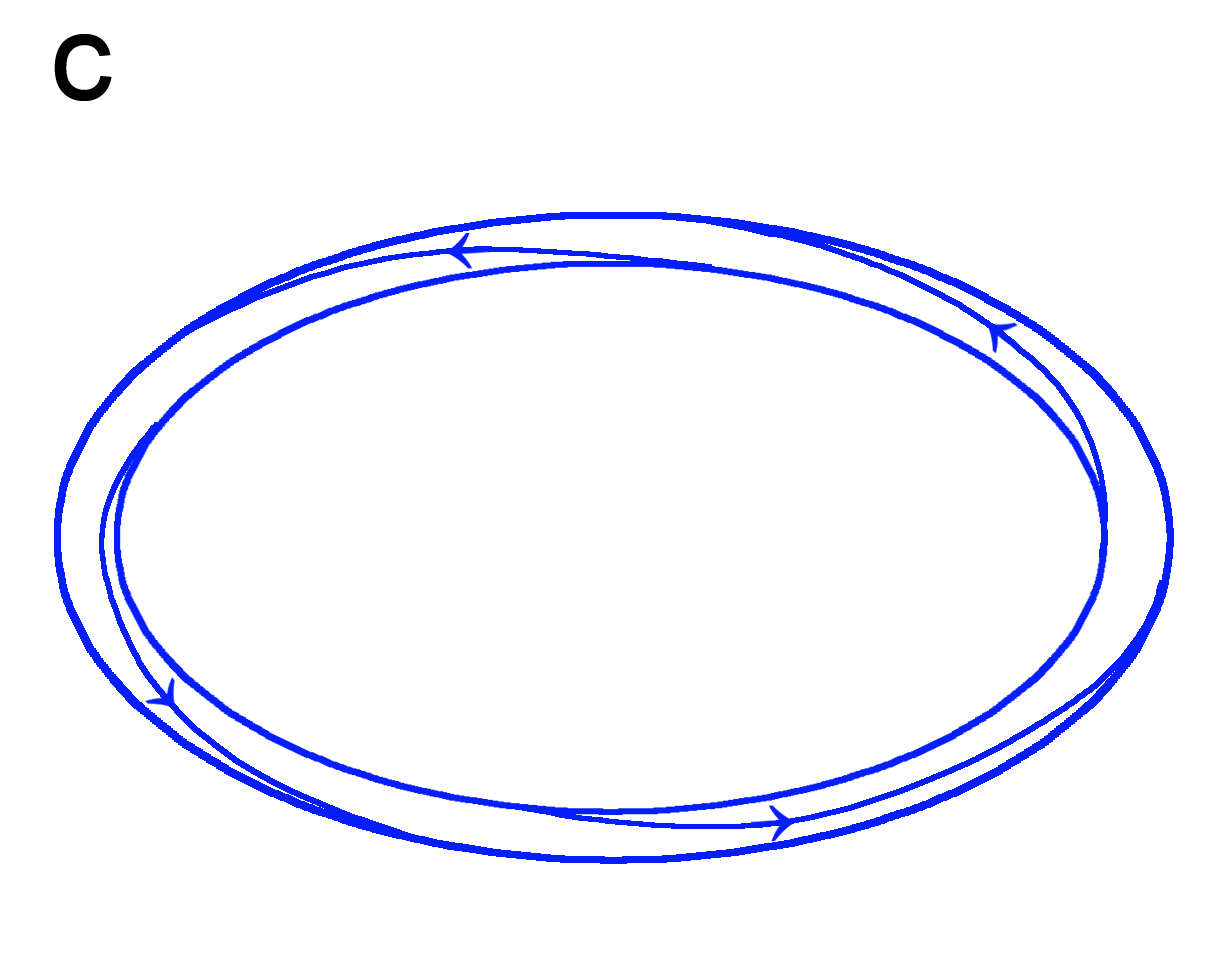}\hfil
\caption{\small Helicity generation by reconnection. Two untwisted loops ({\bf a}) with zero helicity reconnect at the point of closest approach ({\bf b}). The resulting single loop is helical; on folding it out into a plane the helicity appears as a twist of one turn ({\bf c}).}
\label{helic}
\end{figure}

The inverse sequence {\bf c-b-a} in Fig.\,\ref{helic} is of course also possible. One can imagine a twisted loop as in {\bf c} to become unstable, untwisting itself by folding into a figure-eight, which subsequently reconnects into two untwisted O-rings. The sequence in Fig.\,\ref{helic} is more likely in an environment where fluid flows are sufficiently dominant ($v/\va\ga 1$) to bring initially disconnected loops into contact. The inverse could take place in a hydrodynamically more quiescent environment ($v/\va\la 1$), where the sequence  of events is dictated instead by the magnetic instability of the twisted loop itself.

The field configuration of Fig.\,\ref{helic} can be enclosed entirely in a simply connected volume $V$ with ${\bf B}=0$ on the boundary, so its helicity can be defined uniquely (and is conserved until reconnection takes place). Intuitively one feels the need for something that can be defined also for cases where $\bf B\cdot n$ does not necessarily vanish everywhere on the surface of the volume we are interested in. With some restriction, this is possible in terms of a {\em relative helicity}. 

If the normal component $\bf B\cdot n$ does not vanish, but can be {\em kept fixed} on the surface of $V$, we can complement the region outside $V$ by attaching to it an external volume $V_{\rm e}$. In this volume we imagine a fictitious magnetic field ${\bf B}_{\rm e}$ that `connects the ends' of the field lines sticking out out $V$. That is\,: 
a) on the part of the surface shared by $V$ and $V_{\rm e}$ the normal component $\bf B_{\rm e}\cdot n$ matches that of $\bf B$ on the surface of $V$ (so that $\rm{div}\,{\bf B}=0$ is satisfied there, b) $\bf B_{\rm e}\cdot n=0$ on the remaining surface of $V_{\rm e}$, and c) ideal MHD holds inside $V_{\rm e}$. With this artifice, the helicity is still not defined uniquely since it depends on the configuration of ${\bf B}_{\rm e}$, but {\em changes} in helicity are now uniquely defined irrespective of the shape of ${\bf B}_{\rm e}$  (\hyperlink{berg}{Berger \& Field} 1986). That is, the helicity of the entire volume can now be meaningfully attributed, apart from a fixed constant, to the helicity of $\bf B$ in $V$ only. Essential for this to work is that ${\bf B\cdot n}$ does not change on the interface between $V$ and $V_{\rm e}$ and that this interface stays fixed, whatever happens inside $V$ or is imagined to happen in $V_{\rm e}$. 

This is illustrated with the example sketched in Fig.\,\ref{helicr}, a variation on the inverse of the sequence of Fig.\,\ref{helic}. It  shows a simplified view of events that can happen on the surfaces of magnetically active stars like the Sun, or strongly magnetic neutron stars (magnetars). A section of a  twisted magnetic flux bundle initially buried below the surface rises through the surface (by magnetic buoyancy for example, cf.\ {section \ref{buoy}}). Only the erupting segment interests us, but it has open ends. In order to measure the changes in helicity  in this sequence, we attach a fictitious bundle of field lines to it (a `spectator field', dashed lines in the figure) so as to complete the segment into a loop. This configuration can then be embedded in a volume where helicity is defined uniquely.

\begin{figure}[t]
\hfil\includegraphics[width=0.25\hsize]{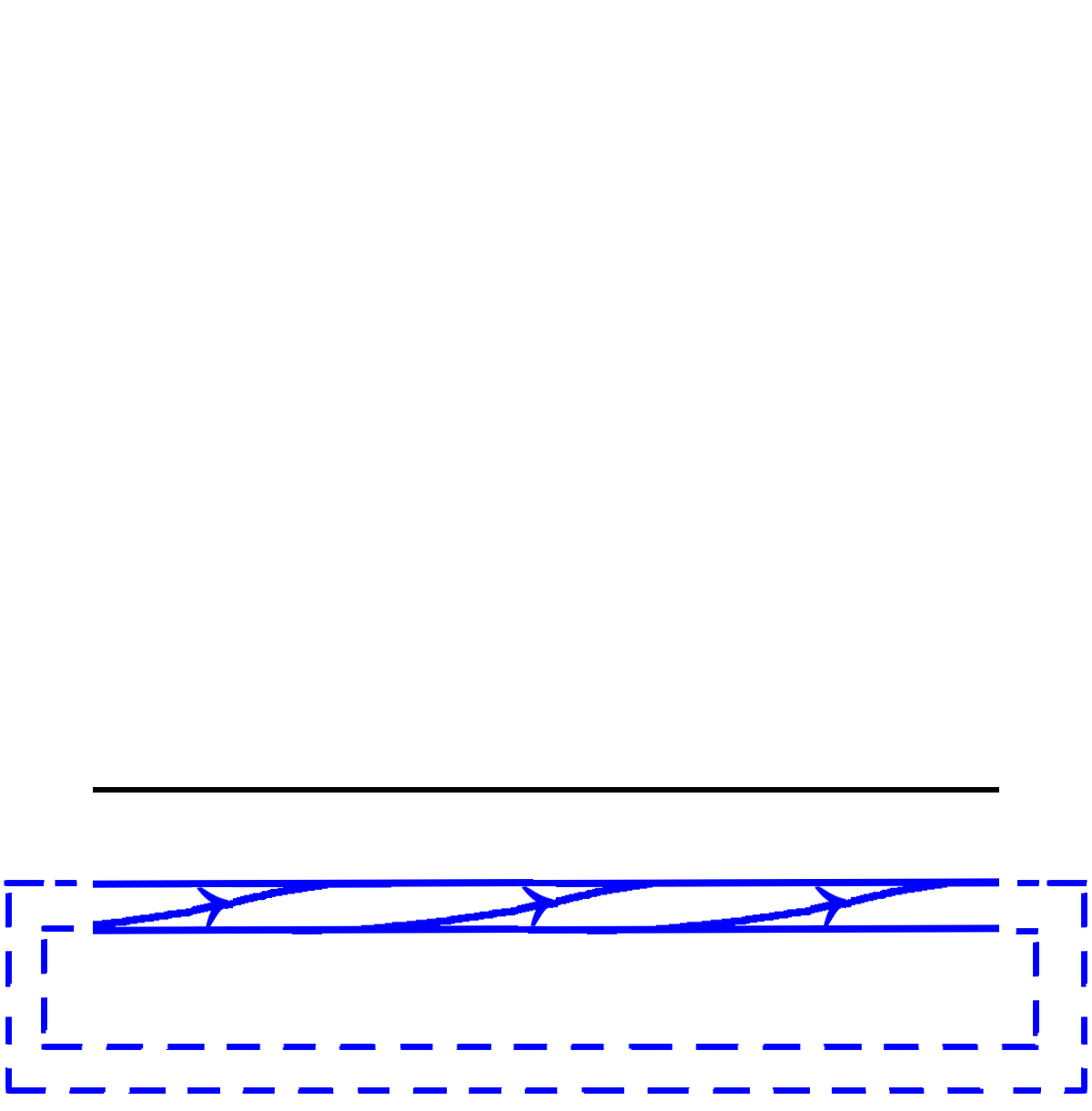}\hfil\includegraphics[width=0.25\hsize]{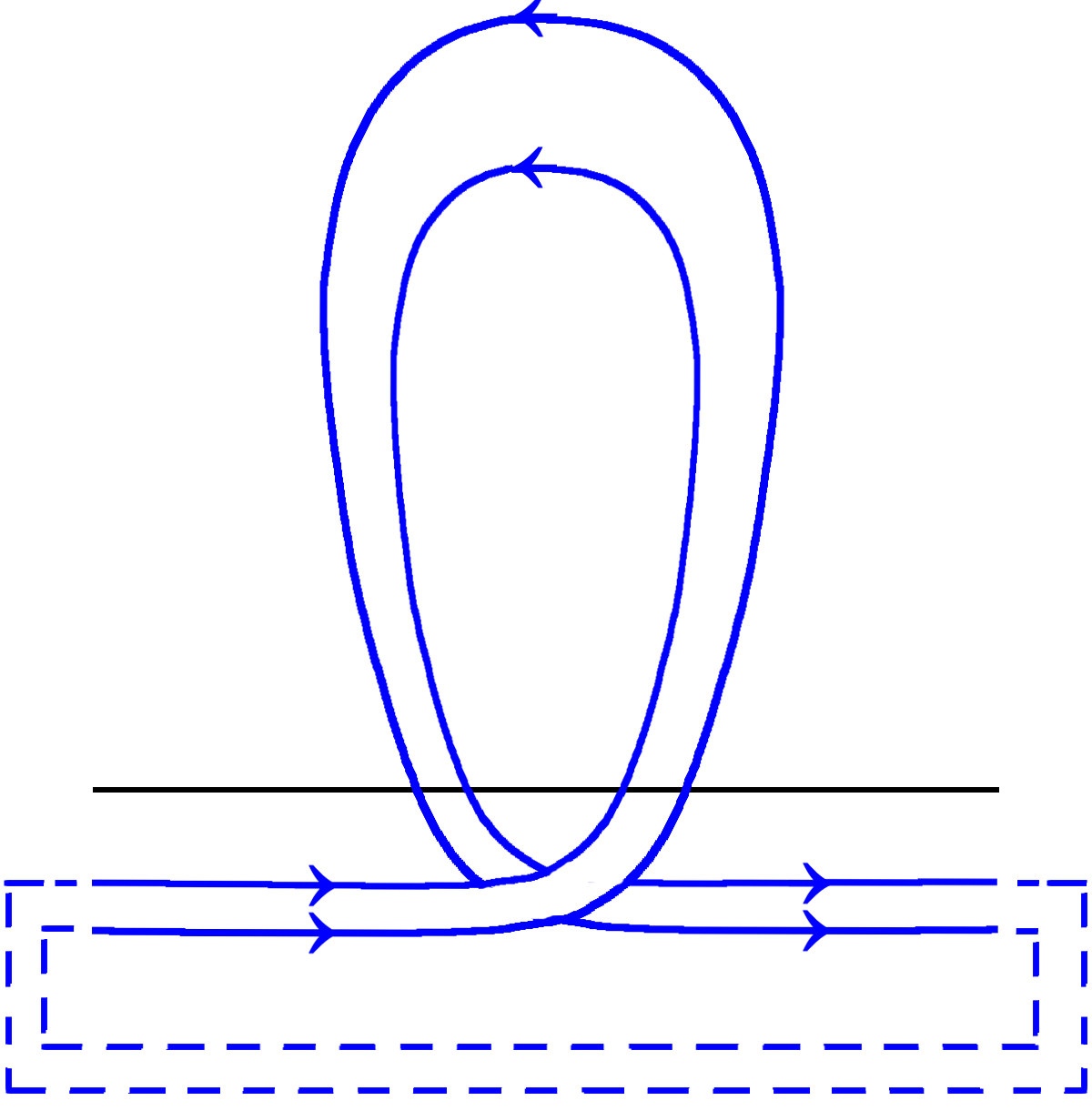}\hfil\includegraphics[width=0.25\hsize]{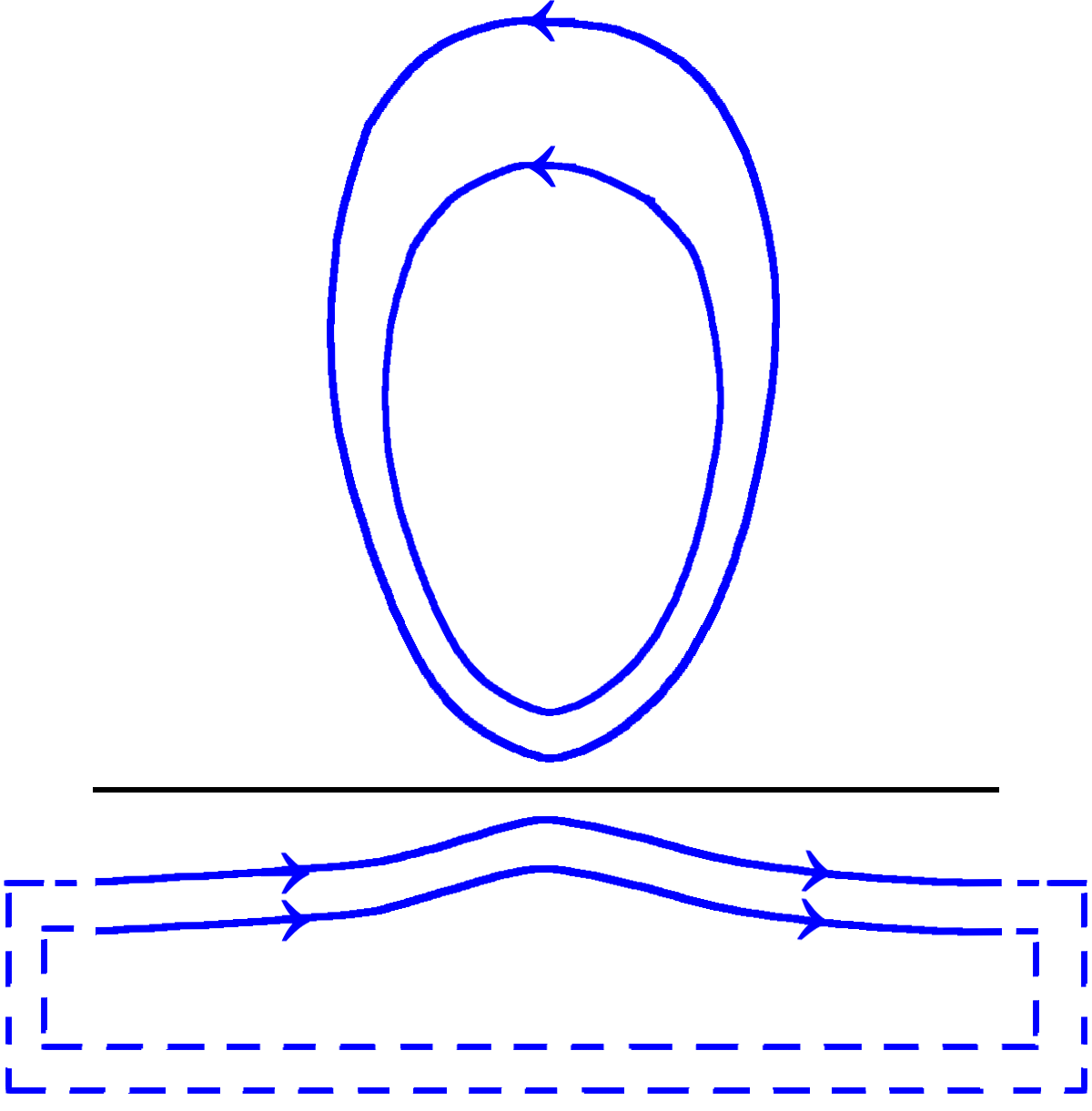}\hfil
\caption{\small Emergence of a twisted field through the surface of a star. Left\,: a section of a strand below the surface (black) containing one turn of twist. Middle\,: it erupts through the surface, while its path makes a half-turn rotation. Right\,: an untwisted loop reconnects from it into the atmosphere while leaving a configuration of zero helicity. Dashed\,:  loop completion used for the definition of helicity.}
\label{helicr}
\end{figure}

The low density environment of the atmosphere of the star does not support twist to the same degree as the high-density interior. The field reduces twist by exchanging it for a loop in its path. Helicity is conserved in this step. By reconnection at the point where the path touches itself (middle panel) a loop of field disconnects, and is ejected from the atmosphere by the pressure gradient in the surrounding magnetic field (see {\cpr problem \ref{diamag}}). The magnetic bundle is now contained below the surface as before, but is untwisted, and the helicity of the configuration has decreased by the equivalent of one turn of twist.
The helicity has disappeared already at the moment of reconnection, and nothing helical has left the star.

 This is an idealized example, but it serves to illustrate that the change of helicity in a magnetic ejection process like Fig.\,\ref{helicr} is not properly described by a term like `helicity ejection'. The concept of \hyperlink{curhel}{current helicity} 
does not fare much better. Unlike magnetic helicity, this is a quantity that does have a local definition, but there is no conservation equation for it like there is for mass or energy. As a result it does not make sense to describe the ejection process in terms of a `flux' of current helicity either. All one can say is that unwinding of twist that was present in an emerging loop of field lines leads to the formation of a plasmoid that is ejected from the star, while helicity decreases in the process. In the Sun, such processes are responsible for the {\em coronal mass ejections} that cause aurorae and disturbances in the Earth's magnetic field. 

For a more on the properties of magnetic helicity, and its relation to vorticity in ordinary fluid mechanics see \hyperlink{moff}{Moffatt} (1985).

\section{Polarization}
\label{polariz}
\subsection{Conducting sphere in a vacuum field}
\label{csphere}
The charge densities discussed in {section \ref{chargedd}} are frame transformation quantities that disappear in the locally comoving, corotating frame of the fluid. At the boundary with a non-conduction region (a vacuum or an unionized gas) they become real, however, in the form of  surface charge densities.

As an example, imagine a laboratory setup containing an initially uniform magnetic field $\bf B$ and a vanishing electric field. The laboratory environment is non-conducting. Then move an electrically conducting sphere of radius $R$  into into this field, at a constant velocity $\bf v$.  Assume that the magnetic diffusivity $\eta$  (see {sect.~\ref{diffusion}}) of the sphere is small but finite, and observe the sequence of events in a frame comoving with the sphere. After its introduction, the magnetic field inside the sphere vanishes initially, shielded by currents on its surface. The field lines accommodate this as in Fig.\,\ref{gap}.

The laboratory field diffuses into the sphere on a time scale  $\tau=R^2/\eta$, until for $t\gg\tau$ the field is again uniform everywhere\footnote{~In the laboratory frame we now have a conducting sphere moving across a magnetic field without distorting it. This has confused some authors into questioning the whole idea of flux freezing of magnetic fields in ideal MHD. Flux freezing and MHD do not apply outside the conductor, however. Key here is the appearance of a surface charge density.}. In the comoving frame the sphere finds itself embedded in an external electric field  ${\bf E}={\bf v\times B}/c$, but inside it the electric field vanishes due to the conductivity of the sphere. This is a classical problem in electrostatics. Let the magnetic field be in the $x$-direction, the velocity $\bf v$ in the $y$-direction.\ {\em Polarization} of the sphere in the electric field displaces the positive and negative charges with respect to each other. Inside the sphere the displacement is uniform hence does not cause a charge density, but on the surface of the sphere it results in a charge distribution $\sigma_{\rm s}=3vB \cos\theta/(4\pi R c)$ (charge per unit area), where $\theta$ is the polar angle with respect to the $z$-axis.  (Exercise\,: verify this with \hyperlink{jack}{Jackson E\&M} Ch.\ 2.5).

Outside the sphere, this surface charge creates an electric field equivalent to that of a point dipole at the center of the sphere, inside it a uniform electric field which just cancels the external field ${\bf v\times B}/c$. On transforming to the observer's frame, the external electric field disappears at large distance from the sphere, and only the dipole field created by the surface charge remains. Inside the sphere the electric field then has the expected frame-transformation value $\bs{E}_{\rm int}=-{\bf v\times B}/c$. (Exercise\,: sketch the electric field lines in both frames).

The polarization charge is the same in the comoving and the laboratory frames [up to relativistic corrections of order $(v/c)^2$ due to Lorentz contraction], but its interpretation is very different.  In the lab frame, the flow $\bf v$ carries two fluids of opposite charge across the magnetic field. The Lorentz force is of opposite sign for the two, causing them to displace from each other and leaving an excess of $+$'s on one side and $-$'s on the other. In the comoving frame of the sphere  these Lorentz forces are absent. In this frame the surface charge is an electrostatic effect, induced on the conducting sphere by an external electric field.

\subsection{Pulsars}\label{pulsar}
An important application of the above concerns the magnetic field of a rotating magnetic neutron star, i.e.\ a pulsar. The surface temperature of a pulsar is so low compared with the virial temperature (the temperature needed to cause the atmosphere to inflate to the size of the star itself) that near-perfect vacuum holds already some 10 meters above its surface. The interior of the star is perfectly conducting in the MHD sense, so that the electric field in a frame corotating with the star vanishes in its interior. As in the simpler example above, a surface charge appears at the boundary with the vacuum. In the corotating frame, it can be seen as polarization resulting from the external electric field $({\bf\Omega\times r)\times B}/c$ in which the conducting star finds itself embedded. In the inertial frame, where this field does not exist, the polarization can be seen as due to the differential Lorentz force acting on the two charged species rotating in the magnetic field of the star.

The surface charge develops at the level in the atmosphere where vacuum conditions take over from the high-conductivity regime  in deeper layers.  This happens at such a low density that the magnetic field completely dominates over the gas pressure there, $\beta\ll1$. Unaffected by fluid pressure, the magnetic field  in the surface charge region is thus the continuation of the vacuum field outside. 

Suppose the magnetic field of the star is stationary in an inertial frame, $\pa{\bf B}/\pa t=0$. It follows that it must be axisymmetric with respect to the axis of rotation, and by the induction equation ({\ref{inM})} the electric field has a potential, ${\bf E}=-\nb\Phi$. It therefore has a potential everywhere, in the conducting region below the surface charge as well as in the surface charge region itself and in the vacuum outside. The radial component of ${\bf E}$ makes a jump across the surface charge, but its potential is continuous. To find the electric field in the vacuum, solve $\nabla^2\Phi=0$ there, with the boundary condition that $\Phi$ is continuous across the surface charge. 

 As an example take for the magnetic field configuration a dipole aligned with the rotation axis and centered on the star. If $B_0$ is the field strength at the rotational north pole of the star, the field is
\beq {\bf B}={B_0 R^3\over 2r^3}(3\cos\theta\,{\bf \hat r}-{\bf \hat z} ),\label{dip}\eeq
where $r, \theta$ are the spherical radius and polar  angle,  ${\bf \hat r}$ and ${\bf \hat z}$ unit vectors  in the (spherical) radial direction and along the rotation axis respectively, and $R$ the radius of the star. The electric field ${\bf E}_{\rm i}$ in the star, as seen in an inertial frame:
\beq {\bf E}_{\rm i}=-{(\bf\Omega\times r)\times B}/c,\eeq
becomes, with (\ref{dip}),
\beq {\bf E}_{\rm i}=B_0{\Omega R\over 2 c}({R\over r})^2[(3\cos^2\theta-1){\bf\hat r}-2\cos\theta\,{\bf\hat z}].\eeq
Its potential $\Phi_{\rm i}$ is (exercise\,: show this)
\beq \Phi_{\rm i}=B_0 R{\Omega R\over 2 c}{R\over r}(\cos^2\theta-1) + k,\eeq
where $k$ is an arbitrary constant. The angular dependence of the field is that of a quadrupole. Since the potential is continuous across the surface charge at $r=R$, the external field has the same angular dependence (but a different dependence on $r$). Matching at $r=R$ yields, (taking the potential to vanish at infinity):
\beq \Phi_{\rm e}=B_0 R{\Omega R\over 2 c}({R\over r})^3(\cos^2\theta-1/3).\eeq
The surface charge density $\sigma_{\rm s}$ is then
\bea 
\sigma_{\rm s}&=&({\bf E}_{\rm e}-{\bf E}_{\rm i})\cdot\hat{\bf r}/4\pi= -\pa_r(\Phi_{\rm e}-\Phi_{\rm i})/4\pi\\
&=&-{B_0\over 4\pi}{\Omega R\over c}\cos^2\theta.
\eea
(\hyperlink{gold1}{Goldreich and Julian} 1969). For the typical magnetic field strengths and rotation rates of pulsars the electric field is so strong that the external volume gets filled with charges pulled from the surface of the neutron star. This leads to much more difficult problems,  the subject of {\em pulsar magnetospheres}.
  
\subsection{Electricity from MHD}
\label{eMHD}
Experimental devices for the direct extraction of electric power from a gas flow (MHD power generation) provide a nice illustration of the discussions in  {sect.~\ref{origj}} and the sections above. A hot gas, made partially conducting by seeding it with atoms of low ionization potential like Potassium, is made to flow with velocity $\bf v$ across an externally applied magnetic field.  On the sides  of the flow channel there are electrodes, oriented parallel to both $\bf B$ and $\bf v$. With the experimental environment being non-conducting, the setup is similar to the polarization {experiment in \ref{csphere}}.
If the electrodes are not connected by an external circuit, an electric field $-{\bf v\times B}/c$ is present in the flow, as seen in the lab frame, and a corresponding polarization charge appears on the electrodes. The current {discussed in \ref{origj}} is absent. The deflection of the charges in their orbits around the field that would drive this current is canceled exactly by the $\bf E\times B$ drift due to the electric field in the lab frame. In the fluid frame, on the other hand, the electric field vanishes. The fluid is not aware that in the lab frame it is seen as flowing across the $B$-field. 

We now change conditions by closing the external circuit with a negligible resistance. The polarization charge is shorted out, and with it the electric field in the lab frame disappears. In the fluid frame there is now an electric field $+{\bf v\times B}/c$, and associated with it is a current. In the lab frame, the source of this current is as described {in \ref{origj}}, in the fluid frame on the other hand it is just a current driven by an electric field, across the internal resistance of the plasma. 
The flow pattern and the distortion of the field lines by this current follow from the MHD equations of motion ({\ref{eqm}}) and induction ({\ref{inddif}}). The flow does work against the curvature force of the distorted field lines. In a short-circuited state like this, the associated power is all dissipated in the flow itself (like in a short-circuited battery). 

As this discussion illustrates, it is important to keep in mind how much the physical description in terms of  microscopic processes can differ in the lab and in the fluid frame.  One has to keep them clearly separate to avoid confusion.

\subsection{Critical ionization velocity}\label{critioni}
An interesting situation arises if the moving conductor of {section \ref{csphere}} is replaced by an insulator in the form of an unionized (atomic or molecular) gas. Assume again that no electric field is present in the laboratory frame. Let the gas consist of atoms A of mass $m$ flowing with speed $v$ across the magnetic field.  Now insert an ion A$^+$ into the gas by hand.  Being charged, it is tied to the magnetic field. Since the electric field vanishes in our reference frame, the guiding center of the ion's orbit is initially at rest. Neutral atoms flowing past it collide with the ion, with a center-of-mass energy of $mv^2/2$ (neglecting the thermal velocity of the orbiting particle). The collision can lead to the further ionization of an atom of the incoming gas if the collision energy exceeds the ionization energy $\epsilon$ of A. The new ion will also be tied to the magnetic field, so there are then additional ion-neutral collisions, each producing more ions. Runaway ionization of the gas will therefore result when the velocity $v$ between the neutrals  A and its ions A$^+$ sufficiently exceeds the {\em critical ionization velocity} $v_{\rm c}$,
\beq v_{\rm c} = (2\epsilon/m)^{1/2}.\label{crition}\eeq
This yields $v_{\rm c}\approx 50$ km/s for A= neutral Hydrogen. 
Since the presence of a stray ion is usually plausible, the conclusion is that a neutral gas moving across a magnetic field will ionize when its velocity (in the frame where $E=0$) exceeds $v_{\rm c}$ (a suggestion due to Alfv\'en). 

An assumption in the simple picture above is that the ions can be considered tied to the field, in the laboratory frame, while the neutrals flow past it. This means that the density must be low enough for ambipolar drift (sect.\ \ref{hall-am}) to be significant:  the drift velocity $v_{\rm a}$ must exceed (\ref{crition}), otherwise collisions between the ions and neutrals do not have enough energy to ionize further atoms. Since  $v_{\rm a} $ decreases with increasing ion density, the runaway is limited\,: the degree of ionization stabilizes when $v_{\rm a}$ drops to $v_{\rm c}$. The process is unlikely to be encountered in practice,  since the high ambipolar drift velocities required for critical ionization limit it to very low density environments.

\bigskip\bigskip

\section{References}\label{refs}

For some older textbooks no links are listed.  For Roberts,  and for  Ferraro \& Plumpton try amazon for used copies. In the case of Landau \& Lifshitz, downloadable scans of the first and second editions have appeared on the internet.\medskip

{\parindent=-1em

\noindent\par

\hypertarget{balb}{Balbus, S.A. 2009}, in {\em Physical Processes in Circumstellar Disks Around Young Stars}, ed. P. Garcia, University of Chicago Press, \url{http://arxiv.org/abs/0906.0854}

\hypertarget{berg}{Berger, M.A., \& Field, G.B. 1984}, J.\ Fluid Mech. 147, 133, \hfill\-
\url{http://dx.doi.org/10.1017/S0022112084002019}

\hypertarget{brai} Braithwaite, J.\ 2015,  Mon.\ Not.\ Roy.\ astron.\ Soc.,  450, 3201 \url{http://dx.doi.org/10.1093/mnras/stv890}

\hypertarget{ferr}{Ferraro, C.V.A., \& Plumpton}, C. 1966, {\em An introduction to magneto-fluid dynamics}, Oxford University Press.

\hypertarget{flow}{Flowers, E., \& Ruderman, M.A.}\ 1977, Astrophys.\ J.\  215, 302,  \hfill\-
\url{http://dx.doi.org/10.1086/155359}

\hypertarget{gals}{Galsgaard, K., \& Nordlund}, \AA. 1996, J.\ Geophys.\ Res.\ 101, 13445, \hfill\- 
\url{http://dx.doi.org/10.1029/96JA00428}

\hypertarget{gold1}{Goldreich, P., \& Julian, W.H.}\ 1969, Astrophys.\ J.\  157, 869,  \hfill\-
\url{http://dx.doi.org/10.1086/150119}

\hypertarget{gold2}{Goldreich, P., \& Reisenegger, A.}\ 1992, Astrophys.\ J.\   395, 250,  \hfill\-
\url{http://dx.doi.org/10.1086/171646}

\hypertarget{jack}{Jackson, J.D.}\ {\em Classical Electrodynamics}, Wiley,\hfill\-  \url{https://archive.org/details/ClassicalElectrodynamics}

\hypertarget{jaro}{Jaroschek, C.H., Treumann, R.A., Lesch, H., Scholer, M.} 2004, Phys.\ Plasmas 11, 1151,   \hfill\- 
\url{http://dx.doi.org/10.1063/1.1644814}

\hypertarget{kipp}{Kippenhahn, R., Weigert, A., \& Weiss, A.}\ 2012, {\em Stellar Structure and Evolution} (2nd edition), Springer,  \url{http://www.springer.com/us/book/9783642302558}

\hypertarget{komm}{Komissarov, S.S.}\ 1999, Mon.\ Not.\ Roy.\ astron.\ Soc., 303, 343, \hfill\- 
\url{http://dx.doi.org/10.1046/j.1365-8711.1999.02244.x}

\hypertarget{kuls}{Kulsrud, R.M.}\ 2005, {\em Plasma Physics for Astrophysics}, Princeton University Press, 
\hfill\- \url{http://www.isbns.net/isbn/9780691102672}

\hypertarget{land}{Landau, L.D. \& Lifshitz, E.M.}\  {\em Fluid Mechanics}, Pergamon Press, Oxford, \hfill\- \url{http://www.isbns.net/isbn/9780750627672}

\hypertarget{mest}{Mestel, L.} 2012, {\em Stellar magnetism}, second edition. Oxford science publications (International series of monographs on physics 154),  \hfill\- e-book:
\url{http://www.lehmanns.ch/shop/naturwissenschaften/28076582-9780191631498-stellar-magnetism-second-edition}

\hypertarget{moff}{Moffatt, H.K.}\ 1985, Journal of Fluid Mechanics, 159, 359, \hfill\- 
\url{http://dx.doi.org/10.1017/S0022112085003251}

\hypertarget{moll}{Moll, R.}\ 2009, Astron.\ Astrophys.\ 507, 1203,  \hfill\-
\url{http://dx.doi.org/10.1051/0004-6361/200912266}

\hypertarget{park2}{Parker, E.N.}\ 1972,  Astrophys.\ J., 174, 499, 
\url{http://dx.doi.org/10.1086/151512}

\hypertarget{oster}{Oster, L.}\ 1968, Solar Physics, 3, 543
\url{http://dx.doi.org/10.1007/BF00151936}

\hypertarget{park1}{Parker, E.N.}\ 1979, {\em Cosmical magnetic fields}, Clarendon, Oxford \hfill\- 
\url{http://www.isbns.net/isbn/9780198512905}

\hypertarget{park3}Parker, E.N.}\ 2007, {\em Conversations on Electric And Magnetic Fields in the Cosmos}, Princeton University Press \url{http://www.isbns.net/isbn/9780691128412}

\parindent=-1em
\hypertarget{prie}{Priest, E.R.}\  2014, {\em Magnetohydrodynamics of the Sun}, CUP, 
 \hfill\- \url{http://www.isbns.net/isbn/9780521854719}
 
\hypertarget{robe}{Roberts, P.H.}\ 1967, {\em  An Introduction to Magnetohydrodynamics}, Longmans, London.

\hypertarget{ster}{Stern, D.P.}\ 1970, American Journal of Physics, 38, 494,  \hfill\-
\url{http://dx.doi.org/10.1119/1.1976373}

\hypertarget{trit}{Tritton, D.J. 1992}, {\em Physical fluid dynamics}, Oxford university Press,  \hfill\- 
\url{https://global.oup.com/academic/product/physical-fluid-dynamics-9780198544937?cc=de&lang=en&}

\hypertarget{uzde}{Uzdensky, D.A., \& Kulsrud, R.M.}\ 2006, Physics of Plasmas, 13, 062305,\hfill\- 
\url{http://arXiv.org/abs/astro-ph/0605309}

\hypertarget{vanb}{van Ballegooijen, A.A.}\ 1986, Astrophys.\ J.\  311, 1001,\hfill\- 
\url{http://adsabs.harvard.edu/abs/1986ApJ...311.1001V}
}

\chapter{Exercises and problems}\label{Exc}
\pb
\section{Currents from flows}\label{pr.0} This exercise gives an example how currents appear and disappear when a flow acts on a magnetic field.  An initially uniform magnetic field ${\bf B}=B\,{\bf\hat z}$ in the $z$-direction is embedded in a flow in the $x$-direction, varying with $z$ as ${\bf v}= \tanh(z)\,{\bf\hat x}$. From the ideal MHD induction equation calculate how the magnetic field has changed after a time $t$. Calculate the current distribution. Now reverse the sign of ${\bf v}$ and notice how the currents have vanished again after another time interval $t$.

\pb\section{Particle orbits}\label{pr.2}
Calculate the orbit of a charged particle, initially at rest, in a magnetic field with $\bf B$ in the $y$-direction and $\bf E$ in the $z$-direction. Show that the motion consists of a circular motion around the $y$-axis superposed on a uniform velocity $v_{\rm d}=cE/B$ in the $x$-direction. Sketch the resulting orbit. The velocity $v_{\rm d}$ is called the {\em ${\bf E\times B}$ drift velocity}. Note that it is identical to the fluid velocity $\bf v$ in (\ref{ee}), hence the same for particles of all mass and charge. 

\pb\section{Displacement current at finite conductivity}\label{curdif}
In {section \ref{ecur}} perfect conduction was assumed. If an Ohm's law conductivity  ({\ref{ohm}}) is assumed instead, verify with (\ref{ohm1}) that Maxwell's equation (\ref{amp}) can be written as
\beq 4\pi{\bf j}+{\pa\over\pa t}({{\bf j}\over\sigma_{\rm c}})-{\pa\over\pa t}({\bf v\times B}/c)=c{\nb\kern-1pt\times\bf B}, \label{maxohm}\eeq
in the nonrelativistic limit. As in {sect.~\ref{ecur}} the third term on the left is negligible compared with the  right hand side. With characteristic values for velocity $V$, length $L$ and time scale $L/V$ as in {section \ref{diffusion}}, show that the LHS of (\ref{maxohm}) is of the order
 \beq 4\pi{\bf j}\,(1+{V\over c} {\eta\over L c}),\eeq
 where $\eta$ is the magnetic diffusivity. If  $l$ and $v$ are the mean free path and velocity of the current carriers, $\eta$ is of the order $\sim l v$. Why can the second term in brackets be ignored? (Hint: section \ref{MHDapp}).

\pb\section{Alternative form of the induction equation}\label{pr.3}
{\parindent=-10pt \noindent\par
{\bf a}.\ Verify eq.~({\ref{brho}}) by using the induction equation in the form ({\ref{ind1}}) and the continuity equation.  \noindent\par
{\bf b}.\ A horizontal magnetic field ${\bf B}=B_0\,{\bf\hat x}$ of $10^5$ G lies in pressure equilibrium at the base of the solar convection zone, where the density is $\rho=0.2$ g/cm$^3$. If this field were to rise to the surface of the Sun ($\rho=2\,10^ {-7}$), and assuming it remains horizontal and in pressure equilibrium  in the process, what would its field strength be? [The fields actually observed at the surface of the Sun are not horizontal, and much larger in strength].}
 
\pb\section{Integrated induction equation}
\label{indint}
Derive {eq.~(\ref{deltaB}}) by integrating the continuity equation from $0$ to $\delta t$, keeping
only first order in small quantities.

\pb\section{Stretching of a thin flux tube}
\label{pr.stretch}
A straight bundle of field lines with strength $B(t)$ oriented along the $x$-axis is embedded in a field-free medium with constant pressure $p_{\rm e}$. A section with initial length $L_0=1$ of the bundle is stretched uniformly in the $x-$direction to a new length $L(t)=at$, with $a={\rm cst.}$ During the stretching, the bundle is in pressure equilibrium with its surroundings, $p_{\rm i}+B^2/8\pi=p_{\rm e}$, where $p_{\rm i}$ is the internal gas pressure, $p_{\rm i}={\cal R}\rho T$. The temperature $T$ is kept constant in time. Using conservation of mass in the lengthening segment, calculate $B(t)$, $\rho(t)$ and the cross section of the bundle, starting from initial conditions $\rho_0$, $B_0$ such that $B_0^2/8\pi\ll p_{\rm e}$. Describe the initial ($t\rightarrow0$) and asymptotic ($t\rightarrow\infty$) dependences of $B$, $\rho$. What is the value of the plasma-$\beta$ at the transition between these regimes? 

\pb\section{Magnetic flux}\label{pr.1}
Using the divergence theorem, show that the magnetic flux passing through a surface $S$ bounded by a given fluid loop, as in {Fig.~\ref{loop}}, is independent of the choice of $S$. (Hint: Stokes' theorem)

\pb\section{Magnetic forces in a monopole field}\label{pr.7}
This exercise illustrates the limitations of identifying the two terms in eq.\ ({\ref{prescurv}}) with a pressure gradient and a curvature force.
In spherical coordinates ($r,\theta,\varphi$), the field of a monopole is
($1/r^2$,0,0).  \vskip 0.5\baselineskip
{\parindent=-10pt \noindent\par
{\bf a}.\ Show that the Lorentz force in this field vanishes except at the origin. \noindent\par
{\bf b}.\ Calculate the `pressure gradient' and `curvature' terms of eq.\ (\ref{prescurv}).} \par\noindent
(A magnetic monopole does not exist, of course, but a physically realizable field can be made by reversing the sign of the field in the southern hemisphere ($\theta>\pi/2$)\,: the `split monopole' configuration. Since the Lorentz force does not depend on the sign of $\bf B$, this makes no difference for the calculation.)

\pb\section{Magnetic forces in an azimuthal magnetic field}\label{pr.5}
In cylindrical coordinates ($\varpi,\varphi,z$), consider a purely azimuthal magnetic field ${\bf B}=B\,{\bs{\hat \varphi}}$, such that $B=B_0\varpi_0/\varpi$, where $B_0$ is a constant. 
Calculate the magnetic pressure gradient, the curvature force, the Lorentz force, and the current along the axis.

\pb\section{The surface force at a change in direction of {\bf B}}
\label{sforce}
A field ${\bf B}=B_0\,\hat{\bf z}$ changes direction by an angle $\alpha$ at a plane surface $z=z_0$. Calculate the force vector exerted on the fluid at this surface. 
Such surface forces occur in MHD shocks).

\pb\section{Magnetic energy and stress}
\label{dipenerg}
Consider a sphere with a uniform magnetic field $\bf B$ in it. Outside the sphere, the field continues as a potential field.\vskip 0.5\baselineskip
{\parindent=-10pt \noindent\par
{\bf a}. Remind yourself that the outside field is that of a point-dipole centered on the sphere (cf.\ \hyperlink{jack}{Jackson E\&M}, Ch.\ 5.10).\noindent\par
{\bf b}.\ Calculate the magnetic stress acting at the surface of the sphere. \noindent\par
{\bf c}.\ Calculate the total magnetic energy of i) the field inside the sphere, ii) the external field.\noindent\par
{\bf d}.\ A field configuration like this in a star is unstable, because displacements inside the star can be found that reduce the energy of the external potential field without changing the energy  the interior (\hyperlink{flow}{Flowers \& Ruderman} 1977). Estimate the growth time scale for this instability in a star of mean density $\bar\rho$.
}

\pb\section{Expanding field loop in a constant density fluid}\label{pr.4}
(Exercise in vector calculus)
A circular loop of field lines of radius $R$ lies in a perfectly conducting fluid of constant density. It is centered on the $z-$axis in the $z=0$ plane of cylindrical coordinates ($\varpi,\varphi,z$). The radius expands by an axially symmetric flow in the ${\varpi}$-direction carrying the loop with it. Using ({\ref{brho}}), show that the field strength of the loop increases as $B\sim R$. Find the same answer with an argument based on conservation of mass in the loop.

\pb\section{Magnetic buoyancy}\label{cbuoy}
A horizontal bundle of magnetic field of diameter $d$ and field strength $B$, in temperature equilibrium with its surroundings, rises by magnetic buoyancy in a plasma with density $\rho$ and pressure $p$. Calculate the speed of rise of the tube in the presence of hydrodynamic drag with drag coefficient $c_{\rm d}$ (see wikipedia). Express the speed in terms of an Alfv\'en speed and the pressure scale height $H=p/(\rho g)$. (In this exercise, ignore the effect of a stabilizing density gradient. See the next problem for the effect of a stable stratification).

\pb\section{Speed of buoyant rise}
\label{drift}
(Exercise in astrophysical order-of-magnitude estimates).\\
Consider a magnetic field in a stably stratified star (a main sequence Ap star, say). As in the physics of the Earth's atmosphere, the stability of the stratification is characterized by the difference between the temperature gradient and an {\em adiabatic} gradient. If $p(z)$ is the (hydrostatic) distribution of pressure with depth $z$ (counted positive in the direction of gravity) these gradients are, in astrophysical notation (see \hyperlink{kipp}{Kippenhahn et al.}\ 2012 for details):
\beq 
\nabla = {{\rm d}\ln T\over{\rm d}\ln p},\qquad \nabla_{\rm a}=({{\rm d}\ln T\over{\rm d}\ln p})_{\rm ad},
\eeq
where $\nabla_{\rm a}$ is the dependence on pressure of the {\em potential temperature}\,: the temperature after adiabatic compression/expansion from a fixed reference pressure to the pressure $p$ of the stratification. It is a function of the thermodynamic state of the gas only. In a stable stratification, $\nabla_{\rm a}>\nabla$.  A fluid element displaced from its equilibrium position in the stratification oscillates about it with the {\em buoyancy frequency} $N$, given by
\beq N^2={g\over H}(\nabla_{\rm a}-\nabla),\eeq
where $g$ the acceleration of gravity, and $H=({\rmd}\ln p/{\rm d}z)^{-1}$ the pressure scale height. If it is displaced by a small amount $\delta z$ (maintaining pressure equilibrium with its surroundings), it develops a density difference 
\beq {\delta\rho\over \rho_{\rm e}} = -{\delta z\over H}(\nabla_{\rm a}-\nabla).\label{bdelrho}\eeq
In this stratification imagine a horizontal flux strand of strength $B$ and radius $r$, initially in temperature equilibrium with its surroundings, so it is buoyant ({section \ref{buoy}}). To simplify the algebra assume $B^2/(8\pi p)$ to be small.
\vskip 0.5\baselineskip
{\parindent=-10pt \noindent\par
{\bf a}.\ From sect.\ {\ref{buoy}} calculate the density difference of the strand with its surroundings, and the displacement $\delta z$  after it has settled to a new position of density equilibrium. Calculate 
the temperature difference with its surroundings, assuming the displacement took place adiabatically.}
Let the temperature difference equalize again. The time scale for this, the thermal diffusion time, is $\tau=r^2/\kappa_{\rm T}$, where $\kappa_{\rm T}$ is the thermal diffusivity of the gas and $r$ the radius of the strand. 
{\parindent=-10pt \noindent\par
{\bf b}.\ Estimate the speed of rise of the tube when thermal diffusion is taken into account (explain why this is a good estimate). \noindent\par
{\bf c}.\ Express the rise time ($t_{\rm r}$) over a distance of order of the radius $R$ of the star in terms of its thermal relaxation time, the Kelvin-Helmholtz time $t_{\rm KH}=R^2/\kappa_T$.\noindent\par
{\bf d}.\ A typical value of $t_{\rm KH}\sim R^2/\kappa_{\rm T}$ for this Ap star is $10^6$ yr, internal pressure of the order $10^{17}$ erg/cm$^3$. Compare $t_{\rm r}$ with the main sequence life time of the star ($t_{\rm MS}\sim 10^9$ yr), assuming the largest field strength observed in an Ap star: $\sim 10^5$ G, a length scale $H/r=0.1$ for the field and $H/R\sim 0.1$. Conclusion?}

\pb\section{Pressure in a twisted flux tube}
\label{twistp}
Take a flux tube like in Fig.~\ref{twistub} with an initially uniform vertical field ${\bf B}_0$:
\beq {\bf B}_0=B_0\,\hat{\bf z}, \quad (\varpi<1),\quad {\bf B}_0=0, \quad (\varpi>1),\eeq
where $\varpi$ is the cylindrical radial coordinate.
We twist it by applying a uniform rotation in opposite directions at top and bottom over an angle $\theta$, such that the displacement vetor  $\bs{\xi}$ is:
\beq \bs{\xi}=\theta\,\hat{\bf z}\times{\bs{\varpi}}\quad (z=1), \quad 
 \bs{\xi}=-\theta\,\hat{\bf z}\times{\bs\varpi}\quad (z=-1). 
 \eeq
To prevent the tube from expanding due to the azimuthal field component that develops, we compensate by  changing the internal gas pressure by an amount $\Delta p$ such as to keep the configuration in pressure equilibrium with $B_z$ and $p_{\rm e}$ unchanged. Assume initial transients have been allowed to settle, so the configuration is in a static equilibrium. Calculate $\Delta p(\varpi)$.

\pb\section{Currents in a twisted flux tube}
\label{ctwist}
Starting with a flux tube {as in \ref{twistp}}, apply  a twist over an angle $\theta$ which here depends on the distance from the axis:
\beq 
\theta=\theta_0\quad (\varpi <0.5),\quad \theta= \theta_0(1-\varpi) 
\quad (0.5<\varpi<1), \quad \theta=0 \quad (\varpi>1).
\eeq
Assume $\theta_0\ll1$ so radial expansion can be neglected. Sketch, as functions of $\varpi$\,: the azimuthal field $B_\varphi$, the vertical current $j_z$ and the total current contained within a circle of radius $\varpi$ (compute these quantities if you are not sure about the sketch). Note that the direction of the current changes across the tube while the direction of twist ($B_\varphi/B_z$) does not.

\section{Magnetic stars}
\label{magstar}
 (From the literature on Ap stars.) Consider a star with a static magnetic field (a magnetic Ap star, for example). Outside the star there is vacuum, the field there is a potential field $\bs{\nb}\times{\bf B}=0$. Some force-free field is constructed in the  interior of the star. The normal component $B_{\rm n}$ of this force-free field is evaluated on the surface of the star. This used as boundary condition to construct the field outside the star, using standard potential field theory. What's wrong\footnote{This mistake has been made a number times in the literature.} with such a model for Ap stars; why is it not a counterexample to the vanishing force-free field theorem?

\pb\section{Magnetic compressibility}\label{pr.6}
(From {section \ref{Bcompress}}). {\parindent=-10pt \noindent\par
{\bf a}.\  Show that  under a uniform expansion, a uniform magnetic field stays uniform. \noindent\par
{\bf b}.\ Next consider a uniform, isotropic expansion, such that the positions of fluid particles initially at ${\bf r}_0$ becomes ${\bf r}_1=K{\bf r}_0$. If $n$ is the density of the fluid, show that the magnetic pressure $p_{\rm m}$ varies like $n^{4/3}$, for arbitrary initial field ${\bf B}_0({\bf r}_0)$. (Be warned, however, of the limited practical significance of this fact.)}

\pb\section{Winding-up of field lines in a differentially rotating star} \label{pr.rotstar}
A spherical fluid star of radius $R$ rotates with an angular velocity  $\Omega(r)$ which is constant on spherical surfaces but varies with distance $r$ from the center, such that
\beq \Omega(r)=\Omega_0 (2-r/R).\eeq
At time $t=0$ the star has  in its interior $r<R$ a uniform magnetic field ${\bf B}_0=B_0\,{\bf\hat z}$ along the rotation axis. Outside  the star is a vacuum magnetic field. 
\vskip 0.5\baselineskip
{\parindent=-10pt \noindent\par
{\bf a}.\ Remind yourself  (or verify with \hyperlink{jack}{Jackson} Ch.\ 5) that the field outside the star that matches to a uniform interior field is a point-dipole of strength $\mu=B_0R^3/2$.  \noindent\par
{\bf b}.\ Using spherical coordinates ($\varpi,\theta,\varphi$) with axis taken along the rotation axis, calculate how the magnetic field changes with time. (If you feel like it, also calculate the current in this field.) \noindent\par
{\bf c}.\ Note that the external vacuum field does not change since the normal component of ${\bf B}$  does not change on the surface of the star. (If in {\bf b} you did, why does the current not change the field outside the star?) \noindent\par
}

\pb\section{Diamagnetic forces}
\label{diamag}
An unmagnetized, perfectly conducting sphere is placed in an initially uniform external field  $\bf B_0$ in the $z-$direction.\vskip 0.5\baselineskip
{\parindent=-10pt \noindent\par
{\bf a}.\ Calculate the change in the external field after introduction of the sphere, assuming the field remains current-free (hint\,: compare with {\cpr problem \ref{dipenerg}}). Sketch the field lines.\noindent\par
{\bf b}.\ (`squeezing' by magnetic forces). Calculate the forces acting on the surface of the sphere. If the sphere consists of a fluid of fixed density $\rho$, estimate (by order of magnitude) the time scale for the sphere to change shape under the influence of these forces.\noindent\par
{\bf c}.\ (`melon seed effect') Replace the initially uniform external field of {\bf a}) by a potential field that decreases in strength along the axis $z$ on a characteristic length scale $L$. Show (qualitatively) that the sphere now experiences a net force acting on it (in which direction?) and estimate the acceleration (of the center of mass) under this force.} Inclusions of low field strength in the fluid thus behave like {\em diamagnetic} material. 

\pb\section{Helicity of linked loops}
\label{link}
Using the conservation of magnetic helicity, show that the helicity of the linked loop configuration in Fig.\,\ref{helicfig} is $2\Phi_1\Phi_2$. (hint\,: deform the loops, without cutting through field lines, to give them shapes for which the calculation becomes easy).

\pb\section{Stream function in a plane}
\label{planar}
Analogous to the axially symmetric case, define a stream function for a magnetic field with planar symmetry (e.g.\ in Cartesian coordinates $x,y,z$ with $\pa_y=0$), and show that in this case it is equivalent to a vector potential of the field.

\pb\section{{\kern -4 em}{\small{\dbend}}{\kern+3 em}Convective flux expulsion}
\label{expul}
As a (very simplified) model for the interaction of a (weak) magnetic field with a convective `eddy', consider a  steady rotating flow in plane geometry, acting on an initially uniform field ${\bf B}_0$ with Cartesian components $B_{0x}=0,\,B_{0y}=B_0$. In polar coordinates ($r,\varphi$) the flow has  components $v_r=0,~v_\varphi=r\Omega(r)$. $\Omega$ varies linearly with $r$: $\Omega=\Omega_0(1-r/R)$ ($r<R$), and $\Omega=0$ ($r>R$).  Compute $B_r (r,\varphi,t)$. From this compute the length scale in $r$ on which $B_r$ changes sign, as a function of time. Next, consider the qualitative effect of a small magnetic diffusion term in the induction equation, such that $\eta/R^2\ll\Omega$. Around $t=\eta/(R^2\Omega)$ diffusion starts canceling the neighboring opposite signs; derive the qualitative time dependence of this process.

\pb\section{Torsional Alfv\'en waves} 
\label{cylwave}
{\parindent=-10pt \noindent\par
{\bf a}.\ Write the linearized equations of  motion and induction in cylindrical coordinates ($\varpi,\varphi,z$), with $\bf B$ along the $z$-axis as before. Show that there exist `cylindrical' Alfv\'en waves, with all perturbations vanishing except $\delta B_\varphi$ and $v_\varphi$. These are called torsional Alfv\'en waves.

\noindent\par
{\bf b}.\ In a perfectly conducting fluid of constant density $\rho$ there is at time $t=0$ a uniform magnetic field parallel to the axis of a cylindrical coordinate frame, ${\bf B}(\varpi,\varphi,z,t=0)=B_0\,{\bf\hat z}$. The field is `anchored' at $z=0$ in a plate of radius $R$ in the ($\varpi,\varphi$) plane. At $t=0$ it starts to rotate uniformly around the $z$-axis with rate $\Omega$. 
Assuming this can be treated as a linear perturbation, what is the field line angle $B_\varphi/B_z$ in the torsional Alfv\'en wave launched by this setup? Calculate the torque $T(\varpi)$ acting on the plate, and the work done by the plate against this torque. }

\pb\section{Currents in an Alfv\'en wave}
\label{Awavej}
Sketch the electric current vectors in the torsional Alfv\'en wave of {Figure \ref{torsA}}. 

\pb\section{Magnetic Reynolds numbers in a star}
\label{Rm}
When the dominant non-ideal effect is electrical resistance due to Coulomb interaction of the electrons with the ions and neutrals, an ionized plasma has a magnetic diffusivity of the order  (the `Spitzer value')
\beq \eta\sim 10^{12}\, T^{-3/2} \quad{\rm cm}^2{\rm s}^{-1},\label{Spitz}\eeq
with temperature $T$ in Kelvin. A magnetic `flux tube' of radius $R=10^8$ cm and field strength $10^5$ G lies at the base of the solar convection zone, where the temperature is $2~10^6$ K, and the density $\rho=0.2$ g~cm$^{-3}$.\hfill\break
Calculate the magnetic Reynolds number for a convective flow of $v\sim 100$ m/s across the tube. Calculate the order of magnitude of the Hall drift velocity $v_{\rm H}$ of the electrons in this tube  ({eq.~\ref{hall}}). Which direction does the Hall drift flow? Compare a dimensionless `Hall number' $R_{\rm H}= v/v_{\rm H}$ with $R_{\rm m}$.\hfill\break
Repeat the calculations for {\bf a)} a magnetic filament of width 300 km and field strength 1000 G in the penumbra of a sunspot, where the temperature is 5000 K, the density $10^{-6}$ g/cm$^{-3}$ and velocities are of the order of 2 km/s, {\bf b)} a magnetic loop of strength 100 G and size 10\,000 km in the solar corona where velocities are of the order 10 km/s, temperatures $\sim10^{6}$ K and densities $\sim10^{-14}$ g\,cm$^{-3}$.\hfill\break
Conclusions?\hfill\break
(Use Eq.\ \ref{Spitz} only for estimates where a couple of orders of magnitude don't matter, and when the degree of ionization is substantial. Quantitative values of the magnetic diffusivity require more detailed consideration of the physical conditions in the plasma, e.g.\ \hyperlink{oster}{Oster 1968}).

\pb\section{Poynting flux in an Alfv\'en wave}
\label{PoyA}
Calculate the vertical component of the Poynting flux vector in the torsional wave of {\cpr problem \ref{cylwave}}. What does the other component of $\bf S$ mean to you?

\pb\section{Electric field of a current wire}
\label{curwire}
Take a copper wire, electrically neutral and at rest in the lab frame, with an electric current flowing in it. Eq.\  (\ref{transsigma}) shows that in a frame moving with velocity $v$ along the wire, the wire is electrically charged. Consequently there is an electric field around it. Why is this field not seen in the lab frame? [This question has sparked confused postings on the internet].

\pb\section{Ambipolar drift}
\label{ambs}
A straight flux tube of circular cross section $R$ is embedded in a field-free environment. Its field strength varies with distance $r$ from its axis as ${\bf B}(r)={\bf B}_0 [1-(r/R)^2]$. Calculate the current density in the tube. 
{\parindent=-10pt \noindent\par
{\bf a}.\ 
Show that the ambipolar drift velocity is irrotational. Which direction do the ions flow? 

\noindent
In pressure equilibrium,  the resulting pile-up leads to a pressure gradient between the charged component and the neutrals.  In a stationary equilibrium, diffusion in this gradient just cancels the ambipolar drift.
{\parindent=-10pt \noindent\par
{\bf b}.\  The friction coefficient $\gamma$ in eq.~({\ref{ambi}}}) is about $3\, 10^{13}$ cm$^3$ s$^{-1}$ g$^{-1}$ for an astrophysical mixture. Calculate the ambipolar drift velocity for the conditions of problem {\cpr\ref{Rm}}. Assume for this that the degree of ionization $\rho_{\rm i}/\rho$ is about 1 at the base of the solar convection zone, $\sim 10^{-3}$ in the sunspot atmosphere, and $\sim1$ in the corona.  (What is the significance of ambipolar drift in a nearly fully ionized gas?)
 }
 
 \pb\section{Conformal mapping of a potential field}
\label{edge}
Simplify an accretion disk as a thin perfectly conducting surface with a circular hole in it. We add a bundle of magnetic field lines passing through the hole, and ask ourselves what configuration this field will have if it is current-free except at the surface of the plate, where the field vanishes. As described in \ref{curvb}, tension in the field lines wrapping around the edge of the hole cause the lines to pile up there. We want to know how the field strength varies close to the edge. Sufficiently close to the edge, the field can be approximated in a  planar ($x,y$) geometry, with  $x$ along the surface of the plate at $x<0$, $y=0$. Let lines $x=$\,cst.\ be field lines of a homogeneous potential field, and $z=x+iy$ a complex variable in the plane $(x,y)$. With the theory of conformal mapping {\url{http://mathworld.wolfram.com/ConformalMapping.html}} show that the function $w=z^2$ transforms the field lines of the homogeneous field in the half-plane $x>0$ into field lines of a potential field wrapping around the plate. From this, show that the field strength increases towards the edge as $B\sim x^{-1/2}$. (Hint: the field strength is inversely proportional to the separation between field lines).

\chapter{Appendix}\label{appen}
\parindent=0pt

\section{Vector identities}
\label{idents}

\beq 
{\bf a\cdot}({\bf b\times c})={\bf c\cdot}({\bf a\times b})={\bf b\cdot}({\bf c\times a})\label{dotcross}
\eeq
\beq
{\bf a\times}({\bf b\times c})=({\bf a\cdot c}){\bf b}-({\bf a\cdot b}){\bf c}
\eeq 
\beq \bs{\nb}\times\bs{\nb}\psi=0 \eeq
\beq {\bs{\nb}\cdot}({\bs{\nb}\bf\times a})=0\eeq
\beq \bs{\nb}\times(\bs{\nb}\bf\times a)={\bs{\nb}}({\bs{\nb}\cdot a})-\nb^2{\bf a}\eeq
\beq 
{\bs{\nb}\cdot}(\psi{\bf a})=\psi{\bs{\nb}\bf\cdot a}+{\bf a\cdot\bs{\nb}}\psi
\eeq
\beq
{\bs{\nb}\times}(\psi{\bf a})=\bs{\nb}\psi{\bf\times a}+\psi{\bs{\nb}\bf\times a}
\eeq
\beq
{\bs{\nb}\cdot}({\bf a\times b})={\bf b\cdot}({\bs{\nb}\bf\times a})-{\bf a\cdot}({\bs{\nb}\bf\times b})
\eeq
\beq 
{\bs{\nb}\times}({\bf a\times b})={\bf a}({\bs{\nb}\bf\cdot b})-{\bf b}({\bs{\nb}\bf\cdot a})+({\bf b\cdot\bs{\nb}}){\bf a}-({\bf a\cdot\bs{\nb}}){\bf b}
\eeq

When memory fails these identities and others are rederived quickly using the properties of the permutation symbol or completely antisymmetric unit tensor\,:  
\beq \epsilon_{ijk},\eeq
where the indices each stand for one of the coordinates $x_1,x_2,x_3$. Conventionally, in Cartesian  coordinates, $x_1=x$, $x_2=y$, $x_3=z$. It has the properties
\beq 
\epsilon_{123}=1,\qquad \epsilon_{jik}=-\epsilon_{ijk},\qquad \epsilon_{jki}=\epsilon_{ijk},\qquad\epsilon_{ijk}=0\quad {\rm if~i=j.}\eeq
It remains unchanged by a circular permutation of the indices. A product of two symbols with one common index has the property
\beq \epsilon_{ijk}\epsilon_{imn}=\delta_{jm}\delta_{kn}-\delta_{jn}\delta_{km},\eeq
where $\delta_{ij}$ is the Kronecker delta, and the sum convention has been used, implying summation over a repeated index:
\beq a_ib_i=\sum_{i=1}^3a_ib_i={\bf a\cdot b}.\eeq
The cross product and the curl operator can be written as
\beq ({\bf a\times b})_i=\epsilon_{ijk}a_jb_k,\qquad ({\bs{\nb}\times{\bf b}})_i=\epsilon_{ijk}{\pa b_k\over\pa x_j}.\eeq

\section{Vector operators in cylindrical and spherical coordinates}\label{cylco}
\begin{sloppypar}
Seen also \url{https://en.wikipedia.org/wiki/Del_in_cylindrical_and_spherical_coordinates} or  \hyperlink{trit}{Tritton} (1992).
\end{sloppypar}

\medskip 
Let ${\bf a,b}$ be vectors and $f$ a scalar quantity.

\medskip
{\leftline {\bf Cylindrical coordinates ($\varpi,\varphi,z$)\,:}}
\beq 
\nabla f={\pa f\over\pa \vp}\bs{\hat \vp}+{1\over r}{\pa f\over\pa \varphi}\bs{\hat\varphi}+{\pa f\over\pa z}{\bf\hat z} \eeq
\beq
\nabla^2 f={1\over \vp}{\pa\over\pa \vp}(\vp{\pa f\over \pa \vp})+{1\over \vp^2}{\pa^2 f\over\pa\vh^2}+
{\pa^2 f\over\pa z^2} \label{cylv}
\eeq
\beq 
\nb\kern -2pt\cdot\kern -2pt{\bf a}={1\over \vp}{\pa\over\pa \vp}(ra_\vp)+
{1\over \vp}{\pa a_\vh\over\pa\vh}+{\pa a_z\over\pa z}
\eeq
\beq (\nb\times {\bf a})_\vp={1\over \vp}{\pa a_z\over \pa \vh}-{\pa a_\vh\over \pa z}\eeq
\beq (\nb\times {\bf a})_\varphi={\pa a_\vp\over \pa z}-{\pa a_z\over \pa \vp}\eeq
\beq (\nb\times {\bf a})_z={1\over \vp}{\pa\over \pa \vp}(ra_\vh)-{1\over \vp}{\pa a_r\over \pa \vh}\eeq
\beq (\nabla^2 a)_\vp=\nabla^2a_\vp-{a_\vp\over \vp^2}-{2\over \vp^2}{\pa a_\vh\over \pa_\vh}\eeq	
\beq 
(\nabla^2 a)_\vh=\nabla^2a_\vh-{a_\vh\over \vp^2}+{2\over \vp^2}{\pa a_\vp\over \pa_\vh}
\eeq
\beq(\nabla^2 a)_z=\nabla^2a_z\eeq
where $\nabla^2 a_i$ denotes the result of the operator $\nabla^2$ acting on $a_i$ regarded as a scalar. 
\beq
({\bf a}\kern -2pt\cdot\kern -2pt\nb\,{\bf b})_\vp=({\bf a}\kern -2pt\cdot\kern -2pt\nb\, b_\vp -b_\vh a_\vh/\vp)\,\bs{\hat \vp}
\eeq
\beq
({\bf a}\kern -2pt\cdot\kern -2pt\nb\,{\bf b})_\vh=({\bf a}\kern -2pt\cdot\kern -2pt\nb\,b_\vh -b_\vh a_\vp/\vp)\,\bs{\hat\vh}
\eeq
\beq ({\bf a}\kern -2pt\cdot\kern -2pt\nb\,{\bf b})_z={\bf a}\kern -2pt\cdot\kern -2pt\nb\,b_z\,{\bf \hat z} \label{adgb} \eeq

\medskip
{\leftline {\bf Spherical coordinates ($r,\theta,\varphi$)\,:}}
\beq
\nabla f={\pa f\over\pa r}{\bf \hat r} +{1\over r}{\pa f\over\pa \theta}\bs{\hat\theta}+{1\over r\sin\theta}{\pa f\over\pa \varphi }\bs{\hat\varphi}
\eeq
\beq
\nabla^2 f={1\over r^2}{\pa \over\pa r}(r^2{\pa f \over\pa r})+ 
{1\over r^2\sin\theta}{\pa f\over\pa\theta}(\sin\theta{\pa f\over\pa\theta})+
{1\over r^2\sin^2\theta}{\pa^2 f\over\pa\varphi^2}
\eeq
\beq
\nb\kern -2pt\cdot\kern -2pt{\bf a}={1\over r^2}{\pa\over\pa r}(r^2a_r)+{1\over r\sin\theta}{\pa\over\pa\theta}(a_\theta\sin\theta)+{1\over r\sin\theta}{\pa a_\varphi\over\pa\varphi}
\eeq
\beq
(\nb\times {\bf a})_r={1\over r\sin\theta}{\pa\over\pa\theta}(a_\varphi\sin\theta)-{\pa a_\theta\over\pa\varphi}
\eeq
\beq
(\nb\times {\bf a})_\theta={1\over r\sin\theta}{\pa a_r\over\pa\varphi}-{1\over r}{\pa\over\pa r}(ra_\varphi)
\eeq
\beq
(\nb\times {\bf a})_\varphi={1\over r}[{\pa\over\pa r}(ra_\theta)-{\pa a_r\over\pa\theta}]
\eeq
\beq
(\nabla^2 {\bf a})_r=\nabla^2 a_r-{2\over r^2}[a_r+{1\over\sin\theta}{\pa\over\pa\theta}(a_\theta\sin\theta)+{1\over\sin\theta}{\pa a_\varphi\over\pa\varphi}]
\eeq
\beq
(\nabla^2 {\bf a})_\theta=\nabla^2 a_\theta+{2\over r^2}[{\pa a_r\over\pa\theta}-{a_\theta\over 2\sin^2\theta}-{\cos\theta\over\sin^2\theta}{\pa a_\varphi\over\pa\varphi}]
\eeq
\beq
(\nabla^2 {\bf a})_\varphi=\nabla^2 a_\varphi+{2\over r^2\sin\theta}[{\pa a_r\over\pa\varphi}+\cot\theta{\pa a_\theta\over\pa\varphi}-{a_\varphi\over 2\sin\theta}]
\eeq
\beq
({\bf a}\kern -2pt\cdot\kern -2pt\nb\,{\bf b})_r=({\bf a}\kern -2pt\cdot\kern -2pt\nb\, b_r-{a_{\theta}b_\theta/r+a_\varphi b_\varphi}/r)\,\bs{\hat r} \label{abspherr}
\eeq
\beq
({\bf a}\kern -2pt\cdot\kern -2pt\nb\,{\bf b})_\theta=({\bf a}\kern -2pt\cdot\kern -2pt\nb\, b_\theta +a_\theta b_r/r-a_\varphi b_\varphi\cot\theta/r)\,\bs{\hat\theta} \label{absphert}
\eeq
\beq
({\bf a}\kern -2pt\cdot\kern -2pt\nb\,{\bf b})_\varphi=({\bf a}\kern -2pt\cdot\kern -2pt\nb\, b_\varphi+a_\varphi b_r/r+a_\varphi b_\theta\cot\theta/r)\,\bs{\hat\varphi} \label{abspherp}
\eeq


\section{Useful numbers in cgs units}\label{cgs}
Abbreviated here to 4 digits. (For recommended accurate values see \hfill\break
{\url{http://physics.nist.gov/cuu/Constants}})
\vspace{1\baselineskip}

\begin{tabular}{lll}
speed of light &$c=2.998~10^{10}$ &cm s$^{-1}$\cr
elementary charge &$e=4.803 ~10^{-10}$  &(erg cm)$^{1/2}=$ g$^{1/2}$  cm$^{3/2}$ s$^{-1}$\cr
gravitational constant & $G=6.674~10^{-8}$ &cm$^3$ s$^{-2}$ g$^{-1}$ \cr
proton mass &$ m_{\rm p} =1.673\,10^{-24}$ &g\cr
electron mass &$m_{\rm e}=0.911~10^{-27}$ &g\cr
Boltzmann constant &$k=1.381~10^{-16}$ &erg K$^{-1}$\cr
gas constant &${\cal R}~ (\approx k/m_{\rm p})=8.314~ 10^7$ &erg K$^{-1}$g$^{-1}$\cr
Stefan-Boltzmann constant & $\sigma=ac/4=5.670~10^{-5}$ &erg s$^{-1}$ cm$^{-2}$\,K$^{-4}$\cr
electron volt  &1 eV$=1.602~10^{-16}$ & erg \cr
&&\cr
astronomical unit & 1 AU $=1.496~10^{13}$ &cm\cr
solar mass & $M_\odot = 1.989~10^{33}$ &g\cr
solar radius & $R_\odot = 6.963~10^{10}$ &cm\cr
& 1 year $\approx\pi\cdot10^7$ &s\cr
\end{tabular}
\section{MKSA and Gaussian units}\label{MKSA}
The choice of Gaussian units, where $\bf E$ and $\bf B$ have the same dimensions, is informed by the knowledge that they are components of a single quantity, the electromagnetic tensor. It is a natural choice in astrophysics, where quantities measured in earth-based engineering units are less likely to be needed. 

The MHD  induction equation is the same in MKSA and in Gaussian quantities. In the equation of motion the only change is a different coefficient in the Lorentz force. In MKSA\,:

\begin{tabular}{ll}
&\cr
Lorentz force\,: & $F_{\rm L}=({\bs{\nb}\bf\times B}){\bf\times B}/\mu_0$\cr
Magnetic energy density\,: & $e_{\rm m}=B^2/(2\mu_0)$\cr
&\cr
\end{tabular}

\noindent
 The MKSA unit for the magnetic induction $B$, the Tesla, is worth $10^4$ Gauss. The MKSA forms of $\bf E$ and $\bf j$ in MHD:

\begin{tabular}{ll}
&\cr
Electric field strength\,: & ${\bf E}=-({\bf v\times B})$\cr
current density\,: &${\bf j}=(\nb{\bf \times B})/\mu_0$\cr
&\cr
\end{tabular}

\noindent
In Gaussian units the difference between magnetic field strength and induction, and between electric field and displacement, are dimensionless factors given by the relative permeability $\mu_{\rm r}$ and relative permittivity $\epsilon_{\rm r}$ of the medium.  In vacuum both are equal to unity, as assumed here. They have to be taken into account in some high-density environments, however.

\vspace{5\baselineskip}\noindent

\leftline{\large\bf Credits}
\vspace{1\baselineskip}\noindent
Most of the figures and all the video clips were made by Merel van 't Hoff. It is a pleasure to thank Eugene Parker for many discussions on magnetohydrodynamics over the past 4 decades, and to thank the colleagues whose comments and corrections have helped to get the text into its present form. Among them Andreas Reisenegger ({sect.\ \ref{hall-am}}), Rainer Moll (sects.\ {\ref{ecur}, \ref{magpres}, Fig.~{\ref{molljet}}), and Irina Thaler ({problem \ref{curdif}})
).

\pagebreak

\chapter{\bf Problem solutions}
\vspace{1\baselineskip}

In the following partial derivatives of a quantity $q$ with respect to a coordinate $x$ are abbreviated as $\partial _x q$. Derivatives with respect to time are written in dot  notation ($\dot q$) where convenient.

\bigskip
{\bf 3.1 Currents from flows}\\
The velocity and the initial field as specified depend only on $z$,  the problem is linear,  and $v_y=v_z=0$. The derivatives  $\partial_x=\partial_y$ then vanish for  all quantities at all $t$.  Using this, the induction equation (\ref{ind}) written in components yields
\beq \partial_t \, B_z=\partial_x({\bf v\times B})_y-\partial_y({\bf v\times B})_x=0, \eeq
so $B_z(t)=B_0$. Using this,
\beq \partial_t B_x=-\partial_z({\bf v\times B})_y=B_0\partial_z v_x. \label{pr1}\eeq
and 
$B_y(t)=0$. With $v_x=v={\rm tanh}\, z$, this yields
\beq B_x=t\,B_0(1-{\rm tanh}^2 z)\eeq
(sketch this as a function of $z$).   In the induction equation, a change of sign in the velocity is equivalent  to a change of direction of time. Flipping the direction of the flow at time $t$ brings the configuration back to its initial state after another interval $t$.

\bigskip 
{\bf 3.2. Particle orbits}\\
The equation of motion of the particle of mass $m$ and charge $q$ is
\beq m\dot{\bf v}=q{\bf E}+q{\bf v\times B}/c.\label{pr2s}\eeq
Let the particle be initially at rest at the origin of Cartesian coordinates ($x,y,z$) with unit vectors ${\bf\hat x,\hat y,\hat z}$. Take the uniform electric field in the $z$-direction, ${\bf E}=E\hat{\bf z}$, the uniform  magnetic field in the $y$-direction, ${\bf B}=B\hat{\bf y}$. In components, eq.\ (\ref{pr2s}) is:
\beq m\dot v_x= -qBv_z /c, \qquad m\dot v_y=0,\qquad m\dot v_z=qBv_x/c+qE.\label{compo}\eeq
Since $v_y=0$ at $t=0$, the particle's path is in the $x-z$-plane.  The general solution to eq.\ (\ref{compo}) is easily found but somewhat cumbersome. Instead, write
\beq v_x=v_{\rm r}-c\,E/B,\label{shift}\eeq
 so the eqs reduce to
\bea 
 m\dot v_{\rm r}&= &-qBv_z /c \cr
 m\dot v_z&= &~~qBv_{\rm r}/c.
\eea
In complex notation the solution is
\beq v_{\rm r}=Ae^{i\omega t} ,\qquad v_z=Aie^{i\omega t},\eeq
were $ \omega={qB\over mc}$.
Taking the real part, and using (\ref{shift}):
\beq v_z=A\sin\omega t,\qquad v_x=A\cos\omega t- cE/B.\eeq
The condition ${\bf v}(0)=0$ yields the amplitude of the orbit: $A=cE/B$. The particle's path $\bs\xi(t)$ is related to the velocity by $\dot {\bf \xi}={\bf v}$:
\beq \xi_z=-{A\over\omega}\cos\omega t+k_z,\qquad \xi_x={A\over\omega}\sin\omega t- A\, t +k_x,\eeq
where $k_x,k_z$ are integration constants to be fixed by the initial conditions. With ${\bs\xi}(0)=0$,
\beq \omega\xi_z=A(1-\cos\omega t),\qquad \omega\xi_x=A(\sin\omega t- \omega t).\eeq
A path of this shape is called a {\it cycloid}.  For a movie see \url{https://en.wikipedia.org/wiki/Cycloid}.

\bigskip 
{\bf 3.3. Displacement current at finite conductivity}\\
If $V$ and $L$ are the characteristic velocity and length scales of the problem under study, the characteristic time scale is $L/V$. With the diffusivity $\eta\sim vl$ resulting from collisions with mean free path $l$ and velocity $v$, and with $\eta=c^2/(4\pi\sigma_{\rm c})$, the second term in \ref{maxohm} is of order $l/L~vV/c^2$ compared with unity. The MHD assumption requires the process to be small scale: $l\ll L$. The second term is then small since $v,V<c$. Note however that diffusion equations are not relativistically correct; the use of an Ohmic diffusion term also requires $v,V\ll c$ (cf. sect. 1.10).

\bigskip 
{\bf 3.4. Alternative form of the induction equation}\\
Write the left hand side of (\ref{brho}) out as
\beq 
{\rd \over \rd t}{{\bf B}\over\rho}={1\over\rho}{\rd {\bf B}\over \rd t}-{{\bf B}\over\rho^2}{\rd\rho\over\rd t}.
\eeq
 Wal\'en's equation follows from this using eq.\ (\ref{ind1}) (with the second term on the right taken to the left)  and the continuity equation in the form (\ref{ctya}).

\bigskip 
{\bf 3.5. Integrated induction equation}\\
Set $\bs \xi=0$ at $t=0$. To first order, ${\bs\xi}\approx {\bf v}\, \delta t$.

\bigskip
{\bf 3.6. Stretching of a thin flux tube}\\
The mass of the stretching segment, $M=AL\rho_{\rm i}$, and the magnetic flux it carries, $\Phi=BA$, where $A(t)$ is the  cross section of the tube, are constant in time. With the isothermal equation of state assumed, the pressure balance condition yields
\beq {M{\cal R}T\over a\Phi}={t\over B}(p_{\rm e}-{B^2\over 8\pi}).\label{stret}\eeq
Introduce a dimensionless field strength $b=B/B_0$ and the dimensionless constant $K={\cal R}TB_0/(a\Phi p_{\rm e})$. Eq.\  (\ref{stret}) then can be written as  $bK=t(1-{b^2/\beta_0})$, where $\beta_0=8\pi p_{\rm e}/B_0^2$ (note that this is only a notional object, as it mixes internal and external physical quantities). Solving for $b$ while introducing the dimensionless time $\tau=t/(K\beta_0^{1/2})$ :
\beq b(\tau)={\beta_0^{1/2}\over 2\tau}[-1+(1+4\tau^2)^{1/2}].\eeq
Initially  ($\tau\ll 1$) the field strength increases linearly.  After $\tau\approx 1$ it saturates to $B\rightarrow(8\pi p_{\rm e})^{1/2}$.  The plasma-beta inside the tube, $\beta_{\rm i}=8\pi p_{\rm i}$, has the asymptotic dependence $\beta_{\rm i}= 1/(2\tau)$ ($\tau\rightarrow\infty$).

\noindent  Continued stretching causes the internal gas pressure to eventually become negligible compared with the magnetic pressure, and the external pressure is then balanced entirely by the magnetic pressure of the tube.

\bigskip 
{\bf 3.7. Magnetic flux}\\
Let $\Phi=\int_S{\bf B}\cdot\rd{\bf S}$ be the magnetic flux though a surface $S$ bounded by a closed path ${\bf s}$. If $\bf A$ is a vector potential of ${\bf B}$, Stokes' theorem says
\beq \int_S\nb\bf{\times A}\cdot\rd {\bf S}=\Phi= \int_s {\bf A}\cdot \rd {\bf s}.\eeq
The second equality shows that $\Phi$ is the same for all surfaces with the same boundary path.
[In a more intuitive way: consider a bundle of field lines passing through 2 surfaces with the same boundary; visualize the number of field lines passing through each. How does ${\rm div}\,{\bf B}=0$ enter in this view?]

\bigskip 
{\bf 3.8. Magnetic forces in a monopole field}\\
{\bf a}. Write $(\nb{\bf \times B)\times B}$  out using the vector identities in Ch.4.\\
{\bf b}. The force exerted by the `magnetic pressure' gradient is $-\nb B^2/8\pi= 1/(2\pi r^5)\,\hat {\bf r}$. Since the Lorentz force vanishes, the curvature force must be the opposite: ${\bf B\cdot}\nb {\bf B}/4\pi=-1/(2\pi r^5)\,\hat {\bf r}$. [Exercise in vector calculus: verify this by calculating the curvature force explicitly (eqs. \ref{abspherr}-\ref{abspherp} will help)].

\bigskip 
{\bf 3.9. Magnetic forces in an azimuthal magnetic field}\\
With ${\bf B}$ varying as $1/\varpi$, the magnetic pressure force is $-\nabla B^2/8\pi=2B^2/(8\pi \varpi)\,\bs{\hat\varpi}$. 
To keep the equations easier to read, write ${\bf B}=B_0\varpi_0 {\bf b}$, so ${\bf b}=\bs{\hat\varphi}/\varpi$, and write the current as ${\bf j}={c\over 4\pi}\,{\bf h}$.  The current has only a $z$-component, 
\beq 
h_z=(\nb{\bf \times b})_z={1\over\varpi}\partial_\varpi(\varpi {\bf b}_\varphi)={1\over\varpi}\partial_\varpi\bs{\hat\varphi}=0,
\eeq
where the last equality holds because the unit vector $\bs{\hat\varphi}$ varies only with azimuth $\varphi$,  not with $\varpi$. The current vanishes, except for a singularity (`$0/0$') on the axis. To find the amplitude of this singular current, integrate $h$ over a circular area $S$ centered on the axis. With Stokes' theorem:
\beq \int_S h=\int_S {\nb\times b}\,\rd S=\oint_{\partial S} b\cdot \rd {\bf l},\eeq
where $\rd {\bf l}=\bs{\hat\varphi}\, \rd l$, and $l$ is the path length around the boundary of the circle. This yields $\int h=2\pi$, independent of the radius of the circle. Reinstating the amplitude factors of current and field:
\beq J\equiv \int j= c{B_0\over 2}\varpi_0.\eeq
Since, apart from the singularity on the axis,  the current vanishes, so does the Lorentz force. Like in problem 3.8, the `hoop stress' is the opposite of the pressure force.

\bigskip 
{\bf 3.10. The surface force at a change in direction of B}\\
Imagine a pizza box of unit surface area and infinitesimal height $2\epsilon$, with top surface ($^+$) at $z_0+\epsilon$, and bottom surface ($^-$) at $-\epsilon$. The surface force (a force per unit area) is the volume integral of the Lorentz force in the box. Using its representation in terms of the magnetic stress tensor (\ref{M}), it is equal to the sum of the surface stress vectors at top and bottom, taking outward normals $\bf n$, so ${\bf n}^-=-{\bf n}^+$.  Take the direction of the inclination angle $\alpha$ along the $+x$ axis. The surface force  ${\bf F}_{\rm s}$ exerted by the field then has components $F_{{\rm s}z}=0$, $F_{{\rm s}x}= \sin\alpha\, B_0^2/4\pi$.

\bigskip 
{\bf 3.11. Magnetic energy and stress}\\
{\bf a}. Let the field inside the radius $R$ of the sphere be oriented along the $z$-axis (in cylindrical coordinates), ${\bf B}=B_0\hat{\bf z}$.  The potential $\Phi$ of the field outside the sphere, in spherical coordinates ($r,\theta,\varphi$) is then (cf.\ Jackson):
\beq \Phi=-{1\over 2}B_0R^3\cos\theta/r^2, \qquad (r>R)\eeq
and the components of the field ${\bf B}=-\nb\Phi$ are
\beq B_r=B_0 ({R\over r})^3\cos\theta,\qquad B_\theta={1\over 2}B_0 ({R\over r})^3\sin\theta.\eeq
You can verify that at the surface $B_r$ matches the radial component of the internal field.\\
{\bf b}. The $\theta$-component of the field jumps across the surface. Similar to problem 3.10, the surface force ${\bf F}_{\rm s}$ exerted by the field is the difference between the magnetic tension vectors across the surface. This yields  ${\bf F}_{\rm s}={3\over 8\pi} B_0^2 \sin\theta\,\bs{\hat\theta}$.\\
{\bf c}. The internal magnetic energy, $E_{\rm int}=B_0^2R^3/6$ is the volume integral of $B_0^2/8\pi$. Since spherical coordinates are orthonormal, the magnetic energy density is $(B_r^2+B_\theta^2)/8\pi$. Integrating this over the external volume yields $E_{\rm ext}={5\over 3}E_{\rm int}$.\\
{\bf d}. A magnetic instability converts magnetic into kinetic energy. Equating the external magnetic energy to kinetic energy ${1\over 2}M v^2$ where $M=(4\pi/3) R^3\bar\rho$ is the mass of the sphere and $v$ a typical (rms) velocity, this yields  a velocity of the order of the mean Alfv\'en speed $v_{\rm A}=B_0/(4\pi\bar\rho)^{1/2}$. The typical instability time scale, $t_{\rm inst}\sim R/v$, is thus  of order $R/v_{\rm A}$. 

\noindent
[This is a crude `astrophysical style' estimate that does not take into account complications like the variation of density in the star or the fraction of external magnetic energy that is converted. The idea is to get the number in the right range on a logarithmic scale, for comparison with other processes that live elsewhere on this scale].

\bigskip 
{\bf 3.12 Expanding field loop in a constant density fluid}\\
With $\rho=$ cst.,\ Wal\'en's equation reduces to $\rd{\bf B}/\rd t=({\bf B\cdot}\nb){\bf v}$. In cylindrical coordinates ($\varpi,\varphi,z$), ${\bf B}=B\bs{\hat\varphi}$, ${\bf v}=v\bs{\hat \varpi}$; this yields
\beq B{\rd\bs{\hat\varphi}\over\rd t}+\bs{\hat\varphi}{\rd B\over d t}={B\over \varpi}\partial_\varphi(v\bs{\hat\varpi}).\label{expa}\eeq
Since $\bs{\hat\varphi}$ does not change in the (radial) direction of the flow, the first term on the left vanishes. By inspection, $\partial_\varphi\bs{ \hat\varpi}=\bs{\hat\varphi}$, so that (\ref{expa}) reduces to $ \rd B/ \rd t=B v/\varpi$.
With $v={\rd R/\rd t}$, and $\varpi=R$ at the location of the loop, this integrates to $B\sim R$. Intuitively: since the flow has been specified as in the $\varpi$-direction, changes in cross section take place in this direction  only. Mass conservation then implies $B\sim R$. 

\bigskip 
{\bf 3.13. Magnetic buoyancy}\\
Pressure balance in temperature equilibrium gives ${\cal R}T(\rho_{\rm e}-\rho_{\rm i})=B^2/8\pi$. The buoyancy force per unit volume of the flux bundle  is $F_{\rm b}=(\rho_{\rm e}-\rho_{\rm i})g=-B^2/(8\pi H)$, where $H={\cal R}T/g$ is the pressure scale height. Per unit length of a bundle of diameter $d$, the buoyancy force is $F_{\rm b}d=\pi (d/2)^2F_{\rm g}=-B^2d^2/(32 H)$. Balancing this with the drag force per unit length $F_{\rm d}=c_{\rm d}\,\rho_{\rm e}v^2\,d$ yields an estimate of the velocity of rise, $v\approx \tilde v_{\rm A}(d/ H)^{1/2}$,
where $\tilde v_{\rm A}=B/(4\pi\rho_{\rm e})^{1/2}$. If the internal and external densities do not differ much (high $\beta$ conditions), this is approximately the Alfv\'en speed inside the tube.

\bigskip 
{\bf  3.14. Speed of buoyant rise}\\
{\bf a}. The fluid element is in buoyant equilibrium when the displacement $\delta z$ is such that the sum of (\ref{bdelrho}) and ({\ref{delrb}}) vanishes. This yields 
\beq {\delta z\over H}=-{1\over 2}{v_{\rm A}^2\over {\cal R} T} {1\over\nabla_{\rm a}-\nabla}.\eeq
Assume that any buoyancy oscillations around the equilibrium are damped out. The temperature difference with the surroundings is $\delta T/T=\delta z/H~(\nabla_{\rm a}-\nabla)$. \\
{\bf b}. The strand rises at the velocity where the rate of temperature change due to displacement matches that due to thermal diffusion:
\beq v=\delta z/\tau=-{1\over 2}{\kappa_{\rm T}\over r^2}H{v_{\rm A}^2\over c_{\rm i}^2}{1\over\nabla{\rm a}-\nabla}.\eeq
{\bf c}. The rise time $\tau_{\rm r}$ over a distance R is
\beq \tau_{\rm r}/\tau_{\rm KH}={r^2\over RH}{8\pi p\over B^2}(\nabla_{\rm a}-\nabla).\eeq
{\bf d}. About $10^{12}$ yr. Under the buoyancy force alone, the fields of Ap stars are unlikely to rise significantly through the star over the age of the universe. Evolution of the field configuration by magnetic diffusion is more important.  

\bigskip 
{\bf  3.15. Pressure in a twisted flux tube.}\\
In order for the configuration to be in equilibrium, and with $B_z$ assumed to be unchanged, the displacement field must vary linearly with $z$ (after Alfv\'en waves  in the assumed settling process have disappeared). The induction equation then yields $\bs{\xi}=\theta z \varpi\bs{\hat\varphi}$ ($-1<z<1$), $B_r=0$, $B_\varphi=\theta\varpi B_z$, and $\Delta p=-(\theta\varpi B_z)^2/8\pi$.

\bigskip 
{\bf 3.16. Currents in a twisted flux tube.}\\
As in (\ref{twistp}), $B_\varphi=\theta\varpi B_z$ ($\varpi<1/2$), and $B_\varphi=\theta(1-\varpi)B_z$ ($1/2<\varpi<1$). The current density is given by  $4\pi j_z/(\theta B_z)=2$ ($\varpi<1/2$), and $1/\varpi-2$  ($1/2<\varpi<1$). The contributions  to the total current of the areas inside and outside of $\varpi=1/2$ are of opposite sign and equal in magnitude.

\bigskip 
{\bf 3.17. Magnetic stars.}\\
In such a construction the normal component to the surface of the star is continuous as required by ${\rm div}\,{\bf B}=0$, but the tangential component is not. As in problem (\ref{dipenerg}), the field configuration implies the presence of a force that keeps the field lines bent against the magnetic tension. In an Ap star, this force is not supplied at the surface (where the density vanishes) but  throughout the interior. This also applies to neutron stars.

\bigskip 
{\bf 3.18. Magnetic compressibility}.\\
{\bf a}. Use eq. \ref{ind1}. Since the expansion is uniform, ${\rm div}{\bf v}$ is a constant, and the first term on the right is proportional to $\bf B$. The second term vanishes since ${\bf B}$ is independent of position. The gradient tensor $\nb{\bf v}$  is also independent of position. The third term is uniform, so $\partial_t{\bf B}$ is uniform. Unless the expansion is also isotropic, however, the  direction of $\bf B$ will change on expansion.\\
{\bf b}. The velocity field in homogeneous isotropic expansion is ${\bf v}=a{\bf r}$, where $\bf r$ is the position vector from the origin of the coordinate system, and $a$ a constant. Eq. (\ref{ind1}) yields $\partial_t{\bf B}=-2a{\bf B}$, reflecting the fact that only flows in the two directions perpendicular to ${\bf B}$ enter in the induction equation. The mass density of the expanding, perfectly conducting fluid decreases with time $t$ as  $(a t)^{-3}$, while the strength of a field embedded in it decreases as $(at)^{-2}$. The magnetic pressure therefore behaves like the pressure in a gas with ratio of specific heats $\gamma=4/3$, $p\sim\rho^{-4/3}$.

\bigskip 
{\bf 3.19. Winding-up of field lines in a differentially rotating star}.\\
{\bf b}. Writing the induction equation in spherical coordinates, yields $\partial_t B_r=\partial_t B_\theta=0$, and $\partial_t B_\varphi=B_0\Omega_0\,(6-4r)\cos\theta$.\\
{\bf c}. The external field does not change because the normal component at the surface is constant. Thinking of the internal currents as the source of the magnetic field in the exterior leads astray.

\bigskip 
{\bf 3.20. Diamagnetic forces}.\\
{\bf a}. At the surface of the sphere, the uniform field has components $B_{r0}=B_0 \cos\theta$, $B_{\theta 0}=B_0\sin\theta$. From 3.11, the dipole field ${\bf B}_{\rm d}$ canceling the radial component of the uniform field is
\beq B_{{\rm d}r}=-B_0\,\cos\theta,\qquad B_{{\rm d}\theta}=-{1\over 2}B_0\,\sin\theta.\eeq
Adding the two gives $B_r=0$, $B_\theta=1/2\,B_0\sin\theta$.\\
{\bf b}. The magnetic pressure acting on the sphere's surface is $1/4\,B_0^2\sin^2\theta$. The pressure squeezes the sphere, elongating it in the direction of ${\bf B}_0$. If the radius of the sphere is $R$, the time scale for this is of the order $R/\tilde v_{\rm A}$, where $\tilde v_{\rm A}\sim 1/2~ B_0/(4\pi\rho)^{1/2}$. [The tilde indicates that $\tilde v_{\rm A}$ is not the Alfv\'en speed at any point, since the interior field vanishes, as well as the external density.]
{\bf c}. The magnetic pressure acting on the upper hemisphere is now lower than on the lower hemisphere, resulting in a net upward force.

\bigskip 
{\bf 3.21. Helicity of linked loops}.\\
(See fig.\ \ref{helicfig}). Under helicity conservation, the two loops can be deformed arbitrarily without changing the helicity, as long as they do not cross. Deform loop 1 into a axisymmetric ring (if that's not what it is already). The vector potential of this ring consists of closed loops around the ring's path. Choose one such loops, with path $C$. Deform loop 2 by squeezing it to infinitesimal cross section, along a path that coincides with $C$, and such that the (absolute value of) the  field strength is constant along it. The contribution of loop 2 to $H$ is then $\Phi_ 2$ times the integral of $A_1$ along the loop, which with Stokes is the magnetic flux of loop 1 passing through loop 2. This yields one contribution $\Phi_1\Phi_2$ to the helicity. The same procedure with loop 1 adds another $\Phi_1\Phi_2$.

\bigskip 
{\bf 3.22. Stream function in a plane}.\\
Select a suitably representative field line $C$ and define a coordinate $s_\parr$ as the path length along this line. Divide the plane into curves that are everywhere perpendicular to ${\bf B}$, and define a second coordinate $s_\perp$ as the path length along these curves $D$, with zero point on $C$. The function ($\Psi(s_\parr,s_\perp)=\int_D B\, \rd s_\perp$  is then a stream function of the field, as can be seen by imagining a coordinate transformation that stretches the curvilinear system ($s_\parr,s_\perp$) to cartesian coordinates ($x,y$) aligned with the field.

\bigskip 
{\bf 3.23. Convective flux expulsion}.\\
With ${\bf v}$ purely azimuthal, steady, and depending on $r$ only, the induction equation yields $ \pa_t B_r=-v_\varphi\pa_\varphi B_r$. 
With  $\rd/\rd t\equiv\pa_t+v_\varphi\pa_\varphi$, this can be written as a conservation equation, $\rd B_r/\rd t=0$. At $t=0$, $B_r=B_0\cos(\varphi)$. The solution is then $ B_r=B_0\cos(\varphi-\alpha)$, where $\alpha=\Omega(r) t$ is the azimuthal displacement angle after time $t$. It changes by an amount $\Delta\alpha=\pi$ over a distance $l$ given by $t\,l\,\vert{\rd\Omega/\rd r}\vert=\pi$ or, for the given profile of $\Omega$:
\beq {l\over R}={\pi\over t\Omega_0}.\eeq
Over the length scale $l$ diffusion acts on a time scale $l^2/\eta$. When this time scale becomes shorter than the winding-up time scale $1/\Omega$, the approximate time dependence changes from winding-up to decay by canceling of nearby polarities,
\beq \pa_t B_r=-{\eta\over l^2} B_r = -k t^2 B_r,\eeq
with $k=\eta\Omega_0^2/(\pi R)^2$. The  time dependence in this diffusive stage is
\beq B_r\sim  e^{-\frac{1}{3}k t^3}. \eeq
The field in the differentially rotating cell disappears from its interior with this approximate time dependence. The field lines  initially present pass around it in a thin layer just inside $r=R$. \\
The assumption of a steady overturning flow is a strong limitation of the model, but the qualitative effect is also present in flows with shorter coherence time.

\bigskip 
{\bf 3.24. Torsional Alfv\'en waves}.\\
{\bf a}. Use the vector formulas in section \ref{idents}. The equations containing $v_\varphi$ and $\delta B_\varphi$ combine into a standard 2nd order wave equation with wave speed $v_{\rm A}$. \\
{\bf b}. $B_\varphi/B_z=-\Omega \varpi\,v_{\rm A}$. The torque is 
\beq 
T=\int_0^R \varpi {B_\varphi B_z\over 4\pi}\, 2\pi\varpi \,{\rd \varpi} =-{1\over 8} B_0^2R^3{\Omega R\over v_{\rm A}}.\eeq
[The minus sign indicates that this is the torque exerted by the magnetic field,  the opposite of the torque the plate exerts on the field.]
The work done by the plate per unit time is $W=-\Omega T$.

\bigskip 
{\bf 3.25. Currents in an Alfv\'en wave}.\\
In the interior there is a volume current flowing  along the axis, turning into a horizontal surface current along the wave front, turning into a surface current along the outer boundary, and closing through the rotating plate. [Or, depending on the signs of  $\Omega$ and the initial field $B_0$, the same pattern flowing in the opposite direction].

\bigskip 
{\bf 3.26. Magnetic Reynolds numbers in a star}.\\
Base convection zone: $R_{\rm m}\sim 3\,10^9$, $v_{\rm H}=j/(e\,n_{\rm e})\sim 10^{-7}$ cm/s, $R_{\rm H}\sim 10^{11}$. {\bf a} (sunspot): $R_{\rm m}\sim 2\, 10^6$, $v_{\rm H}\sim 3$  cm/s, $R_{\rm H}\sim10^5$. {\bf b} (corona):  $R_{\rm m}\sim 3\,10^9$, $v_{\rm H}\sim 100$  cm/s, $R_{\rm H}\sim 10^4$.

\bigskip 
{\bf 3.27. Poynting flux in an Alfv\'en wave}.\\
From (\ref{poymhd}), taking the $z$-axis along the tube, the components of the Poynting flux are $4\pi\,S_\vh=B_z^2v_\vh$, $4\pi\,S_z=-B_\vh B_z v_\vh$, where $v_\vh=\Omega\varpi$. Integrating $S_z$ over the area of the rotating plate the result is the same as the work done by the plate in problem 3.24. The azimuthal component describes an energy flux circulating around the axis, which normally will be of no consequence.

\bigskip 
{\bf 3.28 Electric field of a current wire}.\\
The electric field is absent in the lab frame, but  because of the current the wire also has a magnetic field around it. In the moving frame, there is an electric field $-{\bf v\times B/c}$. Its divergence corresponds to the charge of the wire observed in this frame (cf.\ section \ref{chargedd}).

\bigskip 
{\bf 3.29. Ambipolar drift}.\\
Base of the convection zone: $v_{\rm a}\sim 10^{-11} $ cm/s. Spot: $\sim 0.1$ cm/s. Corona: the expression for $v_{\rm a}$ would yield $\sim 10^{22} $ cm/s. This means that for the Lorentz force to be balanced by friction with the neutrals, an unrealistically high velocity difference would be needed. This shows that ambipolar friction is negligible under coronal conditions. In its absence the magnetic field there must be close to force-free, even though the plasma-$\beta$ is not all that small. (Exceptions occur in high-density inclusions in the corona called prominences, where magnetic curvature suspends plasma against gravity).

\bigskip 
{\bf 3.30. Conformal mapping of a potential field}.\\
The picture for $f=z^2$ in \url{http://mathworld.wolfram.com/ConformalMapping.html} shows how it can be used for the disk edge problem. A uniform vertical field ${\bf B}= \hat{\bf y}$ in the source plane has a stream function $\psi$ such that  ${\bf B}=(0,\pa_x\psi)$, i.e.\ $\psi=x$. It can be used to label field lines. In the image plane, a footpoint on the $x-$axis is moved from $x_{\rm s}$ to $x=x_{\rm s}^2$, so the value of $\psi$ at some $x$ in the image plane is $x^{1/2}$, and the field strength in the image plane is $\pa\psi/\pa x=x^{-1/2}/2$.

\end{document}